\documentclass[twoside,12pt]{article}

\usepackage{epsfig,latexsym}
\usepackage{amsmath}
\usepackage{color}
\usepackage[
   linktocpage=true
  ,pdfproducer=medialab
  ,pdfa=true
]{hyperref}

\let\emph\textit

\usepackage{epsfig,latexsym}
\usepackage{amsmath}
\usepackage{color}
\usepackage{hyperref}

\let\emph\textit

\def\bfnabla{\mbox{\boldmath $\nabla$}}

\def\bfsigma{\mbox{\boldmath $\sigma$}}
\def\bftau{\mbox{\boldmath $\tau$}}

\def\al{\alpha}
\def\lQ{\Lambda_{QCD}}
\newcommand{\nn}{\nonumber}
\newcommand{\be}{\begin{equation}}
\newcommand{\ee}{\end{equation}}
\newcommand{\bea}{\begin{eqnarray}}
\newcommand{\eea}{\end{eqnarray}}
\def\al{\alpha}

\def\siml{{\
    \lower-1.2pt\vbox{\hbox{\rlap{$<$}\lower6pt\vbox{\hbox{$\sim$}}}}\ }} 
\def\simg{{\ \lower-1.2pt\vbox{\hbox{\rlap{$>$}\lower6pt\vbox{\hbox{$\sim$}}}}\ }}

\newcommand{\MS}{\overline{\rm MS}}

\def\dsl{\,\raise.15ex\hbox{/}\mkern-13.5mu D}
\def\dse{\,\raise.12ex\hbox{/}\mkern-10.5mu e}
\def\dsv{\,\raise.12ex\hbox{/}\mkern-10.5mu v}

\def\lQ{\Lambda_{\rm QCD}}

\newcommand{\beqa}{\begin{eqnarray}}
\newcommand{\eeqa}{\end{eqnarray}}

\newcommand{\eq}[1]{Eq.~\eqref{#1}}

\newcommand{\Sec}[1]{Sec.~\ref{#1}}

\def\bdm{\begin{displaymath}}
\def\edm{\end{displaymath}}

\numberwithin{equation}{section}

\begin{document}

\title{ \vspace{1cm} The proton radius (puzzle?) and its relatives}
\author{Clara Peset$^{(a)}$, Antonio Pineda$^{(b,c)}$ and Oleksandr Tomalak$^{(d,e)}$\\
\\
${}^{(a)}$
Technische Universit\"at M\"unchen, 
Physik Department, \\
James-Franck-Strasse 1, 85748 Garching, Germany
\\
${}^{(b)}$ Institut de F\'\i sica d'Altes Energies (IFAE), \\
The Barcelona Institute of Science and Technology, \\
Campus UAB, 08193 Bellaterra (Barcelona), Spain\\
${}^{(c)}$ Grup de F\'{\i}sica Te\`orica, Dept. F\'\i sica, \\
Universitat Aut\`onoma de Barcelona,
E-08193 Bellaterra, Barcelona, Spain\\
$^{(d)}$
Department of Physics and Astronomy, University of Kentucky, \\
Lexington, KY 40506, USA
\\
$^{(e)}$
Theoretical Physics Department, Fermilab, Batavia, IL 60510, USA}

\maketitle
\begin{abstract} 
We review determinations of the electric proton charge radius from a diverse set of low-energy observables. We explore under which conditions it can be related to Wilson coefficients of appropriate effective field theories. This discussion is generalized to other low-energy constants. This provides us with a unified framework to deal with a set of low-energy constants of the proton associated with its electromagnetic interactions. Unambiguous definitions of these objects are given, as well as their relation with expectation values of QCD operators. We show that the proton radius obtained from spectroscopy and lepton-proton scattering (when both the lepton and proton move with nonrelativistic velocities) is related to the same object of the underlying field theory with ${\cal O}(\alpha)$ precision. The model dependence of these analyses is discussed. The prospects of constructing effective field theories valid for the kinematic configuration of present, or near-future, lepton-proton scattering  experiments are explored.
\end{abstract}
\hspace{-1.cm}\rightline{FERMILAB-PUB-21-254-T; TUM-HEP 1340/21}

\eject
\tableofcontents
\vfill
\newpage

\begin{flushright}
{\bf GENESIS 1. The beginning}\\
In the beginning God created quarks,
\\
And made them interact through the strong forces, 
\\
And it was dark ...\\
And God said, ``I do not understand a damn thing'',\\ 
And so he said ``Let there be photons'',\\
 And there was light ...
\end{flushright}
\section{Introduction}

The determination of the electric proton charge radius from the measurement of the Lamb shift of muonic hydrogen by the CREMA collaboration \cite{Pohl:2010zza,Antognini:1900ns} with unprecedented accuracy, and its discrepancy with the, until then, accepted value of the proton radius, obtained as a weighted average of measurements from electron-proton scattering and the regular hydrogen Lamb shift \cite{Mohr:2008fa} (up to some exceptions \cite{Belushkin:2006qa}) produced a shock in the scientific community, shaking the, then accepted, methods and, above all, error analyses of specialized determinations of the proton radius and related quantities. Different branches in high-energy, hadron, nuclear and atomic physics, associated with the physics of the different experiments used for these determinations, turned their attention to this problem producing a flurry of activity. The possibility that the discrepancy was coming from new physics effects, like those that break lepton universality, was a powerful motivation for these studies. Here, we review some of this research. Some earlier reviews on the proton radius puzzle can be found in \cite{Pohl:2013yb,Carlson:2015jba}. In this review we put emphasis in posing the problem in an effective field theory (EFT) context. This allows us to describe the different experiments that yield determinations of the proton radius in an equal footing. In particular, we will make explicit the theoretical expressions that guarantee that the same definition of the proton radius is used for the different observables (with relative ${\cal O}(\alpha)$ precision). This connection is achieved for the different spectroscopy and lepton-proton scattering experiments, but for the latter only in a very specific kinematic region. Nevertheless, we also discuss how to construct EFTs for lepton-proton scattering experiments in an extended kinematics, which still guarantees that the very same proton radius is measured. 

The use of EFTs also allows us to relate the determination of the proton radius with the determinations/definitions of other low-energy observables, providing a unified framework for dealing with low-energy constants that involve a single proton. These low-energy constants can also be related to form factors and structure functions (via expectation values of QCD operators). Expressions showing these relations are also displayed in this review.   
 The fact that they can be understood as Wilson coefficients of an effective theory makes explicit that the form factors should be computed with an infrared cutoff. This infrared cutoff is the ultraviolet cutoff of the effective theory. In some cases, like for the anomalous magnetic moment, this cutoff can be sent smoothly to zero and the Wilson coefficient can be identified with a low-energy constant. In other cases, the cutoff produces logarithmic divergences. Therefore, only the combination of the Wilson coefficient with some Feynman diagrams of the effective theory yields a finite result. This is indeed the case of the proton radius. We will call such Wilson coefficients quasi-low-energy constants. 

The experiments we mainly consider in this review are the elastic electron-proton scattering, the elastic muon-proton scattering, the Lamb shift of regular hydrogen, and the Lamb shift of muonic hydrogen. For these four experiments, we will consider that the transfer momentum between the lepton (either electron or muon) and the proton, $Q^2$, is much smaller than the mass of the pion: $Q^2 \ll m_{\pi}^2$. This will imply that all hadronic effects can be encoded in low-energy constants, which, in the language of EFTs, correspond to Wilson coefficients of the Lagrangian of the effective theory. In order to have a single EFT describing the nonrelativistic bound state and the lepton-proton scattering, we also need the energy of the incoming lepton, $E$, to be nonrelativistic. In this situation, we can guarantee that we are using the same effective theory in all these experiments (or we can connect them by matching) and, thus, the same Wilson coefficients, in particular the same proton radius. 
This situation is clearly fulfilled for the Lamb shift, as we have that $Q \sim m_e \alpha$ and $E-m_e \sim m_e \alpha^2$ for hydrogen, and $Q \sim m_{\mu} \alpha$ and $E -m_{\mu} \sim m_{\mu} \alpha^2$ for muonic hydrogen. For elastic electron-proton or muon-proton scattering, $Q^2$ can be arbitrarily large\footnote{$Q^2$ is bounded from above since it is restricted to be in the interval $ 0 \leq Q^2 \leq 4 M (E^2-m^2)/(M+2E)$. Nevertheless, it can be made arbitrarily large by changing $E$.}  or small. Therefore, we restrict the study of these experiments to kinematic configurations such that $Q^2 \ll m_{\pi}^2$. In order to have the complete connection with spectroscopy, we also require $E -m_{l_i} \ll m_{l_i}$, where $m_{l_i}$ is the mass of the lepton (either muon or electron). In realistic kinematics for present lepton-proton scattering experiments, the lepton is relativistic and the effective theory should be modified. We discuss this further in the main body of the review. 

Another issue raised by the high-precision measurement of the proton radius by the CREMA collaboration was the necessity to fix what had actually been measured in those different experiments. In other words, what the definition of the proton radius was. At leading order in $\alpha$, the electromagnetic coupling, it was known that the definition of the proton radius was $G_E'(0)$, the derivative of the electric Sachs form factor at zero momentum (see for instance the classical reviews ~\cite{Eides:2000xc,Eides:2007exa}). Note, however, this does not mean that the Fourier transform of the Sachs form factor can be interpreted as a density probability, as emphasized in \cite{Miller:2018ybm}. Irrespectively of this discussion, the definition of the proton radius is ambiguous once electromagnetic corrections are incorporated in the observables. The reason is that $G_E'(0)$ is infrared divergent once electromagnetic corrections are included. Actually, this issue had already been discussed in the context of the (muonic) hydrogen Lamb shift \cite{Pachucki:1995cjo,Pineda:2004mx} before the advent of this measurement, but it was this very precise measurement that transformed this issue into an urgent question to be elucidated, as the precision was high enough to discriminate among possible different definitions. Indeed, one point that is often raised, and it is still open to some discussion, is whether the proton radius measured in Lamb shift is the same as the proton radius measured in electron-proton scattering once electromagnetic corrections are incorporated. We clarify this issue in this review. Finally, we also remark that this is a general problem for several (quasi-)low-energy constants\footnote{As stated before, we use the name of quasi-low-energy constants for Wilson coefficients that are logarithmically infrared divergent.} like the electric and magnetic polarizabilities. In any case, for the other quasi-low-energy constants, the present precision is not high enough to require a quantitative study of these effects. 

At this stage, it is worth enumerating the different scales that are at hand  (and the ratios of them that can be generated) for these observables. In the $ep$ and $\mu p$ systems, we are basically testing the proton with
different point-like probes ($e$, $\mu$, $\gamma$) and several different scales are involved in their
dynamics. For the $e p$ system, they are ($M$ is the proton mass): $\dots$, $m_{e}\al^2$ , $m_{e}\al$,
$m_{e}$, $\Delta M=M_n-M$, $m_{\mu}$, $\Delta=M_{\Delta}-M$, $m_{\pi}$,
$M$, $m_\rho$, $\Lambda_{\chi}$, $\dots$, which we group and name in the following way:
\begin{itemize}
\item
$m_{e}\al^2$: ultrasoft (US) scale.
\item
$m_{e}\al$: soft scale.
\item
$m_r^{(e)}= \tfrac{M m_{e}}{ M + m_{e} }$, $\Delta M=M_n-M$, $m_e$:
hard scale. 
\item
$m_{\mu}$, $\Delta=M_{\Delta}-M$, $m_{\pi}$: pion scale.
\item
$M$, $m_\rho$, $\Lambda_{\chi}$: chiral scale.
\end{itemize}

For the $\mu p$ system, they are: $\dots$, $m_{\mu}\al^2$ , $m_{\mu}\al$,
$m_{\mu}$, $\Delta M=M_n-M$, $m_e$, $\Delta=M_{\Delta}-M$,
$m_{\pi}$, $M$, $m_\rho$, $\Lambda_{\chi}$, $\dots$, which we group and name in the following way:
\begin{itemize}
\item
$m_{\mu}\al^2$: US scale.
\item
$\Delta M=M_n-M$, $m_e$, $m_{\mu}\al$: soft scale.
\item
$m_r^{(\mu)}= \tfrac{M m_{\mu}}{ M + m_{\mu} }$, $m_{\mu}$,
$\Delta=M_{\Delta}-M$, $m_{\pi}$: hard/pion scale.
\item
$M$, $m_\rho$, $\Lambda_{\chi}$: chiral scale.
\end{itemize}

Besides all these scales, we have the parameter $Q$, the transferred momentum between the lepton and the proton. For the hydrogen and muonic hydrogen, $Q \sim$ soft scale, whereas for the elastic scattering, we will set $Q$ to fulfill $ Q^2 \ll m_{\mu}^2$ but will otherwise let it be free. 

On top of that, for the case of the lepton-proton scattering, we have to consider the energy of the incoming lepton, $E$, as another free variable. For spectroscopy, the bound-state dynamics fixes $E -m_{l_i} \sim m_{l_i} \alpha^2$, which ensures that the lepton is nonrelativistic. Nevertheless, for lepton-proton scattering, $E$ is not fixed a priori and could be very large. Actually, for nowadays experiments, it is large. 

By doing ratios of the different scales, several small expansion
parameters can be built. Basically, this will mean that the
observables, can be written, up to large
logarithms, as an expansion, in the case of the $ep$, in $\al$, $\tfrac{m_e}{
m_{\pi}}$ and ${m_{\pi} \over M}$, and in the case of the $\mu p$,
in $\al$ and ${m_\mu \over M}$. For the elastic scattering, we will also have the extra ratio: ${Q \over m_{\pi}}$. In some cases, it will also prove convenient to use the reduced mass $m_r^{(\mu/e)}$,\footnote{To avoid producing cumbersome notation, we will just write $m_r$ for the reduced mass, following from the context if we refer to $m_r^{(e)}$ or $m_r^{(\mu)}$.} since it will
allow to keep (some of) the exact mass dependence at each order in
$\al$. 

The main purpose of this review is the determination of the proton radius (though we will also discuss other low-energy constants), which is a quasi-low-energy constant. Therefore, ideally, we should take $Q^2 \rightarrow 0$, or approach this limit as much as possible. According to the typical values of $Q$ mentioned above, the determination of the proton radius from the measurement of the Lamb shift of regular hydrogen would be ideal. Actually, the theoretical expression of the Lamb shift of regular hydrogen has been computed to very high orders (for some reviews see \cite{Eides:2000xc,Eides:2007exa,Yerokhin:2018gna,Karshenboim:2019iuq}). On the other hand, the smaller the value of $Q$, the most precise has to be the measurement, and the theoretical prediction, to determine the slope of the Sachs form factor. At present, the experimental precision of Lamb shift of regular hydrogen is not high enough and the muonic hydrogen represents, at present, the place on which the precision of theory and experiment are optimal. 

The electromagnetic proton radius, $r_p^2$, is a hadronic quantity. Chiral loops give contributions to $r_p^2$ that scale as $\sim 1/\Lambda_{\chi}^2$, up to logarithmically enhanced contributions.  Nevertheless, actual measurements give that the size of the inverse proton radius is of the order of (or slightly smaller than) twice the pion mass. The theoretical expressions that we will use in the determination of the electromagnetic proton radius will have ${\cal O}(\alpha)$ relative accuracy.\footnote{Some logarithmically enhanced ${\cal O}(\alpha^2)$ effects will also be considered.} Nevertheless, with this precision, other hadronic quantities will appear. In principle, these have to be determined too. In practice, the only one that may cause problems is what is called the two-photon exchange (TPE) correction. This correction is proportional to the mass of the lepton. For the case of the electron-proton sector, this introduces an extra suppression factor of order $m_e/m_{\pi}$ that makes such contribution subleading. For the muon-proton sector, there is no such suppression, since $m_{\mu}/m_{\pi} \sim 1$, and such hadronic effect has to be carefully determined, at least with a precision of order ${\cal O}(\al \times \frac{1}{r_p^2\Lambda_{\chi}^2})$. In practice, the accuracy achieved for this quantity fixes the accuracy one achieves in the determination of the proton radius from the muonic hydrogen Lamb shift. We devote some time to study these TPE effects. We also do so for its spin-dependent counterpart, which can be determined from measurements of the hyperfine splitting of regular hydrogen and muonic hydrogen.  

The structure of the review will be as follows. We first discuss the EFTs that describe the observables we use to determine the proton radius. We then discuss the relation of the Wilson coefficients of the effective theory with the proton radius and other low-energy constants, as well as their relation with form factors. We also give the expression for the TPE contribution in terms of structure functions, as well as in terms of dispersion relations.  We then write and discuss the theoretical expressions for the different observables we consider in this review. Afterwards, we review determinations of the proton radius, as well as of some other low-energy constants, including the TPE corrections. Finally, we conclude and summarize the situation of the proton radius puzzle. In the appendix, we give some details of the computation of the soft-photon emission in dimensional regularization. 

\section{Effective Field Theories}

We display the EFTs suitable for the description of the Lamb shift in regular hydrogen and muonic hydrogen. We also show that these EFTs apply to the description of the muon-proton and electron-proton elastic scattering in some specific kinematic region. For the muon-proton sector, the EFT is characterized by being in the kinematic regime with $Q^2 \ll m_{\mu}^2$ and $E - m_{\mu} \ll Q$, where $E$ is the energy of the incoming muon. For the electron-proton sector, the EFT is characterized by being in the kinematic regime with $Q^2 \ll m_{e}^2$ and $E-m_ e  \ll Q$, where $E$ is the energy of the incoming electron. 
We then discuss the relation of the Wilson coefficients of the effective theory with the proton radius and other low-energy constants, as well as with the form factors. We also give the expression for the TPE contributions in terms of structure functions, as well as in terms of dispersion relations. Finally, we discuss how the EFTs should be changed to accommodate different kinematic regimes more relevant for, nowadays or near-to-come, lepton-proton elastic scattering experiments.  

\subsection{NRQED(\texorpdfstring{$\mu p$}{mup})}
\label{NRQED(mu)}

In the muon-proton sector, by integrating out the scale $m_\pi \sim m_{\mu}$ of HBET (see Sec.~\ref{Sec:HBET}), an
EFT for nonrelativistic muons and protons, relativistic electrons and photons appears. In
principle, we should also consider neutrons but they play no role at
the precision we aim for. The effective theory is nothing but NRQED \cite{Caswell:1985ui} applied to this matter sector, as discussed in \cite{Pineda:2002as,Pineda:2004mx,Peset:2015zga}. 
It has a hard
cut-off $\nu \ll m_\pi$ and therefore pion, Delta and higher resonances have been
integrated out. The effective Lagrangian reads
\be
\label{eq:NRQEDmu}
{\cal L}_{\rm NRQED(\mu)}=
{\cal L}_{\gamma}
+
{\cal L}_{e}
+
{\cal L}^{(\rm NR)}_{\mu}
+
{\cal L}_{N}
+
{\cal L}_{Ne}
+
{\cal L}_{N\mu}^{(\rm NR)}
\,.
\ee
The pure photon sector is approximated by the following Lagrangian ($i D_\nu=i\partial_\nu-eA_\nu$)
\be
\label{Lg}
{\cal L}_\gamma=-\frac{1}{4}F^{\mu\nu}F_{\mu \nu} 
+
\left(\frac{d_{2}^{(\mu)} }{ m_{\mu}^2}+\frac{d_{2}}{ M^2}+\frac{d_{2}^{(\tau)} }{ m_{\tau}^2} 
\right)
F_{\mu \nu} D^2 F^{\mu \nu}
\,,
\ee
$d_{2}^{(\mu)}$ and $d_{2}^{(\tau)}$ are generated by the vacuum polarization loops with only muons and taus respectively. 
At ${\cal O}(\al)$ they read
\be
 d_{2}^{(\mu)}=\frac{\alpha}{ {60\pi}}+{\cal O}(\al^2) \,,
 \qquad 
 d_{2}^{(\tau)}=\frac{\alpha}{ {60\pi}}+{\cal O}(\al^2)
 \label{d2}
\,.
\ee
 
The hadronic effects of the vacuum polarization are encoded in $d_{2}$ (where $Z$ is the charge of the nucleus, with  $Z=1$ for the proton):
\be
d_{2}=\frac{M^2 }{4}\Pi^\prime_{h}(0)=\frac{Z^2\al}{60\pi}+d^{\rm had}_2+{\cal O}(\al^2)
\,.\label{d2p}
\ee
$\Pi^\prime_{h}(0)$ is the derivative of the hadronic vacuum polarization
(we have defined $\Pi_{h}(-{\bf k}^2)=-{\bf k}^2\Pi^\prime_{h}(0)+\,.\,.\,.$). The 
experimental figure for the total hadronic contribution reads
$\Pi_{h}^{\prime} \simeq 9.3 \times \, 10^{-3} \, {\rm GeV}^{-2}$ 
\cite{Jegerlehner:1996ab}. 
Following standard practice, we have singled out the contribution due to the loops of protons (assuming them to be point-like) in the second equality of Eq.~(\ref{d2p}). Note though that $d_2^{\rm had}$ is still of order $\al$.

The electron sector reads 
\be
\label{Le}
{\cal L}_e= \bar l_e  (i\dsl-m_e) l_e
\,.
\ee
We do not include the term 
\be
-\frac{e g_{l_e}
}{ m_{\mu}}\bar l_e \sigma_{\mu\nu}l_eF^{\mu\nu}
\,,
\ee
since the coefficient $g_{l_e}$ is suppressed by powers of $\al$ and 
the mass of the lepton. Therefore, it would give contributions beyond the
accuracy we aim for. In any case, any eventual contribution would be
absorbed in a low-energy constant.

The muonic sector reads
\bea
{\cal L}^{(\rm NR)}_{\mu}&=& 
l^\dagger_{\mu} \Biggl\{ i D_{\mu}^0 + \, \frac{{\bf D}_{\mu}^2}{ 2 m_{\mu}} +
 \frac{{\bf D}_{\mu}^4}{
8 m^3_{\mu}} + e \frac{c_F^{(\mu)} }{ 2m_\mu}\, {\bf \bfsigma \cdot B}\nn
 \\
 &&
 + 
e\frac {c_D^{(\mu)} }{ 8m_{\mu}^2}
   \left[{\bf \bfnabla \cdot E }\right]
   + 
 i e \frac{c_S^{(\mu)} }{ 8 m^2_{\mu}}\,  {\bf \bfsigma \cdot \left(D_{\mu} \times
     E -E \times D_{\mu}\right) } 
\Biggr\} l_{\mu},
\label{LNRmu}
\eea
with the following definitions: $ i D^0_\mu=i\partial_0 - eA^0$ and
$i{\bf D}_\mu=i{\bfnabla}+ e{\bf A}$. The Wilson
coefficients $c_X^{(\mu)}$ can be computed order by order in $\al$. They read (where we have used the fact that  
$c_S^{(\mu)}=2c_F^{(\mu)}-1$ \cite{Manohar:1997qy})
\begin{eqnarray}
c_F^{(\mu)}&=&1+\frac{\al}{2\pi} +{\cal O}(\al^2) \,,
\label{cfi}\\
c_S^{(\mu)}&=&1+\frac{\al}{\pi} +{\cal O}(\al^2) \label{csi}\,.
\end{eqnarray}

Taking the values of the form factors for the muon-electron difference computed in \cite{Barbieri:1973kk} and those for the electron computed in \cite{Barbieri:1972as}, we can deduce the following expression for the 
$c_{D,\MS}^{(\mu)}(\nu)$ Wilson coefficient:\footnote{In NRQED($\mu p$), the electron has not been integrated out. Therefore, 
Eq.~(\ref{cDmu}) is not the $c_D^{(\mu)}$ Wilson coefficient of NRQED($\mu p$). 
Eq.~(\ref{cDmu}) will show up after lowering the muon energy cut-off below the electron mass in pNRQED (to be defined later, see Sec.~\ref{Sec:pNRQED}). Still, we choose to present it here as, otherwise, we would be forced to do an extra intermediate 
matching computation that would unnecessarily complicate the derivation of the final result. Since we have integrated out the 
electron, note also that $\al=1/137.14...$ in this equation, i.e., any running associated with the electron is written explicitly in Eq.~(\ref{cDmu}).}
\bea
\label{cDmu}
c_{D,\MS}^{(\mu)}(\nu)&=&
1+\frac{4\alpha}{3\pi}\ln\left(\frac{m_\mu^2}{\nu^2}\right)
\\
&&+\left(\frac{\alpha}{\pi}\right)^2
\left\{
\frac{8}{9} \ln^2\left(\frac{m_\mu}{m_e}\right)
-\frac{40}{27} \ln\left(\frac{m_\mu}{m_e}\right)+\frac{85}{81}+ \frac{4\pi ^2}{27}
\right.
\nn\\
&&
\left.
+
\left[\frac{\pi ^2 }{6}\left(18 \ln 2-\frac{40}{9}\right)
-\frac{1523}{324}
-\frac{9 }{2}\zeta(3)
\right]
+{\cal O}\left(\frac{m_e}{m_{\mu}}\right)\right\}
\nn\\
\nn
&&+{\cal O}\left(\al^3\right)
\,.
\eea

For the determination of the Lamb shift with ${\cal O}(m_{\mu}\al^6)\times$ large logarithms accuracy (where the large logarithms are generated by the ratios of different scales), we only need $c_D^{(\mu)}$ with ${\cal O}(\al^2)\times $ large logarithm accuracy. We also include the finite piece for completeness but neglect ${\cal O}(m_e/m_{\mu})$ terms.
 Note that analogous ${\cal O}(\al^2)$ terms (changing $m_{\mu}$ by $M$ and either keeping 
$m_e$ or changing it by $m_{\mu}$) would exist for $c_D^{(p)}$ if computing the Wilson coefficient as if the proton were point-like at the scale $M$. Even 
if these effects are small, they should be taken into account for eventual comparisons with lattice simulations where, typically, only the hadronic correction is computed.

For the proton sector, we have 
\bea
\label{LNdelta}
 {\cal L}_{N}&=& N^\dagger_{p} \Biggl\{iD_0+ \frac{{\bf
D}^2_p}{2 M} + 
\frac {{\bf D}_p^4}{
8 M^3} - e \frac{c_F^{(p)} }{ 2M}\, {\bf \bfsigma \cdot B}\nn
\\
&& 
-e\frac{c_D^{(p)} }{ 8M^2}
   \left[{\bf \bfnabla \cdot E }\right] 
 - ie \frac{ c_S^{(p)} }{8M^2}\, \bfsigma \cdot \left({\bf D}_p
     \times {\bf E} -{\bf E}    \times {\bf D}_p\right) 
\Biggr\} N_{p}
\,,
\eea
where $ i D^0_p=i\partial_0 +ZeA^0$, $i{\bf D}_p=i{\bfnabla}-Ze{\bf A}$ and for the proton $Z=1$. The Wilson coefficients $c_F$, $c_D$, $\cdots$ are hadronic, non-perturbative quantities. 
In some cases, they can be directly related to low-energy constants, for instance with the anomalous magnetic moment of the proton, $\kappa_p=1.792847356(23)$ \cite{Agashe:2014kda}, but not in other cases, like the proton radius. We ellaborate on this discussion in Sec.~\ref{Sec:formfactors}. Let us note that, with the conventions above, $N_p$ is the field of the proton 
(understood as a particle) with
positive charge if $l_i$ represents the leptons (understood as
particles) with negative charge.
 
 ${\cal L}_{Ne}$ refers to the four-fermion operator made of nucleons and
 (massless) electrons. It does not contribute to the muonic hydrogen spectrum at 
 ${\cal O}(m_r\al^5)$, nor to elastic muon-proton scattering with the required accuracy. Still, we write it for completeness:
\be
\label{LNe}
\delta {\cal L}_{Ne}=\displaystyle\frac{1}{M^2} c_{3,R}^{pl_e}
{\bar N}_p \gamma^0
  N_p \ \bar{l}_e\gamma_0 l_e
+\displaystyle\frac{1}{M^2} c_{4,R}^{pl_e}{\bar N}_p \gamma^j \gamma_5
N_p 
  \ \bar{l}_e\gamma_j \gamma_5 l_e
\,.
\ee

Finally, we consider the four-fermion operators:\footnote{The coefficients $c_{3}$ 
and $c_{4}$ should actually read $c_{3}^{pl_\mu}$ and $c_{4}^{pl_\mu}$, as they actually depend on the
nucleon and lepton the four-fermion operator is made of. Nevertheless, to ease the notation we eliminate those 
indices when it can be deduced from the context.}
\be
\label{4F}
{\cal L}_{N\mu}^{\rm NR}=
\displaystyle\frac{c_{3}}{M^2} N_p^{\dagger}
  N_p \ {l}^{\dagger}_\mu l_\mu
-\displaystyle\frac{c_{4}}{M^2} N_p^{\dagger}{\bfsigma}
  N_p \ {l}^{\dagger}_\mu{\bfsigma} l_\mu
\,.
\ee
The discussion of $c_3$ and $c_4$ is postponed to Sec.~\ref{Sec:Structurefunctions}.

This EFT (and consequently the very same Wilson coefficients) can also be applied to the elastic muon-proton scattering in the kinematic situation with $(E-m_{\mu})^2 \,, Q^2 \ll m_{\mu}^2$.

\subsection{NRQED(\texorpdfstring{$e p$}{ep})}
\label{NRQED(e)}

The effective Lagrangian relevant for the hydrogen Lamb shift reads
\be
{\cal L}_{\rm NRQED(e)}=
{\cal L}_{\gamma}
+
{\cal L}^{(\rm NR)}_{e}
+
{\cal L}_{N}
+
{\cal L}_{N e}^{(\rm NR)}
\,.
\ee
This effective theory is nothing but NRQED \cite{Caswell:1985ui} applied to this matter sector, as discussed in \cite{Pineda:2002as,Pineda:2004mx,Peset:2015zga}. 
It has a hard
cut-off $\nu \ll m_e$. 

The different terms of the Lagrangian have the same form as in the previous section but changing the Wilson coefficients. 
${\cal L}_{\gamma}$ is as in the previous section (Eq.~(\ref{Lg})) but adding the electron vacuum polarization correction.   ${\cal L}^{(\rm NR)}_{e}$ is as ${\cal L}^{(\rm NR)}_{\mu}$ in the previous section (Eq.~(\ref{LNRmu})) changing the muon by the electron. 
For the hydrogen Lamb shift, we need more precision than for the muonic hydrogen Lamb shift. This means that more terms in the $1/m_e$ expansion of the Lagrangian density than in the analogous $\mu p$ case need to be considered. Nowadays, the NRQED Lagrangian for the electron-proton sector is known to ${\cal O}(1/m^4)$:  \cite{Caswell:1985ui,Manohar:1997qy,Hill:2012rh}. We refer to the last reference for the explicit expressions. Also the Wilson coefficients change. The three-loop expression for $c_F^{(e)}$ was computed in \cite{Laporta:1996mq}:
\bea
&&
c_F^{(e)}=1+\frac{\alpha}{\pi}\frac{1}{2}+\frac{\alpha^2}{\pi^2}\left(\frac{3}{4}\zeta(3)-\frac{\pi^2}{2}\ln 2+\frac{\pi^2}{12}+\frac{197}{144}\right)
\nn
\\
\nn
&&
+\frac{\alpha^3}{\pi^3}
\left(
\frac{83}{72}\pi^2\zeta(3)-\frac{215}{24}\zeta(5)+\frac{100}{3}\left(a_4+\frac{\ln^2 2}{24}\left(\ln^2 2-\pi^2\right)\right) -\frac{239}{2160}\pi^4+\frac{139}{18}\zeta(3)-\frac{298}{9}\pi^2\ln 2
\right.
\\
&&
\left.
+ \frac{17101}{810}\pi^2+\frac{28259}{5184}
\right)
\,,
\label{cFe}
\eea
where $\displaystyle{a_4=\sum_{n=1}^{\infty}\frac{1}{2^n n^4}}$. Combining this result with the three-loop expression of the derivative of $F_1'$ at zero momentum \cite{Melnikov:1999xp}, we can deduce $c_D^{(e)}$, which reads
\bea
\nn
&&
c_{D,\MS}^{(e)}(\nu)=
1+\frac{4\alpha}{3\pi}\ln\left(\frac{m_e^2}{\nu^2}\right)
\\
&&+\frac{\alpha^2}{\pi^2}\left(-\frac{9}{2}\zeta(3)+3 \pi^2\ln 2-\pi^2\frac{20}{27}-\frac{1523}{324}\right)
\nn
\\
\nn
&&
+\frac{\alpha^3}{\pi^3}
\left(\frac{85}{12}\zeta(5)-
\frac{121}{36}\pi^2\zeta(3)-\frac{791}{12}\zeta(3)-a_4\frac{1136}{9}-\ln^4 2 \frac{142}{27}
-\pi^2 \ln^2 2 \frac{478}{135}+\pi^2 \ln 2 \frac{23791}{270}
\right.
\\
&&
\left.
+\pi^4\frac{1591}{1620}- \frac{249767}{4860}\pi^2+\frac{88409}{11664}
\right)
\,.
\label{cDe}
\eea

The important point for us is that the Wilson coefficients of ${\cal L}_{N}$ do not change with respect to the muon-proton case, which is where we have the proton radius. 
${\cal L}_{N e}^{(\rm NR)}$ has the same form as ${\cal L}_{N \mu}^{(\rm NR)}$ (see Eq.~(\ref{4F})) in the previous section (replacing the muon by the electron) but the Wilson coefficients are different. The leading contribution to $c_3$ is proportional to the mass of the lepton (see the discussion in \cite{Pineda:2004mx} and \eq{c3} below). This makes this contribution to be suppressed by an extra $\frac{m_e}{m_{\pi}}$ factor for the Lamb shift of regular hydrogen. Finally, for completeness, we also give $d_2^{(e)}$ to three loops \cite{Baikov:1995ui}:
\be
\label{d2e}
d_2^{(e)}=\frac{\al}{60\pi}+\frac{\al^2}{\pi^2}\frac{41}{648}+
\frac{\al^3}{\pi^3}
\left(
\frac{8135}{36864}\zeta(3)+\frac{23}{1440}\pi^2-\frac{325805}{1492992}-\frac{1}{60}\pi^2\ln 2
\right)
\,.
\ee

This EFT can also be used for the description of the elastic scattering of the electron and proton in the kinematic 
condition $(E-m_e)^2\,, Q^2 \ll m_e^2$, and, therefore, the very same Wilson coefficients (in particular the proton radius and the TPE Wilson coefficient, $c_3$) appear.

\subsection{Relativistic muon with \texorpdfstring{$Q^2 \sim m_{\mu}^2 \sim m_{\pi}^2$}{Q}}
\label{Sec:HBET}
The kinematic constraints of Sec.~\ref{NRQED(mu)} can be relaxed to ease the connection with the kinematics of forthcoming experiments. Therefore, we consider the situation $Q^2 \sim m_{\mu}^2$ and $E^2\sim m_{\mu}^2$, where $E$ is the energy of the incoming electron. In this situation, we have $Q^2 \sim m_{\pi}^2$, and it is natural to consider HBET \cite{Jenkins:1990jv} as the effective theory to describe this kinematic regime. Indeed, this is a situation that could be realized in the MUSE experiment~\cite{Talukdar:2018hia,Talukdar:2019dko,Talukdar:2020aui}. It corresponds to a hard cut-off $\nu << M$,
$\Lambda_{\chi}$, and much larger than any other scale in the problem.

The HBET Lagrangian applied to this specific matter sector has been considered in Ref. \cite{Pineda:2002as}. 
The starting point is the SU(2) version of HBET coupled to leptons,
where the Delta is also kept as an explicit degree of freedom  (based on large $N_c$ arguments).  The degrees
of freedom of this theory are the proton, neutron and Delta, for which
the nonrelativistic approximation can be taken, pions and leptons
(muons and electrons), which will be taken relativistic, and photons.

The Lagrangian can be structured as follows
\be
\label{eq:HBET}
{\cal L}_{\rm HBET}=
{\cal L}_{\gamma}
+
{\cal L}_{l}
+
{\cal L}_{\pi}
+
{\cal L}_{l\pi}
+
{\cal L}_{N}
+
{\cal L}_{(N,\Delta)}
+
{\cal L}_{(N,\Delta)l}
+
{\cal L}_{(N,\Delta)\pi}
+
{\cal L}_{(N,\Delta)l\pi},
\ee
representing the different sectors of the theory. In particular, the
$\Delta$ stands for the spin-3/2 baryon multiplet (we also use
$\Delta=M_{\Delta}-M$, the specific meaning in each case should be
clear from the context).

The Lagrangian can be written as an expansion in $e$ and $1/M$ ($M$ is of the order of the scales that have been integrated out). Let us
consider the different pieces of the Lagrangian more in detail.

${\cal L}_\gamma$ has the same form as Eq.~(\ref{Lg}) but without including the vacuum polarization of the particles that are still relativistic in the theory. Similarly, the leptonic sector reads ($i D_\nu=i\partial_\nu-eA_\nu$)
\be
\label{Ll}
{\cal L}_l=\sum_i \bar l_i  (i\dsl-m_{l_i}) l_i
\,,
\ee
where now $i=e,\mu$. We do not include the term 
\be
-{e g_{l_i}
\over M}\bar l_i \sigma_{\mu\nu}l_iF^{\mu\nu}
\,,
\ee
since the coefficient $g_{l_i}$ is suppressed by powers of $\al$ and 
the mass of the lepton.  In any case, any eventual contribution would be
absorbed in a low-energy constant.

The pionic Lagrangian ${\cal L}_{\pi}$ is usually organized in the
chiral counting. For the chiral computations that appear in this review, the free pion propagator provides with the necessary precision.
Therefore, we only need the free-particle pionic Lagrangian:
\be
{\cal L}_\pi=(\partial_\mu \pi^+)(\partial^\mu \pi^-)-m_\pi^2\pi^+\pi^-
+{1 \over 2}(\partial_\mu \pi^0)(\partial^\mu \pi^0)-{1 \over
2}m_\pi^2\pi^0\pi^0 
\,.
\ee

The one-baryon Lagrangian ${\cal L}_{(N,\Delta)\pi}$ is needed at
${\cal O}(1/M^2)$.  Nevertheless, a closer inspection simplifies the
problem. A chiral loop produces a factor $1/(4\pi F_0)^2 \sim
1/M^2$.  Therefore, the pion-baryon interactions are only needed at
${\cal O}(m_\pi)$, the leading order, which is known
\cite{Jenkins:1990jv,Jenkins:1991es,Hemmert:1996xg}:\footnote{Actually, terms that go into the physical
mass of the proton and into the physical value of the anomalous
magnetic moment of the proton $\kappa_p=c_F^{(p)}-Z$ should also be
included (at least in the pure QED computations) and that will be 
assumed in what follows. For the computations reviewed in this paper, these effects would be 
formally subleading. In any case, their
role is just to bring the {\it bare} values of $M_0$ and $\kappa_p^0$ to
their physical values. Therefore, once the values of $M$ and $\kappa_p$
are measured by different experiments, they can be distinguished from
the effects explicitly considered in this review.} 
 \be
{\cal L}_{ (N,\Delta)\pi}= \bar N \left(i \Gamma_0+g_A u\cdot S\right)N+g_{\pi N\Delta}\left(\bar T_a^\mu w_\mu^a N+\mathrm{h.c.}\right),
\ee
where
\bea
U&=&u^2=e^{i {\bf {\tau\cdot \pi}}/F_{\pi}},\\
\Gamma_\mu&=&\frac{1}{2}\left\{u^\dagger\partial_\mu u+u\partial_\mu u^\dagger-i\frac{e}{2}A_\mu\left(u^\dagger\tau^3 u+u\tau^3 u^\dagger\right)\right\},\\
u_\mu&=&iu^\dagger\nabla_\mu U u^\dagger,\\
w_\mu^a&=&\frac{1}{2}\mathrm{Tr}[\tau^a u_\mu]=-\frac{1}{F_\pi}\partial_\mu \pi^a-\frac{e}{F_\pi}A_\mu \epsilon^{a3b}\pi^b+...
\,,
\eea
and $T_a^\mu $ is the Rarita-Schwinger spin-3/2 field and 
$S_{\mu}=\frac{i}{2} \gamma_5 \sigma_{\mu\nu}v^\nu$ is the spin operator 
(where we take $v_{\mu}=(1,{\bf 0})$).

Therefore, we only need the one-baryon Lagrangian ${\cal
L}_{N}$ at ${\cal O}(1/M^2)$ coupled to electromagnetism. This
would be a NRQED-like Lagrangian for the proton, neutron (of spin 1/2)
and the Delta (of spin 3/2). The neutron is actually not needed at
this stage and ${\cal
L}_{N}$ has the same form as Eq.~(\ref{LNdelta}). It is very important to emphasize that the hadronic Wilson coefficients of ${\cal L}_N$ do not yet incorporate effects associated with the pion and/or Delta particle. 

As for the Delta (of spin 3/2), it mixes with the nucleons at
${\cal O}(1/M)$ (${\cal O}(1/M^2)$ are not needed in our case). The only
relevant interaction in our case is the $p$-$\Delta^+$-$\gamma$ term,
which is encoded in 
\be
 {\cal L}_{(N,\Delta)}
= 
T^{\dagger}(i\partial_0-\Delta)T
+{eb_{1,F} \over 2M}
\left(
T^{\dagger}\bfsigma ^{(3/2)}_{(1/2)}\cdot {\bf
B}\,\bftau^{3(3/2)}_{(1/2)} N + h.c.
\right)
\,,
\ee  
where $T$ stands for the delta 3/2 isospin multiplet and $N$ for the nucleon 1/2
isospin multiplet. The transition spin/isospin matrix elements fulfill
(see \cite{Ericson:1988gk})
\be
\bfsigma^{i(1/2)}_{(3/2)}\bfsigma^{j(3/2)}_{(1/2)}
={1 \over 3}(2\delta^{ij}-i\epsilon^{ijk}\bfsigma^k),
\qquad
\bftau^{a(1/2)}_{(3/2)}\bftau^{b(3/2)}_{(1/2)}
={1 \over 3}(2\delta^{ab}-i\epsilon^{abc}\bftau^c).
\ee

The baryon-lepton Lagrangian provides new terms that are not usually considered 
in HBET. The relevant term in our case is the interaction between 
the leptons and the nucleons (actually only the proton):
\be
\label{LNl}
 {\cal L}_{(N,\Delta)l}=\displaystyle\frac{1}{M^2}\sum_i c_{3,R}^{pl_i}
{\bar N}_p \gamma^0
  N_p \ \bar{l}_i\gamma_0 l_i
+\displaystyle\frac{1}{M^2}\sum_i c_{4,R}^{pl_i}{\bar N}_p \gamma^j \gamma_5
N_p 
  \ \bar{l}_i\gamma_j \gamma_5 l_i
\,.
\ee
The above matching coefficients fulfill 
$c_{3,R}^{pl_i}=c_{3,R}^{p}$ and $c_{4,R}^{pl_i}=c_{4,R}^{p}$ up to terms
suppressed by $m_{l_i}/M$. Note that these Wilson coefficients are different from those that appear in Eq.~(\ref{4F}), since dynamical effects associated with the pion and Delta particle are not incorporated in $c_{3,R}$ and $c_{4,R}$. We do not consider possible extra dimension six four-fermion operators because they are suppressed by an extra factor of $m_{l_i}/\Lambda_{\chi}$ due to chiral symmetry.

Finally, the remaining terms in the HBET Lagrangian in Eq.~(\ref{eq:HBET}) can be neglected for the purposes of this review. 

\subsection{Relativistic muon with \texorpdfstring{$Q^2 \ll m_{\mu}^2$}{Q}}
\label{Sec:Remuon}
An intermediate situation between the one discussed in the previous section and the one discussed in Sec.~\ref{NRQED(mu)} is to set $Q^2 \ll m_{\mu}^2$, $E^2\sim m_{\mu}^2$ and $(E-m_{\mu})^2\sim m_{\mu}^2$. This is a situation that could be realized for a certain range of parameters in the MUSE experiment \cite{Gilman:2017hdr}. This situation could still be described using HBET. Nevertheless, such effective theory does not profit from the kinematic constraint that $Q^2 \ll m_{\pi}^2$, since, in this situation, the pion and Delta could be integrated out, as they do not appear as asymptotic states. Nevertheless, the fact that $E$ and $E-m_{\mu}$ are of the same order as the pion mass complicates the construction of an efficient EFT. $E$ cannot be approximated by $m_{\mu}$ because $E-m_{\mu}$ is of the order of the muon mass. A natural way to deal with this situation with EFTs is to consider $E$ as a fixed scale, in the spirit of LEET \cite{Dugan:1990de} (or of its more modern SCET versions), and treat it at the same level as the muon mass or other scales that have been integrated out and are not dynamical anymore. Actually, one could also consider the situation when $E^2 \gg m_{\mu}^2$ (such kinematics cannot be described with the HBET presented in the previous section), which would also naturally be described by a LEET-like effective theory. This kinematics will be realized in COMPASS \cite{Denisov:2018unj}. This would be a very interesting line of research to be pursued  (indeed the elastic electron-proton scattering in the situation $Q^2 \gg m_e^2$ has been studied using SCET in \cite{Hill:2016gdf} and used to incorporate the resummation of large logarithms).  

We emphasize that trying to describe this kinematic regime with an EFT Lagrangian like 
\be
{\cal L}_{\rm QED(\mu)}=
{\cal L}_{\gamma}
+
{\cal L}_{e}
+
{\cal L}_{\mu}
+
{\cal L}_{N}
+
{\cal L}_{N\mu}
+
{\cal L}_{Ne}
\,,
\ee
would require the four-fermion vertex to be a complicated function. The complication appears in the four-fermion sector due to the dependence on $E$, the lepton energy. Still, it is very important to emphasize that ${\cal L}_N$ and the hadronic Wilson coefficients of ${\cal L}_N$ remain equal, 
since the pion and Delta can be integrated out, and, in the proton bilinear sector, the relevant scale for matching is $Q^2$ rather than $E$, the former being invariant under changes of reference frame. 

\subsection{Boosted NRQED(\texorpdfstring{$\mu p$}{mup})}
\label{Sec:boosted}
A possible way out to the problem posed in the previous section is to boost the proton to the reference frame where the incoming muon is at rest, i.e., such that $E=m_{\mu}$. This would guarantee that the outcoming muon is still nonrelativistic ($E'-m_{\mu} \ll m_{\mu}$), since we restrict to the kinematic situation where $Q^2 \ll m_{\mu}^2$. Then, matching computations, and computations of observables, would be like for standard NRQED($\mu$p) but with a boosted proton: 
$P^{\mu}=Mv^{\mu}+k^{\mu}$, where $v^2=1$ and $k^{\mu}$ is small. Note that with this EFT we can study, on an equal footing, the situation (in the reference frame where the incoming proton is at rest) when $E \sim m_{\mu}$ or when $E \gg m_{\mu}$, as far as $Q^2 \ll m_{\mu}^2$. This was the situation studied in the previous section. Note that, in the EFT that we have in this section, the difference between having $E \sim m_{\mu}$ or $E \gg m_{\mu}$, discussed in the previous section, would just reflect in a different value of $v$, wherever it appears. 

By working with this EFT, we can then do matching computations and integrate out pions and Deltas without problems. For the bilinear proton part of the Lagrangian, the Wilson coefficients do not change, since the internal scale in the matching computations is $Q^2 $, which does not change after boosts. The difference appears in the four-fermion sector. The computation would be similar to the one made in \cite{Nevado:2007dd,Peset:2014jxa} but with a boosted proton. A new vector $v$ appears. In a way, compared with the previous section, we trade $E$ by $v$. The advantage is that it is easier to construct the effective theory, it is just NRQED($\mu$p) but with a boosted proton. The Lagrangian would read as follows
\be
{\cal L}_{\rm QED(\mu)}=
{\cal L}_{\gamma}
+
{\cal L}_{e}
+
{\cal L}_{\mu}
+
{\cal L}_{N_v}
+
{\cal L}_{N_v\mu}
+
{\cal L}_{N_v e}
\,,
\ee
where we use $N_v$ to emphasize that the proton is boosted. The different terms that appear in this Lagrangian are equal to those that also appear in Eq.~(\ref{eq:HBET}), except for the pieces of the Lagrangian with $N_v$ field content. These now read
\bea
\label{LNv}
 {\cal L}_{N_v}&=& N^\dagger_{p} \Biggl\{iv \cdot D+ \frac{(v \cdot D)^2-D^2}{2 M} + 
\frac {((v \cdot D)^2-D^2)^2}{
8 M^3} - e \frac{c_F^{(p)} }{ 4M}\, \sigma_{\mu\nu}G^{\mu\nu}\nn
\\
&& 
-e\frac{c_D^{(p)} }{ 8M^2}
   \left[D_{\mu},G^{\mu\nu}\right] 
 - ie \frac{ c_S^{(p)} }{8M^2}\, \sigma_{\mu\nu}\{D^{\mu},G^{\rho\nu}\}v_{\rho} 
\Biggr\} N_{p}
\,,
\eea
\bea
\delta {\cal L}_{N_v\mu}&=&
\displaystyle\frac{1}{M^2} c_{3,R}^{pl_\mu} v\cdot v'
{\bar N}_p \dsv 
  N_p \ \bar{l}_\mu \dsv' l_\mu
+
\sum_{ij}e^j \cdot e'_i \displaystyle\frac{1}{M^2} c_{4,R}^{pl_\mu}{\bar N}_p \dse_j \gamma_5 N_p 
  \ \bar{l}_\mu \gamma \dse'^i \gamma_5 l_\mu
\nn\\
&&+
\displaystyle\frac{c_{5,R}^{pl_\mu}}{M^2} \sum_j v\cdot e'^j
{\bar N}_p  \dsv 
  N_p \ \bar{l}_\mu \dse'_j \gamma_5 l_\mu
+\sum_je^j \cdot v' \displaystyle\frac{c_{6,R}^{pl_\mu}}{M^2} {\bar N}_p \dse_j \gamma_5
N_p 
  \ \bar{l}_\mu  \dsv' \gamma_5 l_\mu
\,,
\eea

\be
\delta {\cal L}_{N_v e}=\displaystyle\frac{1}{M^2} c_{3,R}^{pl_e}
{\bar N}_p \dsv
  N_p \ \bar{l}_e\dsv l_e
+\displaystyle\frac{1}{M^2}\sum_j c_{4,R}^{pl_e}{\bar N}_p \dse^j \gamma_5
N_p 
  \ \bar{l}_e\dse_j \gamma_5 l_e
\,,
\ee
where, for completeness, we also include the four-fermion operators made of proton and lepton, even though they are negligible for the experiments at hand.

To write these terms as general as possible, we have introduced a $v'$ for the muon, even if we set $v'=(1,{\bf 0})$ for the reference frame of the muon at rest. The vectors $e^i_{\mu}$ and $e^{'i}_{\mu}$ are space-like vectors such that: $v \cdot e^i=0$ and $e^i \cdot e^j=-\delta^{ij}$; and $v' \cdot e'^i=0$ and $e'^i \cdot e'^j=-\delta^{ij}$. Note that the set $\{v,e^i\}$ (and also $\{v',e'^i\}$) form a basis of the space-time manifold. 

The four-fermion operators with muon-proton content now have potentially four Wilson coefficients that should be determined. It would be very interesting to do the matching computation and try to determine them. 
After the computation is done, transforming to the frame where the experiment was actually made would be a trivial thing. 

As we have already mentioned, the Wilson coefficients of the bilinear term do not change with respect to those one has in the rest frame, as they only depend on $Q^2$, which is Lorentz invariant. Note also that with this trick, we could study in a controlled way the experiments MUSE and COMPASS, as far as $Q^2  \ll m_{\pi}^2$. Going from one experimental setup to the other could be done by a change of $v$. 

This EFT could also be useful for a possible experiment of the scattering of a proton on the muonic hydrogen at rest. This is  a kind of gedanken experiment for muons but has been considered in the case of the scattering of protons on hydrogen \cite{Gakh:2016xby}. 

\subsection{Relativistic electron with \texorpdfstring{$Q^2 \ll m_{\pi}^2 \,, m_{\mu}^2$}{Q2}}
\label{Sec:rele}

If $Q^2 \ll m_{\mu}^2$ and $E-m_e \ll m_{e}$, we are in the kinematic configuration where we can apply the NRQED($e p$) Lagrangian described in Sec.~\ref{NRQED(e)}. This guarantees that the very same Wilson coefficients as in hydrogen are used. It also guarantees that the same proton radius is measured as in hydrogen and muonic hydrogen (at least with relative ${\cal O}(\alpha)$ precision). On the other hand, such kinematic constraints are very restrictive and are not satisfied by the experimental setup of present, and near-future, electron-proton scattering experiments. We can relax the previous conditions to $Q^2 \ll m_{\mu}^2$ and $E^2\ll m_{\mu}^2$. Nevertheless, here we cannot play the same trick as before. Even if we put the initial electron at rest, after the collision the electron is scattered ultrarelativistically. The following Lagrangian could be considered if $E^2 \ll m_{\pi}^2$, $m_{\mu}^2$ 
\be
{\cal L}=
{\cal L}_{\gamma}
+
{\cal L}_{e}
+
{\cal L}_{N}
+
{\cal L}_{Ne}
\,.
\ee
In other words, it is the same Lagrangian of HBET (see Eq.~(\ref{eq:HBET})) but without pions, muons, and Deltas. 
If we are in the situation where the incoming electron is more energetic, with $E \sim m_{\mu}$, we can use the HBET Lagrangian of Eq.~(\ref{eq:HBET}) but without muons. If we want to profit from the kinematic constraint $Q^2 \ll m_{\pi}^2$, we have the same problems as in Sec.~\ref{Sec:Remuon}, and the same discussion applies. Also, if we consider the situation $m_{\mu} \siml E$ for electron-proton scattering (which is realistic), we should treat the electron as an ultra-relativistic particle, even if we restrict to $Q^2 \ll m_{\mu}^2$, and the same discussion as in  Sec.~\ref{Sec:Remuon} applies. 

 In any case, irrespective of this discussion, it is very important to emphasize that the hadronic Wilson coefficients of ${\cal L}_N$ remain equal, as far as the energy and three-momentum of the photon is much smaller than the mass of the pion. 

\subsection{Low-energy constants, Wilson coefficients and form factors}
\label{Sec:formfactors}

In this section, we show the relation between the Wilson coefficients of the effective theory, the form factors, and the low-energy constants. 

\subsubsection{Form factors}\label{sec:formfactors}
We define the form factors as ($J^{\mu}=\bar N_p \gamma^{\mu} N_p$) 
\be
\langle {p^\prime,s}|J^\mu|{p,s}\rangle
=
\bar u(p^\prime) \left[ F_1(-q^2) \gamma^\mu +
i F_2(-q^2){\sigma^{\mu \nu} q_\nu\over 2 M} \right]u(p)
\label{current}
\,,
\ee
and Taylor expand $F_i$ in powers of $Q^2\equiv -q^2$: 
\be
F_i(Q^2)=F_i(0)+{Q^2 \over M^2}F_i^{\prime}+...\,,
\ee 
and we define
\begin{align}
\tilde{F_i}\equiv F_i(0),\quad \quad \tilde{F}^{(n)}_i \equiv \frac{d^n F_i}{(d(Q^2/M^2))^n}\bigg|_{Q^2=0}.
\end{align}
Assuming the analyticity of form factors in the complex $q^2$ plane and the appropriate high-energy behavior, i.e., $|F_1(Q^2)/Q^2|,~|F_2(Q^2)| \underset{Q^2 \to \infty}{<} |\mathrm{const}|$, we write down the subtracted and unsubtracted dispersion relations for the proton form factors $F_1$ and $F_2$ respectively (for a discussion about their validity see, for instance, \cite{Oehme:1992py}):
\be
\label{eq:F1F2DR}
 F_1 \left( Q^2 \right) = 1 - \frac{Q^2}{\pi}  \int \limits_{4 m^2_\pi}^{\infty} \frac{\mathrm{Im} F_1 \left( -t - i \varepsilon \right) \mathrm{d} t}{t \left(t+Q^2  \right)} , \qquad F_2 \left( Q^2 \right) = \frac{1}{\pi} \int \limits_{4 m^2_\pi}^{\infty} \frac{\mathrm{Im} F_2 \left( - t - i \varepsilon \right) \mathrm{d} t}{t+Q^2}.
\ee
Dispersive integrals start from the 2-pion production threshold in the time-like region ($Q^2 < 0$) corresponding to  $p\bar{p}$ production or annihilation. Lepton-proton scattering kinematics is given by spacelike momentum transfer $Q^2 > 0$.

One important issue that is often raised is whether electromagnetic corrections are included in the matrix element of Eq.~(\ref{current}). We choose them to be included, otherwise one should include more correlators of ${\cal O}(\al)$, besides the TPE correction, in the observables we consider. 

The Sachs form factors disentangle the interaction of the proton with the electric and magnetic field in the nonrelativistic limit. Therefore, it is natural to use them for the interaction of photons with protons at low energies. The relation between Dirac-Pauli and Sachs form factors (for the proton) reads
\begin{align}
G_E(Q^2)=F_1(Q^2)-\tau F_2(Q^2),\quad \quad G_M(Q^2)=F_1(Q^2)+F_2(Q^2),
\end{align}
with
\begin{align}
\tau=\frac{Q^2}{4M^2}
\,,
\end{align}
and the values at zero-momentum transfer are
\begin{align}
F_1(0)=Z,\quad\quad F_2(0)=\kappa_p.
\end{align}

The low momentum transfer expansion of the Sachs form factors relates to the different radii as
\be
\label{GEfit}
G_{\rm E}(Q^2) = Z - \frac{r_p^2}{3!} Q^2 
+ \frac{\langle r^4 \rangle_{\rm E}}{5!} Q^4 
-  \frac{\langle r^6 \rangle_{\rm E}}{7!} Q^6
+ . . . 
\,,
\ee
and
\be
\label{GMfit}
\frac{G_{\rm M}(Q^2)}{Z+\kappa_p} = 1 - \frac{\langle r^2 \rangle_{\rm M}}{3!} Q^2 
+ \frac{\langle r^4 \rangle_{\rm M}}{5!} Q^4
-  \frac{\langle r^6 \rangle_{\rm M}}{7!} Q^6
+ . . .
  \,.
\ee

The form factors are the scalar components of the matrix element of the electromagnetic current sandwiched between proton states. These matrix elements include loop corrections (both hadronic and electromagnetic). When relating these matrix elements with the Wilson coefficients of the EFT, it should be understood that these loops have an infrared cutoff and that this infrared cutoff corresponds to the ultraviolet cutoff of the effective theory. As the effective theories we consider have an ultraviolet cutoff much smaller than the hadronic scale, all hadronic effects are encoded in the Wilson coefficients. The only loops for which a cutoff has to be introduced are of electromagnetic origin and produce that some $\langle r^n \rangle$ are $\nu$ dependent, where $\nu$ is the cutoff of the effective theory. This dependence is logarithmic if working with dimensional regularization. If there is no logarithmic dependence on the factorization scale, like for $\kappa_p$ and others, they can be associated with low-energy constants. Rewriting them in terms of low-energy constants one has 
\begin{eqnarray}
c_F^{(p)}&=&\tilde{F_1}+\tilde{F_2}=Z+\kappa_p =Z+\kappa_p^{\rm had}+\frac{Z^3\al}{2\pi}+{\cal O}(\al^2)
\label{cfp},\\
c_S^{(p)}&=&\tilde{F_1}+2\tilde{F_2}=Z+2\kappa_p  =Z+2\kappa_p^{\rm had}+\frac{Z^3\al}{\pi}+{\cal O}(\al^2)\label{csp}
\,.
\end{eqnarray}
Note that $\kappa_p$ includes ${\cal O} (\alpha)$ effects. In principle, this is also so for $\kappa_p^{\rm had}$, to which we have 
subtracted the proton-associated point-like contribution to the anomalous magnetic moment 
(note that the point-like result is a bad approximation for $c_F^{(p)}$, even though it gives the right order of magnitude). 

When the radii are factorization scale dependent, one cannot assign low-energy constants to the Wilson coefficients (or the radii).\footnote{Actually, for a unique relation between low-energy constants and Wilson coefficients, the Lagrangian density has to be expanded in a minimal basis of operators. This is true irrespectively of working with low-energy or quasi-low-energy constants.} Instead, as we have already mentioned, we name them quasi-low-energy constants (as they have to be combined with a loop computation in the effective theory to yield the theory prediction for the observable). The most paradigmatic example is the proton charge radius (see also the discussion in Ref.~\cite{Pineda:2004mx}). 
It can be written in the following way in terms of the electromagnetic current form factors at zero momentum:
\be
c_D^{(p)}(\nu)=\tilde{F_1}+2\tilde{F}_2-8\tilde{F'_1}=Z+\frac{4}{3}M^2 r_p^2(\nu)
=Z-8M^2\left.\frac{d G_{E}(Q^2) }{ d\,Q^2}\right|_{Q^2=0}
\,.
\ee
 This object is infrared divergent, which makes it scale and 
scheme dependent. 
This is not a problem from the EFT point of view but makes the 
definition of the proton radius ambiguous. 
The standard practice is to make explicit the proton-associated point-like contributions in the computation. 
This means using the following definition for the proton radius
\be
c_{D,\MS}^{(p)}(\nu)\equiv Z+\frac{4}{3}\frac{Z^3\alpha}{\pi}\ln\left(\frac{M^2}{\nu^2}\right)
+\frac{4}{3} r_p^2 M^2+{\cal O}(\al^2).
\ee
In other words, the standard definition (which corresponds to the experimental number) of the proton radius 
reads (up to ${\cal O}(\al^2)$ corrections):
\be
r_p^2=\frac{3}{4}\frac{1}{M^2}
\left(c^{(p)}_{D,\MS}(M)-Z\right)
\,.
\ee
Note that $r_p$ includes ${\cal O}(\al)$ terms in its definition. This should be kept in mind when comparing with lattice determinations. 
 Note, also, that this definition sets $\nu=M$. This is not natural, since this assumes that the proton is point-like up to (and beyond) the scales of the proton mass. This is a bad approximation, as we can see comparing the piece of  $c_{D,\MS}^{(p)}$ associated to the structure of the proton: 
$\frac{4}{3} r_p^2 M^2 \simeq 21.3$, with "Z=1" for a point-like particle. This illustrates that the point-like result does not even give the right order of
magnitude for $c_D$.\footnote{Note that this 
also happens for the Wilson coefficients $c_{A_1}$ and $c_{A_2}$ (for their definition, see Ref.~\cite{Pineda:2004mx}), for which their physical values are far from zero: $c_{A_1}\simeq 12$ and $c_{A_2}\simeq -72$, even though for a point-like particle their values would be "1" and "0" respectively (up to $O(\al)$ corrections).}

After the discussion of those particular examples, we now discuss the other Wilson coefficients that appear at low orders in the $1/M$ expansion. From \cite{Manohar:1997qy,Hill:2012rh}, we can relate the NRQED Wilson coefficients with the form factors, and therefore the radii as (we make explicit if they depend on the factorization scale)
\begin{align}
c_{W_1}^{(p)}(\nu)&=\tilde{F_1}+\frac{1}{2}\tilde{F_2}-4\tilde{F'_1}-4\tilde{F'_2}=
c_F^{(p)}\left(1+\frac{2}{3}M^2\langle r^2\rangle_M(\nu)\right)-\frac{\kappa_p}{2},\nn\\
c_{W_2}^{(p)}(\nu)&=\frac{1}{2}\tilde{F_2}-4\tilde{F'_1}-4\tilde{F'_2}=c_{W_1}^{(p)}(\nu)-Z,\nn\\
c_{p'p}^{(p)}&=\tilde{F_2}=\kappa_p,\nn\\
c_{M}^{(p)}(\nu)&=\frac{1}{2}F_2-4\tilde{F'_1}=-\frac{1}{2}(Z+\kappa_p-c_D^{(p)}(\nu)),\nn\\
c_{X_1}^{(p)}(\nu)&=\frac{5}{128}\tilde{F_1}+\frac{1}{32}\tilde{F_2}-\frac{1}{4}\tilde{F'_1}=\frac{1}{32}\left(c_D^{(p)}(\nu)-\kappa_p+\frac{Z}{4}\right),\nn\\
c_{X_2}^{(p)}&=\frac{3}{64}\tilde{F_1}+\frac{1}{16}\tilde{F_2}=\frac{1}{64}(3Z+4\kappa_p),\nn\\
c_{X_3}^{(p)}(\nu)&=-\frac{1}{8}\tilde{F'_1}-\frac{1}{4}\tilde{F'_2}+\frac{1}{2}\tilde{F}''_1=\frac{1}{64}\left(-Z-2\kappa_p+c_D^{(p)}(\nu)+\frac{8}{15}M^4\langle r^4\rangle_E(\nu)\right),\nn\\
c_{X_5}^{(p)}&=\frac{3}{32}\tilde{F_1}+\frac{1}{8}\tilde{F_2}=\frac{1}{32}(3Z+4\kappa_p),\nn\\
c_{X_6}^{(p)}(\nu)&=-\frac{3}{32}\tilde{F_1}-\frac{1}{8}\tilde{F_2}+\frac{1}{4}\tilde{F'_1}+\frac{1}{2}\tilde{F'_2}=\frac{1}{8}\left(Z-c_F^{(p)}+\frac{c_D^{(p)}(\nu)}{4}-c_{W_1}^{(p)}(\nu)\right).
\end{align}
Then, we can define the following radii in terms of the NRQED Wilson coefficients as
\begin{align}
\langle r^2\rangle_M(\nu)&=\frac{3}{2c_F^{(p)}M^2}\left(c_{W_1}^{(p)}(\nu)-Z-\frac{\kappa_p}{2}\right),\\
\langle r^4\rangle_E(\nu)&=\frac{15}{8M^4}(64c_{X_3}^{(p)}(\nu)-c_D^{(p)}(\nu)+Z+2\kappa_p).
\end{align}
It is quite remarkable that they are scale dependent, as these radii are of fundamental importance in determinations of the proton radius from lepton-proton scattering \cite{Horbatsch:2016ilr}. Nevertheless, with the present level of precision, such ambiguity should not affect present determinations of the proton radius. In the future, as the precision increases, this discussion will be relevant if we aim to give unambiguous determinations of these quantities with ${\cal O}(\alpha)$ precision (as we do already for $r_p^2$). Therefore, let us discuss what is known at present for the ${\cal O}(\alpha)$ running of these Wilson coefficients. One can read the point-like contribution at one loop in the $\MS$ from \cite{Manohar:1997qy} (see also \cite{Pokorski:1987ed})
\begin{eqnarray}
F_1^{\rm point-like} 
&=& 1 - {\alpha\over \pi}{Q^2\over M^2} \left(  -{1\over 8} +{1\over 3}
\log {M \over \mu} \right)+{\alpha\over \pi}{Q^4\over M^4} 
\left(
 -{11\over 240} +{1\over 20}
\log {M \over \mu}
\right), \\
F_2^{\rm point-like} &=& {\alpha\over \pi}
\left[ {1\over 2} - {Q^2\over M^2}{1\over 12}\right]. 
\end{eqnarray}
One could then consider factoring these contributions out of the radii. One would then have
\begin{align}
\langle r^2\rangle_M(\nu)&=\frac{1}{M^2}\frac{\alpha}{\pi}\frac{1}{c_F^{(p)}}
\left(\ln \frac{M^2}{\nu^2}-\frac{1}{4}\right)+\langle r^2\rangle_M^{(\rm had)}(\nu)
,\\
\langle r^4\rangle_E(\nu)&=\frac{3}{M^4}\frac{\alpha}{\pi}\left(\ln \frac{M^2}{\nu^2}-1
\right)+
\langle r^4\rangle^{(\rm had)}_E(\nu).
\end{align}
Note that, unlike what happens for the proton radius, what we call the pure hadronic quantity may still have $\nu$ scale dependence. This happens when the Wilson coefficients that contribute to the renormalization group equation of a given Wilson coefficient are different from the point-like expressions. This is indeed the case for $c_{W_1}$. Its running was computed in \cite{Lobregat:2018tmn} (for the running including light fermions, see Eq.~(45) of \cite{Moreno:2018lbo}, which produces logarithmically enhanced corrections at higher powers in $\alpha$):
\be
\nu \frac{d}{d\nu}c_{W_1}^{(p)}=-\frac{4}{3}c_F^{(p)}\frac{\al}{\pi}
\,,
\ee
 and does not coincide with the point-like running (since $c_F^{(p)}\not= 1$).\footnote{As an aside, note that $c_{W_1}^{(p)}$ alone does not appear in observables. The combination of Wilson coefficients that appears in observables is $c_{W_1}^{(p)}-c_{W_2}^{(p)}$, which indeed does not run \cite{Lobregat:2018tmn}.} On the other hand, the running of $c_{X_3}^{(p)}$ is not known. Therefore, the running of $\langle r^4\rangle^{(\rm had)}_E(\nu)$ cannot be determined at present. 

It would also be interesting to study the general (to-all-orders in $\alpha$) renormalization properties of the Wilson coefficients and the associated radii. For point-like computations, $c_F^{(p)}$ is renormalization group independent, as it is expected for a low-energy constant. The same should happen after the inclusion of hadronic corrections. For $c_D^{(p)}$, it seems that there is no renormalization after one loop. It would be interesting to see if this could be proven to all orders. 

\subsubsection{Structure functions}
\label{Sec:Structurefunctions}

We also consider matrix elements of the time-ordered product of two electromagnetic currents. For on-shell photons, they are of relevance for the determination of the electric and magnetic polarizabilities of the proton, related to the Wilson coefficients $c_{A_1}^{(p)}$ and $c_{A_2}^{(p)}$ of the EFT. They can be determined from accurate measurements of the Compton scattering. Nevertheless, this is beyond the purpose of this review. Here we only mention that their definition may suffer from some ambiguities, and that they are infrared divergent at ${\cal O}(\al)$ (see, for instance, \cite{Hill:2012rh,Moreno:2017sgd}).  

For the case of off-shell photons, the 
 forward doubly virtual Compton tensor is of particular interest to this review,
\begin{equation} 
 T^{\mu\nu} = i\!\int\! d^4x\, e^{iq\cdot x}
  \langle {p,s}| T \{J^\mu(x)J^\nu(0)\} |{p,s}\rangle
\,,
\end{equation}
which has the following structure ($\rho=q\cdot p/M\equiv v \cdot q$, although we will usually work in the 
rest frame of the proton, where $\rho=q^0$):
\bea \label{inv-dec}
 T^{\mu\nu} &=
  &\left( -g^{\mu\nu} + \frac{q^\mu q^\nu}{q^2}\right) S_1(\rho,Q^2) 
  + \frac1{M^2} \left( p^\mu - \frac{M\rho}{q^2} q^\mu \right)
    \left( p^\nu - \frac{M\rho}{q^2} q^\nu \right) S_2(\rho,Q^2) 
	\nn\\*
  && - \frac i{M}\, \epsilon^{\mu\nu\rho\sigma} q_\rho s_\sigma A_1(\rho,Q^2)
	- \frac i{M^3}\, \epsilon^{\mu\nu\rho\sigma} q_\rho
   \bigl( (M\rho) s_\sigma - (q\cdot s) p_\sigma \bigr) A_2(\rho,Q^2)
   \equiv T_S^{\mu\nu}+T_A^{\mu\nu}
   \,.
\eea
It depends on four scalar functions, which we call structure functions. 
We split the tensor into the symmetric (spin-independent),
$T_S^{\mu\nu}=T_S^{\nu\mu}$ (the first two terms of Eq.~(\ref{inv-dec})), 
and antisymmetric (spin-dependent) pieces, $T_A^{\mu\nu}=-T_A^{\nu\mu}$ (the last two terms of Eq.~(\ref{inv-dec})). 

Forward doubly virtual Compton scattering amplitudes have the following crossing properties:
\begin{align}
S_1(-\rho,Q^2) &= S_1(\rho,Q^2), \qquad \qquad ~S_2(-\rho,Q^2) = S_2(\rho,Q^2), \\
A_1(-\rho,Q^2) &=  A_1(\rho,Q^2),  \qquad \qquad A_2(-\rho,Q^2) =  -A_2(\rho,Q^2).
\end{align}

By means of dispersion relations, real parts of the forward doubly virtual Compton scattering amplitudes $S_1,~S_2$, $A_1,~A_2$ can be expressed in terms of imaginary parts up to the subtraction function in the amplitude $S_1$ ($S^\mathrm{subt}_1(0,Q^2)$) to ensure the convergence of the corresponding integral:
\begin{align}
\mathrm{Re}~S_1(\rho,Q^2) &= \mathrm{Re}~S^\mathrm{pole}_1(\rho,Q^2) + S^\mathrm{subt}_1(0,Q^2) + \frac{2}{\pi} \int \limits_{\nu^\mathrm{inel}_\mathrm{thr} }^{\infty} \frac{\rho^2 \mathrm{Im}~S_1(\rho' + i \varepsilon ,Q^2)}{\rho' \left( \rho'^2 - \rho^2 \right)} \mathrm{d} \rho', \\
\mathrm{Re}~S_2(\rho,Q^2) &= \mathrm{Re}~S^\mathrm{pole}_2(\rho,Q^2) +  \frac{2}{\pi} \int \limits_{\nu^\mathrm{inel}_\mathrm{thr} }^{\infty} \frac{\rho' \mathrm{Im}~S_2\left( \rho' + i \varepsilon, Q^2 \right)}{\rho'^2 - \rho^2} \mathrm{d} \rho', \\
\mathrm{Re}~A_1(\rho,Q^2) &= \mathrm{Re}~A^\mathrm{pole}_1(\rho,Q^2) + \frac{2}{\pi} \int \limits_{\nu^\mathrm{inel}_\mathrm{thr} }^{\infty} \frac{\rho' \mathrm{Im}~A_1 \left( \rho' + i \varepsilon, Q^2 \right)}{\rho'^2 - \rho^2} \mathrm{d} \rho', \\
\mathrm{Re}~A_2(\rho,Q^2) &= \mathrm{Re}~A^\mathrm{pole}_2(\rho,Q^2) + \frac{2}{\pi} \int \limits_{\nu^\mathrm{inel}_\mathrm{thr} }^{\infty} \frac{\rho \mathrm{Im}~A_2 \left( \rho' + i \varepsilon, Q^2 \right)}{ \rho'^2 - \rho^2} \mathrm{d} \rho',
\end{align}
where we have separated the proton pole contributions to the forward doubly virtual Compton scattering amplitudes
\begin{align}
S^\mathrm{pole}_1(\rho,Q^2) &= \frac{2 Q^4 G_M^2 (Q^2)}{Q^4 - 4 M^2 \rho^2 - i \varepsilon}, \\
S^\mathrm{pole}_2(\rho,Q^2) &= 8 M^2 Q^2 \frac{F^2_1 (Q^2) + \frac{Q^2}{4M^2} F^2_2 (Q^2)}{Q^4 - 4 M^2 \rho^2 - i \varepsilon}, \\
A^\mathrm{pole}_1(\rho,Q^2) &= \frac{4M^2 Q^2 F_1 (Q^2) G_M(Q^2)}{Q^4 - 4 M^2 \rho^2 - i \varepsilon}, \\
A^\mathrm{pole}_2(\rho,Q^2) &= - \frac{4M^3 \rho F_2 (Q^2) G_M(Q^2)}{Q^4 - 4 M^2 \rho^2 - i \varepsilon}, 
\end{align}
and have assumed the sufficiently vanishing high-energy behavior in the complex $\rho$ plane:
\begin{equation} \label{DR_condition}
|S_1(\rho,Q^2)/\rho^2|,~| S_2(\rho,Q^2)|,~| A_1(\rho,Q^2) |,~|A_2(\rho,Q^2)/\rho |\underset{\rho \to \infty}{<} \mathrm{const}.
\end{equation}
All integrals start from the pion production threshold $\nu^\mathrm{inel}_\mathrm{thr} = m_\pi + \frac{m^2_\pi-q^2}{2M} $. Assuming the Regge behavior $|S_1| \to \rho^{\alpha_0}, |S_2| \to \rho^{\alpha_0 - 2}, |A_1| \to \rho^{\alpha_0 - 1}, |A_2| \to \rho^{\alpha_0 - 1}$ with the leading Pomeron intercept $\alpha_0=1.08$~\cite{Abarbanel:1967zza,Damashek:1969xj,Donnachie:1998gm,Donnachie:2004pi,Abramowicz:2010asa}, the condition on the vanishing high-energy behavior of Eq.~(\ref{DR_condition}) is satisfied. 

Note that the subtraction function here does not vanish at $Q^2 \to 0$, rather
\begin{equation}
S^\mathrm{subt}_1(0,Q^2) \underset{q^2 \to 0}{\longrightarrow}  \mathrm{Re}~S^\mathrm{Born}_1(\rho,Q^2) - \mathrm{Re}~S^\mathrm{pole}_1(\rho,Q^2) = - 2 F_1^2 \left( Q^2 \right) ,
\end{equation}
due to nonvanishing difference between Born $S^\mathrm{Born}_1(\rho,Q^2)$ and pole contributions.

The proton forward doubly virtual Compton tensor appears convoluted with its lepton analog in the Wilson coefficients of the four-fermion operators made by two lepton fields and two proton fields. The explicit expressions for the spin-independent and spin-dependent Wilson coefficients read \cite{Bernabeu:1973uf}
\bea
\label{c3}
c_{3}^{pl_i}
&=&
- e^4 Mm_{l_i}\int {d^4k_E \over (2\pi)^4}{1 \over k_E^4}{1 \over
k_E^4+4m_{l_i}^2k_{0,E}^2 }
\nn
\left\{
(3k_{0,E}^2+{\bf k}^2)S_1(ik_{0,E},k_E^2)-{\bf k}^2S_2(ik_{0,E},k_E^2)
\right\}
\\
&&
+{\cal O}(\al^3)
\,,
\eea
and
\bea
\nn
c_{4}^{pl_i}&=&
\int {d^4k_E  \over 3\pi^2}
{1 \over k_E^2}{1 \over
k^4_E+4m_{l_i}^2k_{0,E}^2 }
\left\{
(k_{0,E}^2+2k_E^2)A_1(ik_{0,E},k_E^2)+i3k_E^2{k_{0,E} \over M}A_2(ik_{0,E},k_E^2)
\right\}
\\
&&
+{\cal O}(\al^3)
\,,
\label{c4pe}
\eea
consistent with the expression obtained long ago in Ref. \cite{DS}. 

These expressions keep the complete dependence on $m_{l_i}$ and are valid both for NRQED($\mu  p$) and NRQED($e p$). Again, it is common practice to single-out the proton-associated point-like contribution. Note that this assumes that one can treat the proton as point-like at energies of the order of the proton mass. We have already seen that this is a bad approximation for $c_D$ and other Wilson coefficients. Nevertheless, we keep this procedure for the sake of 
comparison.\footnote{In this expression, we have computed the loop with the proton being relativistic to follow common practice. 
Nevertheless, this assumes that one can consider the proton to be point-like at the scales of the proton mass. 
To stick to a standard EFT approach one should consider the proton to be nonrelativistic. Then one would obtain
\be
\label{c3pointlike}
 d_{s}^{\MS}(\nu)
=-Z^2\al^2\left(\ln{m_{l_i}^2 \over \nu^2}+{1 \over 3}\right)
\,.
\ee
The difference between both results is of the order of $\al^2\frac{m_{l_i}^2}{M^2}$, and gets absorbed into $c_3^{\rm had}$ (which we do not know with such precision anyhow). Therefore, the value of $c_{3}^{pl_i}$, will be the same no matter the prescription used. In practice there could be some difference due to truncation, but always of the order of the error of the computation.
} Therefore (we write the expression for a generic lepton $l_i$), 
\bea
\label{c3nu}
c_{3}(\nu)&\equiv&-\frac{M}{m_{l_i}}d_s(\nu)+c_{3}^{\rm had}+{\cal O}(\al^3)
\,,
\\
c_{4}&\equiv&-\frac{M}{m_{l_i}}d_v+c_{4}^{\rm had}+{\cal O}(\al^3)
\,,
\eea
where the point-like Wilson coefficients read as follows:
\begin{eqnarray}
\label{eq:ds}
d_s(\nu)&=&-\frac{Z^2\alpha^2}{m_{l_i}^2-M^2}\left[m_{l_i}^2\left(\ln \frac{M^2}{\nu^2}+\frac{1}{3}\right)-M^2\left(\ln \frac{m_{l_i}^2}{\nu^2}+\frac{1}{3}\right)\right],\\
\label{eq:dv}
d_v&=&\frac{Z^2\alpha^2}{m_{l_i}^2-M^2}m_{l_i} M \ln\frac{m_{l_i}^2}{M^2}.
\end{eqnarray}
The expression of $d_s$ should be understood in the $\MS$ scheme, 
$d_v$ on the other hand is finite. $d_s$ was computed in Ref.~\cite{Pineda:1998kj} and $d_v$ in Ref.~\cite{Caswell:1985ui}.

\subsection{potential NRQED}
\label{Sec:pNRQED}

To deal with the case when both the proton and the lepton are nonrelativistic, it is convenient to use the EFT named potential NRQED (pNRQED) \cite{Pineda:1997bj}. It combines quantum mechanics perturbation theory for the nonrelativistic bound state (or for the fermion-antifermion pair above threshold) and quantum field theory for the relativistic massless modes: photons (and light fermions, if they exist in the effective theory). This EFT is optimal if compared with NRQED because it makes explicit that for nonrelativistic modes (leptons) there are no physical degrees of freedom (as asymptotic states) with energy of order $m_{l_i} v$, where $v$ is the velocity of the lepton. This is achieved by integrating out scales of ${\cal O}(m_{l_i} v)$. Thus, all the dynamical degrees of freedom in the EFT have ultrasoft energy. 
We next discuss the main features of this theory when applied to the muon-proton sector and when applied to the electron-proton sector. A more detailed exposition of the former can be found in \cite{Pineda:2002as,Pineda:2004mx,Peset:2015zga} and of the latter in \cite{Pineda:1997ie,Pineda:1998kn}.

\subsubsection{pNRQED(\texorpdfstring{$\mu p$}{mup})}
\label{Sec:pNRQEDmup}

After integrating out scales of ${\cal O}(m_{\mu}\al \sim m_e)$ in NRQED($\mu p$), 
the resulting effective theory is pNRQED($\mu p$). 
This EFT naturally gives a Schr\"odinger-like formulation of the bound-state problem but still keeping the quantum field theory nature of the interaction with ultrasoft photons, as well as keeping the information due to high-energy modes (of a quantum field theory nature) in the Wilson coefficients of the theory. Up to ${\cal O}(m_r\al^5)$, the effective Lagrangian reads
\bea
&&L_{\rm pNRQED} =
\int d^3{\bf x} d^3{\bf X} S^{\dagger}({\bf x}, {\bf X}, t)
                \Biggl\{
i\partial_0 - \frac{ {\bf p}^2 }{2m_{r}} + \frac{ {\bf p}^4 }{ 8m_{\mu}^3}+ 
\frac{ {\bf p}^4 }{ 8M^3} - \frac{ {\bf P}^2}{ 2(m_{\mu}+M)}
\nn
\\
&&
- V ({\bf x}, {\bf p}, {\bfsigma}_1,{\bfsigma}_2) + e 
\left(
\frac{ M+Z m_{\mu}}{M+m_{\mu}}
\right)
{\bf x} \cdot {\bf E} ({\bf X},t)
\Biggr\}
S ({\bf x}, {\bf X}, t)- \int d^3{\bf X} \frac{1}{ 4} F_{\mu \nu} F^{\mu \nu}
\,,
\eea
where $m_r= \frac{m_{\mu}M }{ m_{\mu}+M}$, ${\bf x}$ and 
${\bf X}$, and
${\bf p}$ and ${\bf P}$ are the relative and center-of-mass coordinate and momentum
respectively.

The potential $V$ can be written as an expansion in $1/m_{\mu}$, $1/M$, $\al$, ... 
We will assume $1/r \sim m_e$ (which is realistic for the case at hand) 
and that $m_{\mu} \ll M$. We then organize the potential as an expansion in $1/m_{\mu}$:
\be
V ({\bf x}, {\bf p}, {\bfsigma}_1,{\bfsigma}_2)
=
V^{(0)}(r)+{V^{(1)}(r) }+{V^{(2)}(r) }+\cdots\,,
\ee
where
\be
V^{(n)} \propto \frac{1}{m_{\mu}^n}.
\ee
We will also make the expansion in powers of $\al$ explicit. This means that 
\be
V^{(n,r)} \propto \frac{1}{m_{\mu}^n}\al^r.
\ee

$V^{(0,1)}=-\frac{Z\al}{r}$ has to be included exactly in the leading order Hamiltonian to yield the leading-order solution 
to the bound or near-threshold state problem:
\be\label{eq:defH}
h=\frac{{\bf p}^2}{2m_r}-\frac{Z\al}{r}
\,.
\ee
Thus, the contribution to the energy of a given potential is (for near-threshold states we take $v \sim \alpha$, though this could be relaxed)
\be
\langle {V^{(n,r)} }\rangle \sim m_{\mu}\al^{1+n+r}
\,,
\ee
up to large logarithms or potential suppression factors due to powers of $1/M$. Iterations of the potential are dealt with using standard quantum mechanics perturbation theory producing corrections such as:
\be
\langle V^{(n,r)}\cdots V^{(m,s)} \rangle \sim m_{\mu}\al^{1+n+r+(1+m+s-2)}
\,,
\ee
and alike. Therefore, in order to reach the desired ${\cal O}(m\al^5)$ accuracy, $V^{(0)}$ 
has to be computed up to $O(\al^4)$, $V^{(1)}$ up to $O(\al^3)$, 
$V^{(2)}$ up to $O(\al^2)$ and $V^{(3)}$ up to $O(\al)$. 

\begin{figure}[ht]
  \begin{center}
   \includegraphics[width=0.25\textwidth]{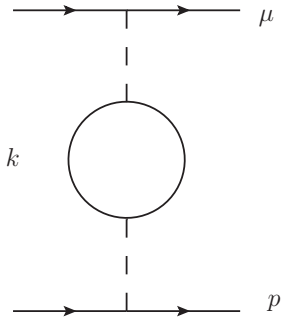}
  \end{center}
  \caption{ One-loop electron vacuum polarization contribution to the static potential.}
  \label{fig:VPelectron}
\end{figure}

\begin{figure}[ht]
  \begin{center}
  \includegraphics[width=0.55\textwidth]{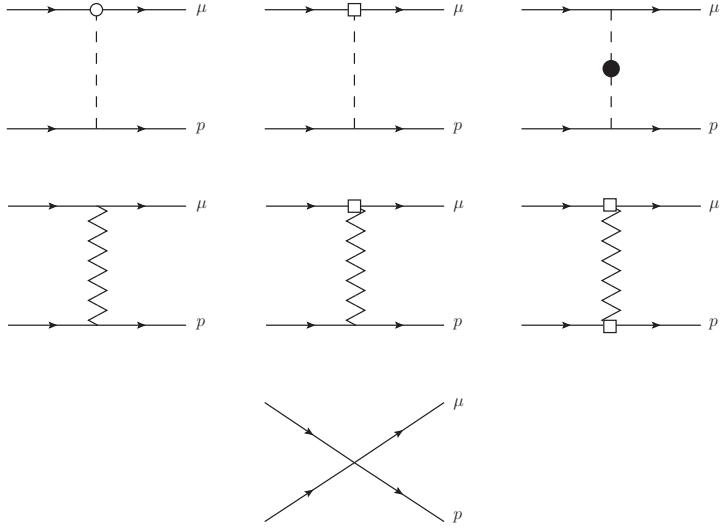}
  \end{center}
\caption{ The non-zero relevant diagrams for the matching at tree level
       in the Coulomb gauge. The dashed and zigzag lines represent the
       $A_0$ and ${\bf A}$ fields respectively, while the continuous
       lines represent the fermion and antifermion fields. For the $A_0$, the circle is the vertex
       proportional to $c_D$, the square to $c_S$ (spin dependent), and
       the black dot to $d_2$, while for
      ${\bf A}$, the square is the vertex proportional to $c_F$ and
       the other vertex appears from the covariant derivative in the
       kinetic term. The last diagram is proportional to $c_3$ and
       $c_4$. The symmetric diagrams are not displayed.}
\label{figtree}
\end{figure}

The different potentials are obtained by matching the NRQED($\mu p$) diagrams (using static propagators for the proton and/or muon) to the diagrams in pNRQED($\mu p$) in terms of potentials (see Ref. \cite{Peset:2015zga} for details).  In this way, the Coulomb potential and its corrections are generated. In particular, the leading correction to the Coulomb potential is generated by the vacuum polarization of the photon (see Fig.~\ref{fig:VPelectron}):
\be
\tilde V^{(0,2)}_{\rm VP}(k)= 4\pi Z\frac{\alpha^2}{\pi}\frac{\Pi_1(-{\bf k}^2) }{ {\bf k}^2},
\ee
where
\begin{equation}
\Pi_{1}(k^2)=k^2\int_4^\infty d q^2\frac{1}{{q^2(m_e^2q^2-k^2)}}u(q^2),
\end{equation}
and
\be
u(q^2)=\frac{1 }{ 3}\sqrt{1-\frac{4 }{ q^2}}\left(1+\frac{2 }{ q^2}\right)
\,.
\ee
This term indeed produces the bulk of the Lamb shift in muonic hydrogen. The next-to-next-to-leading order term of the static potential can be 
understood as a correction to the vacuum polarization. It was computed by K\"allen and Sabry \cite{Kallen:1955fb}.
The next-to-next-to-next-to-leading order term of the static potential coming from the vacuum polarization has been computed in Ref.~\cite{Kinoshita:1979dy}, see also \cite{Kinoshita:1998jf} where the complete set of 
diagrams can be found.
The remaining next-to-next-to-leading order contribution to the static potential is generated by diagrams that cannot be completely associated with the vacuum polarization. This object could be deduced from the computation in Ref.~\cite{Karshenboim:2010cq}. For the other corrections to the Coulomb potential, see \cite{Pineda:2004mx,Peset:2015zga}. The matching at tree level is shown in Fig.~\ref{figtree}. For us, the main concern are the hadronic corrections. The leading hadronic corrections always appear in the same combination 
\be
\delta V_{\rm had}=\frac{1}{M^2}D_d^{\rm had}\delta ^{(3)}({\bf r})
\,,
\ee
\be
\label{Dd}
D_d^{\rm had}=-c_3^{\rm had}-16\pi\alpha d_2^{\rm had}+\frac{2\pi\alpha}{3}r_p^2M^2
\;.
\ee
This contribution is generated by matching the first, third and last diagram of Fig.~\ref{figtree} to a Dirac-delta potential. Therefore, low-energy experiments cannot disentangle the three different Wilson coefficients, but only measure the specific combination in Eq.~(\ref{Dd}). If one wants to determine the proton radius, one needs to determine the other Wilson coefficients from different sources with enough precision. For illustration, if one aims to determine the proton radius with 1 per mille precision, one would need to know the TPE contribution with 10\% accuracy, since the size of the TPE contribution is around 1\% the size of the proton radius contribution, and the hadronic vacuum polarization with 30\% accuracy, since the hadronic vacuum polarization contribution is around 0.3\% the size of the proton radius contribution. 
 
\subsubsection{pNRQED(\texorpdfstring{$ep$}{ep})}
\label{Sec:pNRQED(e)}
One can obtain the pNRQED($ep$) effective Lagrangian from the one of the previous section by switching off the electron contribution, changing the muon by the electron, and then adding the muon vacuum polarization correction to $d_2$. To ${\cal O}(m_e \al^5)$, such Lagrangian can be found in the appendix of \cite{Pineda:1998kn}. Nevertheless, for nowadays precision of regular hydrogen Lamb shift, one should write the Lagrangian to higher order in both the $1/m_e$ and the $\alpha$ expansion. The explicit expression of such Lagrangian is not known at present. Here, we only sketch how existing computations would be encoded in the EFT computation in Sec.~\ref{Sec:regularhyd}. In any case, note that this effective theory is not suitable for the elastic electron-proton scattering case at $Q^2 \gg m_e^2$ but requires $Q^2 \ll m_e^2$. 

\section{Observables. Spectroscopy}

\subsection{Muonic hydrogen Lamb shift: \texorpdfstring{$E(2P_{1/2})-E(2S_{1/2})$}{ELS}}
\label{Sec:muonichyd}

The only experimental measurements of the muonic hydrogen Lamb shift \cite{Pohl:2010zza,Antognini:1900ns} correspond to the energy difference $E(2P_{3/2})-E(2S_{1/2})$ and $E(2P_{1/2})-E(2S_{1/2})$. For theoretical discussion, we focus on the latter. Its determination with EFTs was worked out in \cite{Peset:2014yha,Peset:2015zga}. It uses pNRQED applied to the muon-proton sector (see Sec. \ref{Sec:pNRQEDmup}). It has embedded the Wilson coefficients of the NRQED Lagrangian (see Sec. \ref{NRQED(mu)}) in the potentials. The determination of the spectrum is achieved by a combination of quantum mechanics perturbation theory and the interaction of the bound state with ultrasoft photons, which is computed using quantum field theory techniques. All these computations are made using dimensional regularization. 

\begin{figure}[ht]
  \begin{center}
   \includegraphics[width=0.45\textwidth]{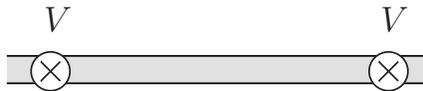}
  \end{center}
  \caption{ 2nd order perturbation theory of the bound-state Green function generated by a generic potential $V$.}
  \label{fig:doubleVP}
\end{figure}

\begin{figure}[ht]
  \begin{center}
    \includegraphics[width=0.5\textwidth]{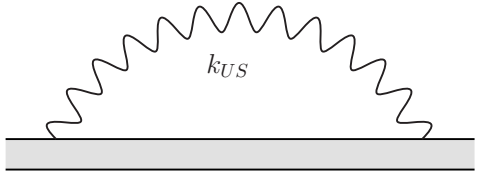}
  \end{center}
  \caption{ Correction due to ultrasoft photons.}
  \label{fig:ultrasoft}
\end{figure}

Illustrative diagrams are shown in Fig.~\ref{fig:doubleVP} for a nonrelativistic quantum mechanics perturbation theory computation, and in Fig.~\ref{fig:ultrasoft} for the leading contribution to the Lamb shift generated by the interaction of ultrasoft photons with the bound state. 

We represent the 2nd order perturbation theory correction to the bound-state Green function generated by the potential $V^{(n,m)}$ (with $(n,m) \not=(0,1)$) by Fig.~\ref{fig:doubleVP}, 
where the double line represents the bound state and the vertices (local in time) 
the potentials. In order to obtain the associated energy shift, we will compute objects like (and analogous expressions in case of different potentials (including permutations))
\bea
\nn
\delta E_{n\ell j}^{V^{(n,m)}\times V^{(n,m)}}&=&
\langle \psi_{n \ell j}|V^{(n,m)}\frac{1}{{(E_{n}-h)'}}V^{(n,m)}|\psi_{n \ell j}\rangle
\\
&=&
\int \mathrm{d}{\bf r_2} \mathrm{d}{\bf r_1} \psi_{n \ell j}^*({\bf r_2})V^{(n,m)}({\bf r_2})
G'_{n \ell}({\bf r_2},{\bf r_1})V^{(n,m)}({\bf r_1})\psi_{nlj}({\bf r_1})
\label{nloV},
\eea
where 
\be
\frac{1}{ (E_{n}-h)'}=\lim_{E \rightarrow E_n}\left(\frac{1}{E-h}-\frac{1}{E-E_n}\right)
\,,
\ee
\begin{equation}
G'_{n \ell}({\bf r_1},{\bf r_2})
\equiv 
\langle {\bf r}_1| \frac{1}{{(E_{n \ell}-h)'}}|{\bf r}_2 \rangle
\equiv
\lim_{E\rightarrow E_n}\left(G({\bf r_1},{\bf r_2}; E)-\frac{ \psi^*_{n\ell}({\bf r}_1) 
\psi_{n\ell} ({\bf r}_2)}{E-E_n}\right),
\end{equation}
$\psi_{n \ell}({\bf r})$ is the bound state wave function of the ($n l$)-state and $E_n$ is the energy of the state with Hamiltonian $h$ as in Eq.~\eqref{eq:defH}, 
and $G({\bf r_1},{\bf r_2}; E)$ is the Coulomb Green function (see for instance, the appendix of Ref.~\cite{Pineda:2011dg}). 

The leading contribution to the bound state energy generated by the interaction of the ultrasoft photons with the bound state  (see Fig. \ref{fig:ultrasoft}) reads (in $\MS$ scheme)
\begin{eqnarray}
\delta E^{\rm US}_{nl}&=&\frac{2}{3}
\left(\frac{ M+Z m_{\mu}}{M+m_{\mu}}\right)^2
\frac{\alpha}{\pi}\left(\left(\ln \frac{\nu}{m_r}+\frac{5}{6}-\ln 2\right)\left(\frac{Ze^2}{2}\right)\frac{|\psi_{n \ell}({\bf 0})|^2}{m_r^2}\right.\nonumber\\
&-&\left.\sum_{n'\neq n}|\langle n|\frac{p}{m_r}|n'\rangle|^2(E_n-E_{n'})\ln\frac{m_r}{|E_n-E_{n'}|}\right)
\nn
\\
&\equiv&\frac{ m_rZ^4\alpha ^5}{n^3\pi } 
\left(\frac{ M+Z m_{\mu}}{M+m_{\mu}}\right)^2
\left(\delta_{\ell,0}\left(-\frac{4}{3}\left(\ln R(n,\ell)+\ln\frac{m_rZ^2\al^2}{\nu}
\right)+\frac{10}{9}\right)\right.\nonumber\\
&-&\left.(1-\delta_{\ell,0})\frac{4}{3}\ln R(n,\ell)\right),
\label{Eus1}\end{eqnarray}
where $|\psi_{n\ell}({\bf 0})|^2=\frac{\delta_{\ell0}}{\pi}\left(\frac{m_rZ\al}{n}\right)^3$. $\ln R(n,l)$ are the Bethe logarithms and are implicitly defined by the equality with Eq.~(\ref{Eus1}). For their numerical values for the $2S$ and $2P$ states, we have used the values quoted in \cite{Pachucki1}.

Most of the contributions to the Lamb shift generated from nonrelativistic quantum mechanics perturbation theory were known prior to the EFT computation, see \cite{Pachucki1,Jentschura:2011nx,Kinoshita:1998jf,Ivanov:2009aa,Karshenboim:2010cq}. These contributions can be easily incorporated into the EFT framework. Some of them were checked in \cite{Peset:2015zga}, where one can find the complete set of contributions to this order. Note that, even though most of the contributions can be 
associated with a pure QED calculation, the hadronic effects are also included in this computation. Their effects are included in the NRQED Wilson coefficients, and are encoded in the delta potential in the Lagrangian of pNRQED (see Eq.~(\ref{Dd}). Finally, the theoretical expression for the Lamb shift in terms of the proton radius and the other hadronic contribution reads \cite{Peset:2015zga}
\begin{eqnarray}
\nn
&&\Delta E_L=
206.0243\,\mathrm{meV}\\
&&
-\left[\frac{1}{\pi}\frac{m_r^3\al^3}{8}\right]\frac{\al}{M^2}\frac{ r_p^2}{\mathrm{fm}^2}
\left[47.3525
+35.1491\al+47.3525\alpha^2\ln(1/\alpha)\right]\nonumber\\
&&
\nn+\left[\frac{1}{\pi}\frac{m_r^3\al^3}{8}\right]\frac{1}{M^2}
\left[
c_{3}^{\rm {had}}+
16\pi \alpha d_2^{\rm {had}}\right]
\\
&&+\mathcal{O}(m_r\alpha^6)
\,.
\label{El3}
\end{eqnarray}
Note that since $c_3^{\rm had} \sim \al^2$ and $\alpha d_2^{\rm {had}}\sim \al^2$, the third line of the previous 
equation encodes all the hadronic effects of order $\al^5$ that are not related to the proton radius.  
This presentation of the result where $r_p$ and $c_3^{\rm had}$ are kept explicit could be important for the future.
In the long term (once the origin of the proton radius puzzle is clarified) the natural place from where to obtain the proton radius is the hydrogen Lamb shift and $c_3^{\rm had}$ (once the radius has been obtained) from the muonic hydrogen, since $c_3^{\rm had}$ is suppressed by an extra factor of the lepton mass. In this scenario, a complete 
evaluation of the $\mathcal{O}(m_r\alpha^6)$ term may improve the precision of an eventual experimental determination of $c_3^{\rm had}$. Note that, in this discussion, we assume that we can determine $d_2^{\rm {had}}$ from alternative methods, like dispersion relations.

\subsection{Regular hydrogen Lamb shift}
\label{Sec:regularhyd}

We now discuss the theoretical determination of the Lamb shift for regular hydrogen. Reviews of theoretical expressions for different energy splittings can be found in \cite{Eides:2000xc,Eides:2007exa} and \cite{PhysRevA.93.022513,Mohr:2012tt,Yerokhin:2018gna,Karshenboim:2019iuq}. The first two give a detailed account for the different contributions to the Lamb shift. The last four references give more updated reviews including the most up-to-date computations. We will use them for the discussion in this section.

The mass of a $(n\ell)^f_j$ state of hydrogen\footnote{We use here the most common notation in atomic physics, where ${\bf F}={\bf L}+{\bf S}$, ${\bf J}={\bf L}+{\bf S_e}$, and the lower case parameters are their corresponding quantum numbers, where ${\bf L}$ is the angular momentum, ${\bf S_e}$ is the spin of the electron and ${\bf S}={\bf S_e}+{\bf S_p}$ is the total spin. A different basis is used in other references, as, e.g., in \cite{Peset:2015zga}, where the notation would change $j\to j_e$ (actually $j_\mu$ because in that case the muonic hydrogen case was under study), $f\to j$ and a redundant basis with $s$ as a quantum number is kept.} can be written as (we use the notation from \cite{PhysRevA.93.022513})
\begin{equation}
E_{n\ell jf}=m_e+M+
E_n^{\rm (g)}
+E^{\rm (fs)}_{n\ell j}
+E^{\rm (hfs)}_{n\ell jf},
\label{eq:Henergy}
\end{equation}
where 
\begin{equation}
E_n^{\rm (g)}=-\frac{(Z\alpha)^2 m_r}{2 n^2}
\label{eq:Eg}
\end{equation}
is the gross structure, 
$E^{\rm (fs)}_{n\ell j}$ is named the fine structure contribution, 
and $E^{\rm (hfs)}_{n\ell jf}$ the hyperfine structure contribution.\footnote{The hyperfine splitting for the ground state of hydrogen is studied in the following section and for the muonic hydrogen in the next-to-following section.} 
Here, 
$m_{r}$=$m_{e}M/(m_{e}+M)$ is the reduced mass.

We focus here on the evaluation of $E^{\rm (fs)}_{n\ell j}$.
Its contributions can be organized in the following way:
\begin{eqnarray}
E^{\rm (fs)}_{n\ell j}&=&m_r[f_{nj}-1+\frac{(Z\alpha)^2}{2 n^2}]-\frac{m_r^2}{2(m_e+M)}[f_{nj}-1]^2\!+\!
 E^{\rm EFT}_{n\ell j}\!+\!E^{(6)}_{n\ell j}
\!+\!E^{(7)}_{n\ell j}\!+\!E^{(8)}_{n\ell j}
\,,
\label{eq:Efs}
\end{eqnarray}
where 
${f_{nj}=[1 + (Z\alpha)^2(n - \delta)^{-2}]^{-1_{\!}/_{\!}2}}$ 
(with $\delta$ = ${j+\frac{1}{2}-[(j+\frac{1}{2})^2-(Z\alpha)^2]^{1_{\!}/_{\!}2}}$).

As it is customary, we include the exact solution of the Dirac equation with the Coulomb potential in the first term in \eqref{eq:Efs} (after subtracting the leading term). This solution includes an infinite set of relativistic corrections in the infinite proton mass limit. They are organized in even powers of $Z\alpha$. This result should be possible to obtain from an EFT computation. One should use the NRQED Lagrangian (see Sec. \ref{NRQED(e)}) to all orders in the $1/m_e$ expansion but setting the transverse photon components to zero and only including the covariant time derivative (i.e., the longitudinal photon in the Coulomb gauge), setting the mass of the proton to infinity, and all the radiative corrections to the Wilson coefficients to zero. At low orders, one has the Balmer formula (which we explicitly subtract) and the first relativistic corrections are traditionally associated with the Breit-Fermi potential (but setting the proton mass to infinity). 

Some recoil corrections associated with the finite mass nature of the proton are included in the second term of \eqref{eq:Efs}. This term is generated by considering some relativistic corrections in $1/M$ but computing the matrix elements using the solution of the Dirac equation with a Coulomb potential. 

The term $E_{\rm EFT}$ incorporates all the contributions that have been computed using the pNRQED($ep$) EFT defined in Sec. \ref{Sec:pNRQED(e)} up to $\mathcal{O}(m_r\alpha^5)$ plus higher-order contributions from the Wilson coefficients in Eqs.~(\ref{cFe}-\ref{d2e}). The total contribution reads\footnote{ Contributions to the hyperfine splittings have been omitted from the expression in \eqref{eq:EEFT}. These are given by expectation values of operators $\langle {\bf L}{\bf S}_p\rangle$, $\langle \hat {\bf S}_{pe}\rangle$ and $\langle {\bf S}_p{\bf S}_e\rangle$. They read
\begin{align}
E^{\rm EFT,hfs}_{n\ell j f}&=\left[\frac{8 \pi  \alpha  c_F^{(e)} c_F^{(p)}}{3 m_e M}\right]\frac{ (\alpha  m_r Z)^3 }{2\pi  n^3}\left(f(f+1)-\frac{3}{2}\right) \delta _{0\ell}\nn\\
&+\left[\frac{D_s c_F^{(e)} c_F^{(p)}}{2 m_e M}+X_{LS_p} \left(\frac{c_F^{(p)}}{m_e M}+\frac{c_S^{(p)}}{2 M^2}\right)\right]\frac{(1-\delta _{\ell 0})}{\ell(\ell+1) (2 \ell+1) }\frac{2 \alpha (Z\alpha  m_r )^3}{n^3},
\end{align}
where
\begin{align}
X_{LS_p}&= \frac{\ell (\ell+1)\left(f (f+1)-j (j+1)-\frac{3}{4}\right) }{2 j (j+1)}\left(1-\frac{1}{2 (2 j+1) (\ell-j)}\right),\quad D_s=\frac{\ell (\ell+1) \left(f(f+1)-j(j+1)-\frac{3}{4}\right)}{2 j (j+1) (2 j+1) (\ell-j)}
\,.
\end{align}}  
\begin{align}\label{eq:EEFT}
E^{\rm EFT}_{n\ell j}&=\frac{m_r (Z\alpha  )^4 }{2 n^3}\left(\frac{1}{j+\frac{1}{2}}-\frac{3}{4 n}+\frac{m_r}{ (m_e+M)}\frac{1}{4n}\right)\nn\\
&+\frac{(Z\alpha )^4 }{8 n^3}\left[m_r \left(\frac{3}{n}-\frac{8}{2 \ell+1}\right)-\frac{m_r^3 }{m_e M}\left(\frac{1}{n}+\frac{8}{2 \ell+1}+\frac{32 \alpha }{3 \pi }\frac{ (m_e Z+M)^2 }{ m_e M}\ln R(n,\ell)\right)\right]\nn\\
&+ \left[\frac{\pi \alpha  }{2 m_e M}\left(\frac{Z c_D^{(e)} M}{m_e}+\frac{c_D^{(p)} m_e}{M} \right)-16 \pi  Z\alpha  \left(\frac{d_2^{(e)}}{m_e^2}+\frac{d_2^{(\mu)}}{m_
\mu^2}+\frac{d_2}{M^2}\right)-\frac{c_3}{  M^2}\right.\nn\\
&\left.+\frac{2 \pi  Z\alpha }{m_e M}\left(1+\frac{4 \alpha  Z}{3 \pi }-\frac{7 Z\alpha  }{3 \pi }\left(\frac{1}{2 n}-H_n+\ln \frac{\nu  n}{2 \alpha  m_r Z}\right)\right)\right.\nn\\
&\left.+  Z\alpha^2 \left(\frac{1}{m_e}+\frac{Z}{M}\right)^2 \left(\frac{10}{9}-\frac{4}{3} \ln \frac{\alpha ^2 m_r Z^2}{\nu }\right)\right]\frac{  (\alpha  m_r Z)^3}{\pi  n^3}\delta _{\ell 0}\nn\\
&+ \left[X_{LS_e} \left(\frac{ Zc_F^{(e)}}{m_e M}+\frac{ Zc_S^{(e)}}{2 m_e^2}\right)+\frac{Z }{2 m_e M}\left(\ell(\ell+1)-\frac{7 Z\alpha }{3 \pi }\right)\right]  \frac{(1-\delta _{\ell 0})}{\ell(\ell+1) (2 \ell+1) }\frac{2 \alpha (Z\alpha  m_r )^3}{n^3}
\,,
\end{align}
where $H_n$ is the $n$-th harmonic number, $\delta_{\ell 0}$ is the Kronecker delta,
\begin{align}
X_{LS_e}&=\frac{1}{2}(j(j+1)-\ell(\ell+1)-3/4)
\,,
\end{align}
$\ln R(n,\ell)$ are the
Bethe logarithms for hydrogen 
(tabulated in 
Ref.~\cite{Drake:1990zz}), and 
\be
c_3|_{{\cal O}(\al^2)}=-\frac{M}{m_{e}}d_s(\nu)+c_{3, \rm TPE}^{\rm had}
\,,
\ee
(see Eq.~(\ref{c3nu})). $c_{3, \rm TPE}^{\rm had}$ is of relative order ${\cal O}(\alpha \times \frac{m_e}{m_{\pi}})$ compared to the proton radius, albeit logarithmically enhanced by a $\ln (m_e/m_{\pi})$ factor. The logarithmic enhanced contribution can be obtained using chiral perturbation theory. It can be found in Eq.~(46) of \cite{Pineda:2004mx}, and can be written in terms of the proton polarizabilities (see \cite{Friar:1997tr,Khriplovich:1997fi}). For the pure pion cloud, the polarizabilities were computed in Ref. \cite{Bernard:1992qa}. The contribution due to the Delta particle can be found in Ref. \cite{Hemmert:1999pz}. The numerical impact of these logarithmic enhanced terms to the Lamb shift is of order $\sim -\frac{\delta_{l0}}{n^3}0.08$ kHz (see \cite{Pineda:2004mx}). What it has been missing this far was the computation of $c_{3, \rm TPE}^{\rm had}$ using chiral perturbation theory beyond the logarithmic approximation. We profit this review to do such computation and fill this gap. We give in Eq.~\eqref{eq:c3TPEep} the full correction from the chiral theory following the analysis of Ref.~\cite{Peset:2014jxa}, both, for the pure pion case, and by also including the Delta particle. The numerical impact of $c_{3, \rm TPE}^{\rm had}$ is $\sim -\frac{\delta_{l0}}{n^3}0.124(42)$ kHz (see Table~\eqref{tab:ETPEep}). 

Note that there are no $m_e (Z\al)^5$ effects in the infinite proton mass limit (they only appear after including $1/M$ corrections) in Eq.~(\ref{eq:EEFT}). 
Note also that the first two terms in \eqref{eq:EEFT} are introduced to compensate for double counting with the first two terms in Eq.~\eqref{eq:Efs}. Finally, it can be seen that the contributions of ${\cal O}(m_r\alpha^6)$ and ${\cal O}(m_r\alpha^7)$ in \eqref{eq:EEFT}, which are associated with the hard scale, correspond to the contributions from the functions $B_{40}$ and $C_{40}$ in \cite{Mohr:2012tt}, respectively.
 
The terms $E^{(n)}_{n\ell j}$ in \eqref{eq:Efs} contain the remaining known terms of $\mathcal{O}(m_r\alpha^n)$. $E^{(6)}_{n\ell j}$ is fully known in the static limit and at leading recoil order. It reads
\begin{align}\label{eq:E6}
E^{(6)}_{n\ell j}&=\frac{m_r^3 (Z\alpha)^6}{m_e M n^3}\frac{3  \left( 1-\delta_{\ell 0}\right)}{4 \left(\ell-\frac{1}{2}\right) \left(\ell+\frac{1}{2}\right) \left(\ell+\frac{3}{2}\right)}\left(1-\frac{\ell (\ell+1)}{3 n^2}\right)\nn\\
&+  \left[\frac{m_r^3(Z\alpha)^6}{3 n^3} r_p^2 \ln \frac{\nu }{m_r(Z\alpha)^{2}}
-\frac{\delta c_3^{(1)}+\delta c_3^{(1),\text{had}}}{M^2}\frac{  (\alpha  m_r Z)^3}{\pi  n^3}\right]\delta _{\ell 0}
\,,
\end{align}
where
\begin{align}
\delta c_3^{(1)}&=\frac{\alpha(Z\alpha)^2\pi  M^2} {m_e^2}\left(2 \ln 2-\frac{427}{96}\right)+\frac{\alpha (Z\alpha)^2 M}{108 \pi  m_e} \left(1792+3 \pi ^2 (72 \ln 2-35)-648 \zeta (3)\right)\nn\\
&-\frac{ (Z\alpha)^3 \pi  M}{m_e}\left(4 \ln 2-\frac{7}{2}\right),\quad \delta c_3^{(1),\text{had}}=-\frac{ \pi  (Z\alpha)^3}{3}M^2 r_p^2 \ln \frac{\Lambda}{\nu}. 
\end{align}
The first term of Eq.~(\ref{eq:E6}) was computed in e.g. \cite{Jentschura:1996zz} (it corresponds to Eq.~(4.23) in \cite{Eides:2007exa}). It could be obtained from the determination of ${\cal O}(1/M)$ tree-level diagrams in NRQED and correspondingly matching them to potentials in pNRQED (see for instance \cite{Anzai:2018eua}). The logarithmic dependent term in \eqref{eq:E6}, proportional to $r_p^2$, comes from second order perturbation theory of the delta-potential and was originally computed in \cite{Friar:1978wv} (see \cite{Pineda:2001et} for the computation in the setup of the pNRQED). 
Note that we only keep the logarithmic term with this precision, as finite corrections will be of the same order as contributions in the Wilson coefficient $\delta c_3^{(1),\text{had}}$, which are model dependent.  ${\cal O}(\alpha^3)$ corrections to $c_3$ are encoded in $\delta c_3^{(1)}$ and $\delta c_3^{(1),\text{had}}$. The first term encodes the contributions that are independent of the structure of the proton. They scale like $1/m_e^2$ and $1/m_e$. They would be generated by a matching computation after integrating out the hard scale $\sim m_e$ (see, for instance, Figs.~3.9 and 3.10 in \cite{Eides:2007exa}, which would produce contributions that would be encoded in the coefficients $A_{50}$ and $V_{50}$ in the notation of \cite{Mohr:2012tt}). The second term encodes the corrections that are sensitive to the structure of the proton, of these we only keep the leading logarithmic-dependent terms, as it is the only piece that is model independent.\footnote{Contrary to \cite{Eides:2007exa,Mohr:2012tt}, we do not include the non-logarithmic $r_p$-dependent contribution of order $(Z\alpha)^6$ as its computation contains certain model dependence.} Their role is to cancel the scale dependence of the second order perturbation theory computation in the EFT. One thing that has to be determined is the scale at which the divergence gets regulated. Typically, it is assumed that it is regulated at some hadronic scale, i.e., $\Lambda \sim m_{\pi}$ or higher. For instance, in \cite{PhysRevA.93.022513}, the inverse of the proton radius was taken. This is something that should be more deeply investigated. In this respect, note that there are no $m_e\al^6 \ln (1/\alpha)$ corrections to the Lamb shift if one considers the proton to be point-like. The effect in the energy shift is however small. 

We now summarize the remaining $\mathcal{O}(m_e\alpha^7)$ and $\mathcal{O}(m_e\alpha^8)$ corrections:
\begin{align}\label{eq:E7}
E^{(7)}_{n\ell j}=
&\frac{\alpha}{\pi}\frac{(Z\alpha)^6 }{n^3}\frac{m_r^3}{m_e^3}
m_e(A_{62}^{( \ell)}\lambda^2+A_{61}^{(n \ell j)}\lambda +G_{\rm SE}^{(n \ell j)}(Z\alpha))\nn\\
&+\frac{\alpha}{\pi}\frac{(Z\alpha)^6}{n^3}\frac{m_r^3}{m_e^3}m_e(V_{61}^{( \ell)}\lambda+G_{\rm VP}^{(n \ell j)}(Z\alpha))\nn\\
&+\left(\frac{\alpha}{\pi}\right)^2\frac{(Z\alpha)^5}{n^3}\frac{m_r^3}{m_e^3}m_e B_{50}^{( \ell)}+\frac{m_e}{M}\frac{\alpha}{\pi}\frac{(Z\alpha)^6}{n^3}m_e \frac{m_r^3}{m_e^3}\left(\frac{2}{3}\lambda^2+\frac{\Delta_{RR}}{\pi}\lambda\right)\delta_{\ell 0}\nn\\
&+\frac{m_e}{M}\frac{(Z \alpha)^7}{\pi n^3}m_e\left(D_{72}^{( \ell )}\lambda^2+D_{71}^{(n \ell j)}\lambda\right)
\,,
\end{align}
and
\begin{align}\label{eq:E8}
E^{(8)}_{n\ell j}&=\left(\frac{\alpha}{\pi}\right)^2\frac{(Z\alpha)^6}{n^3}m_e (B_{63}^{(n\ell)}\lambda^3+B_{62}^{(n\ell)}\lambda^2+B_{61}^{(n\ell)}\lambda+
G_{\text{TPC}}^{(n\ell j)}(Z\alpha))\nn\\
&+\left(\frac{\alpha}{\pi}\right)^3\frac{(Z\alpha)^5}{n^3}m_e C_{50}^{(n\ell ) }
\,.
\end{align}
Note that in \eqref{eq:E7} there are no corrections of ${\cal O}(m_e(Z\alpha)^4\alpha^3)$, as they are all encoded in the Wilson coefficients in \eqref{eq:EEFT}. Actually, it is a general pattern that the corrections of order $m_r(Z\al)^4\al^n$ can be encoded in $E^{\rm EFT}_{n\ell j}$, either from the ultrasoft ${\cal O}(\al^5)$ correction or from radiative corrections associated with the hard scale (the mass of the electron) encoded in the Wilson coefficients. This discussion shows that, at present, there are uncomputed ${\cal O}(m_r(Z\alpha)^4\alpha^4)$ contributions to the Lamb shift. To obtain those, one would need to compute $c_D^{(e)}$ and $c_F^{(e)}$ to ${\cal O}(\alpha^4)$. In Ref. \cite{Mohr:2012tt}, the estimated error associated with these contributions was assumed to be
\be
\left(
\frac{\al}{\pi}
\right)^4
\frac{(Z\al)^4}{n^3}m_e \sim \frac{10\; {\rm Hz}}{n^3}
\,.
\ee
This estimate assumes that the coefficient multiplying this correction is of ${\cal O}(1)$.

In Eqs.~(\ref{eq:E7}) and (\ref{eq:E8}), we follow the standard notation used in \cite{Mohr:2012tt} (but also compare with \cite{Eides:2007exa}), except for $G_{\text{TPC}}^{(n\ell j)}(Z\alpha)$, which we take from \cite{Karshenboim:2019iuq}, and replaces $B_{60}^{nl}$. In the notation of these papers, $D$ functions come from relativistic recoil corrections, $A$'s from self energy, $V$'s from vacuum polarization, $B$'s from two-photon corrections and $C$'s from 3-photon corrections.\footnote{Note that in the previously discussed $m_r(Z\alpha)^4\alpha^n$ corrections, all the effects encoded in the $A$, $B$, ... functions are encoded in the Wilson coefficients of the effective theory.}
The different terms of these corrections read,  
$\lambda$=$\ln ( m_{e}/(m_{r}(Z\alpha)^{2}) )$,
\begin{align}
&A_{62}^{( \ell)}\!=-\delta_{\ell 0}
\label{eq:A62},\\
&A_{61}^{(n \ell j)}\!=
\big[
4H_n
+\frac{28}{3} \ln 2
-4 \ln n
-\frac{601}{180}
-\frac{77}{45 n^2}
\big] \delta_{\ell 0}
\!+(1\!-\!\frac{1}{n^2})(\frac{2}{15}\!+\!\frac{\delta_{j \frac{1}{2}}}{3})
\delta_{\ell 1}
\!+\!\frac{8(3n^2\!-\!\ell(\ell\!+\!1))(1\!-\!\delta_{\ell 0})}
{3n^2\ell(4\ell^2\!-\!1)(\ell\!+\!1)(2\ell\!+\!3)}\label{eq:A61},
\end{align}
\begin{align}
B_{50}^{(\ell)}&\!\!=\!\! -21.55447(13)\delta_{\ell 0}
\,,
\label{eq:B50}\\
B_{63}^{(n\ell)}&\!\!=\!\! -\frac{8}{27}\delta_{\ell 0}
\,,
\label{eq:B63}\\
B_{62}^{(n\ell)}&\!\!=\!\!
\frac{16}{9}
\big[
\frac{71}{60}-\ln 2+\gamma
\!+\!\psi(n)
\!-\!\ln n
\!-\!\frac{1}{n}
\!+\!\frac{1}{4n^2}
\big]\delta_{\ell 0}
\!+\!\frac{4 \delta_{\ell 1}}{27}\frac{n^2\!-\!1}{n^2},\label{eq:B62}\\
B_{61}^{(n\ell)}&\!\!=\!\! \left[\frac{413581}{64800}+\frac{4N(nS)}{3}+\frac{2027\pi^2}{864}-\frac{616\ln 2}{135}-\frac{2\pi^2\ln 2}{3}+\frac{40\ln^2 2}{9}+\zeta(3)\right.\nn\\
&\left.
-\frac{43}{36}+\frac{709\pi^2}{3456}
+\left(\frac{304}{135}-\frac{32 \ln 2}{9}\right)\left(\frac{3}{4}+\gamma+\psi(n)-\ln n-\frac{1}{n}+\frac{1}{4n^2}\right)\right]\delta_{\ell 0}\nn\\
&+\left[\frac{4}{3}N(nP)+\frac{n^2-1}{n^2}\left(\frac{467}{810}-\frac{j}{3}-\frac{8}{27}\ln 2\right)\right]\delta_{\ell 1}
\,,
\label{eq:B61}
\end{align}
with $\psi$ being the digamma function and $N(n\ell)$ is the term generated by the Dirac delta correction to the Bethe logarithm tabulated in \cite{Jentschura:2003we},
\begin{align}
D_{72}^{(\ell )}&=-\frac{11}{60}\delta_{\ell 0}\,,
\label{eq:D72}
\end{align}
and
\begin{align}
V_{61}^{(\ell )}\!=-\frac{2}{15}\delta_{\ell 0}\label{eq:V61}
\,.
\end{align}

As we said, we follow the notation of \cite{Mohr:2012tt}, but these contributions can also be found in \cite{Eides:2007exa}. We give here the relation of our equations to those of this last reference for ease of reference: \eqref{eq:V61} corresponds to Eq.~(3.66).  \eqref{eq:A62} to the double log in Eq.~(3.53). \eqref{eq:A61} to the sum of the single log in Eq.~(3.53) and~(3.55). \eqref{eq:D72} is the recoil correction in Eq.~(4.24). $B_{50}^{(n\ell)}$ has been computed numerically in \cite{Dowling:2009md}. Comparing to the results in \cite{Eides:2007exa}, we find that \eqref{eq:B50} is the sum of Eqs.~(3.41)-(3.43) and (3.46)-(3.48).
For the other coefficients the equivalence is the following:  \eqref{eq:B63} is Eq.~(3.75), \eqref{eq:B62} corresponds to Eqs.~(3.77),~(3.78), (3.86) and~(3.92), while  \eqref{eq:B61} can be obtained from Eqs.~(3.79),~(3.80), (3.83),
(3.87), (3.88) and (3.93)-(3.94). To $B_{61}^{(n\ell)}$ in \eqref{eq:B61} we have also added the missing light-by-light contribution computed in \cite{Czarnecki:2016lzl,Szafron:2019tho}.
The coefficients $A_{61}$ and $A_{60}$ ($G_{\rm SE}^{(n \ell j)}(0)$), 
$B_{62}$, $B_{61}$ and $B_{60}$ have been computed in \cite{Czarnecki:2016lzl,Szafron:2019tho,Jentschura:2005xu} in the context of NRQED.  
The values of 
$G_{\rm SE}^{(n \ell j)}(Z\alpha)$ and
$G_{\rm VP}^{(1)(n \ell j)}(Z\alpha)$ of Eq.~(\ref{eq:E7}),
are discussed in 
Ref.~\cite{Mohr:2012tt}, where one can find tabulated values (see also 
Refs.~\cite{Jentschura:2000xd,Jentschura:2004uz,Jentschura:2005xu}). $G_{\rm TPC}^{(1)(n \ell j)}(Z\alpha)$ is discussed in \cite{Yerokhin:2018gna, Karshenboim:2019iuq}. Note that their determinations formally obtain higher orders in $\alpha$, including some logarithms of ${\cal O}(m_r\alpha^8)$. Expanding in $Z\alpha$,
\begin{align}
G_{\rm SE}^{(n \ell j)}(Z\alpha)&=A_{60}^{(n \ell j)}+A_{71}^{(n \ell j)}(Z\alpha)\lambda+\cdots,\label{eq:GSEexp}\\
G_{\rm VP}^{(n \ell j)}(Z\alpha)&=V_{60}^{(n \ell j)}+V_{71}^{(n \ell j)}(Z\alpha)\lambda+\cdots,\\
G_{\rm TPC}^{(n \ell j)}(Z\alpha)&=B_{60}^{(n \ell j)}+B_{71}^{(n \ell j)}(Z\alpha)\lambda+\cdots,
\end{align}
and, from Eqs.~(3.97) and~(3.100) in \cite{Eides:2007exa}, one obtains
\begin{align}
A_{71}^{(n \ell j)}&=\pi\left(\frac{139}{64}-\ln 2\right)\delta_{\ell 0}, \quad V_{71}^{(n \ell j)}=\frac{5\pi}{96}\delta_{\ell 0}.
\end{align}
Values for $A_{60}^{(n \ell j)}$ can also be found in Table~3.4 of \cite{Eides:2007exa} and $V_{60}$ is obtained from Eqs.~(3.64),~(3.68) and~(3.70) of this reference. It is worth noting that the difference between the all-order computation of $G_{\rm SE}^{(n \ell j)}(Z\alpha)$ and the expansion in \eqref{eq:GSEexp} is $\sim 13$ kHz for 1S hydrogen. The attitude towards this problem is to take the numerical determination as the correct one (assuming that the uncomputed terms would make up for the difference). The claimed error of the numerical evaluation of $G_{\rm SE}^{(n \ell j)}(Z\alpha)$ is enough for nowadays required precision. $G_{\rm TPC}^{(n \ell j)}(Z\alpha)$ is computed in \cite{Karshenboim:2019iuq} for the $1S$ state with an error of $\sim 7$\%. For higher excitations, only $B_{60}$ is computed numerically (with the additional light-by-light contribution found in \cite{Szafron:2019tho}), and the error quoted in \cite{Mohr:2012tt} is of the order of 30\%.

The function $C_{50}^{(n\ell)}$ in \eqref{eq:E8} is not well known. The value quoted in \cite{Mohr:2012tt} corresponds to the partial results found in \cite{Eides:2006hg} (they correspond to Eqs.~(3.51)-(3.52) in \cite{Eides:2007exa}). A more recent and precise estimate was obtained in \cite{Karshenboim:2019iuq} from where we take our result. 

The values of $D_{71}^{(n \ell j)}$ have been recently calculated numerically in \cite{PhysRevLett.115.233002}, where they also give an estimate for $D_{70}^{(n \ell j)}$. Other functions can be found in CODATA \cite{Mohr:2012tt} and agree with \cite{Eides:2007exa}. 

The identification of Eqs.~(\ref{eq:E7}) and (\ref{eq:E8}) with a computation in the effective theory is more complicated, except for the analysis performed in \cite{Jentschura:2005xu}. One can also see that the coefficient $B_{50}$ could be associated with a hard contribution, and similarly for $C_{50}$, though this coefficient is not known completely. Other contributions can be thought of as combinations of perturbation theory of potentials and ultrasoft loops. This is particularly so for those that have a nontrivial (different from $\delta_{l0}/n^3$) dependence on the quantum numbers of the state. 

We now enumerate the dominant theoretical uncertainties to the fine-structure energy contributions $E^{\rm (fs)}_{n\ell j}$ (a more detailed account can be found in Ref.~\cite{Mohr:2012tt}). The first comes from the error of the coefficient $G_{\rm TPC}^{(n \ell j)}$ ($B_{60}^{(n \ell)}$), which leads to an error of order $\delta_{\ell 0}$(0.7~kHz)/$n^3$ for the $1S$ energy level and $\delta_{\ell 0}$(2.0~kHz)/$n^3$ for higher $nS$ states  (for other energy levels the error is smaller).
The second comes from the uncertainty in $r_{\rm p}$. For instance, a 1 per mille error in the determination of the proton radius produces a shift of order  
$\delta_{\ell 0}$(2.0~kHz)/$n^3$ in the energy levels.
The third comes from the coefficient $C_{50}$
of Eq.~(\ref{eq:E6}),  which, following the computation in \cite{Karshenboim:2019iuq}, yields
$-3.3(10.5)\delta_{\ell 0}$. This leads to an uncertainty of 
$\delta_{\ell 0}$(0.3~kHz)/$n^3$.
The fourth comes from uncomputed terms of order $m_e \alpha^7 \ln 1/\alpha \times \frac{m_e}{M}$. Following \cite{Mohr:2012tt,PhysRevA.93.022513}, the error associated with this term is assumed to yield an  uncertainty of order  
$\delta_{\ell 0}$(0.7~kHz)/$n^3$.
An additional uncertainty associated with $D_{71}$ 
has now been resolved by Ref.~\cite{PhysRevLett.115.233002}
and does not need to be included.
All other uncertainties listed in 
Ref.~\cite{Mohr:2012tt} are more than one order of magnitude
smaller. 

Finally, it is worth emphasizing that the theoretical expressions collected in this review are different with respect to those in Refs.~\cite{Mohr:2012tt,PhysRevA.93.022513}, as we neglect terms that may introduce some model dependence, or that are of the same order as these. Numerically, the differences are irrelevant (of the order of 0.4 kHz, 0.02 kHz, and 0.002 kHz for 1S, 2S and 2P states), showing that this is a safer way to proceed. 

\subsection{Hydrogen hyperfine splitting}

The hyperfine splitting of the ground state of hydrogen is one of the most accurate measurements made by mankind~\cite{exp1,exp2,exp3,exp4,exp5,exp6,exp7,exp8} with the result
\be
E_{\rm hyd,HF}^{\rm exp}(1S)=1420.405751768(1)\;{\rm MHz} \label{eq:Hexpt}
\,,
\ee
which we take from the average of Ref.~\cite{Karshenboim:2005iy}.

Since then, there has been a continuous effort to derive this number from theory.   
The first five digits of this number can be reproduced by the theory of an infinitely massive proton (except for the  
$1/M$ prefactor of the Fermi term incorporating the anomalous magnetic moment of the proton) and a nonrelativistic lepton, systematically incorporating the relativistic corrections of the latter. A summary of these pure QED computations can be found in Refs.~\cite{BY,Karshenboim:2005iy,Eides:2000xc,Mohr:2012tt}. Particularly detailed is the account of Refs.~\cite{Eides:2000xc,Eides:2007exa}, which we take for reference. Such computation has reached (partial) ${\cal O}(m_e\al^8)$ precision and reads 
\bea
\nn
&&
\delta E_{\rm hyd,HF}^{\rm QED}(1S)=
\frac{8m_r^3\alpha(Z\alpha)^3}{3Mm_e} c_F^{(p)}c_F^{(e)}
+
\frac{8m_r^3\alpha(Z\alpha)^3}{3Mm_e} c_F^{(p)}Z\al^2
\left\{ \frac{3}{2}Z +\left[\ln 2-\frac{5}{2}\right]
\right.
\\
&+& \frac{Z\alpha}{\pi}\left[-\frac{2}{3} \ln ^2\frac{1}{(\alpha  Z)^2}-\left(\frac{8 \ln 2}{3}-\frac{281}{360}\right) \ln \frac{1}{(\alpha  Z)^2}+\frac{34}{225}+17.1227(11)-\frac{8 \ln 2}{15}\right]\nn\\
&+&
\frac{\alpha}{\pi}\left[\frac{\pi ^2}{9}-\frac{38 \pi }{15}+\frac{91639}{37800}-1.456(1)
-\frac{4}{3}\left(\ln\frac{1+\sqrt{5}}{2}+\frac{5}{3}\sqrt{5}\right)  \ln \frac{1+\sqrt{5}}{2} +\frac{608 \ln 2}{45}\right]
\nn\\
&+&
\frac{Z\alpha^2}{\pi^2}\left[10(2.5)-\frac{1}{3} \ln ^2\frac{1}{(\alpha  Z)^2}+\left(\frac{27 \zeta (3)}{16}+\frac{25187}{8640}+\pi ^2 \left(\frac{133}{432}-\frac{9 \ln 2}{8}\right)-\frac{4 \ln 2}{3}\right) \ln\frac{1}{(\alpha  Z)^2}\right]
\nn\\
&+& Z^2\alpha^2\left[\left(\frac{5 \ln 2}{2}-\frac{547}{96}\right) \ln 
\frac{1}{(\alpha  Z)^2}-3.82(63)\right]
\nn\\
&+& Z^3\alpha^2\left.\left[\frac{17}{8}\right]+{\cal O}(Z^0\al^2)\right\}
\,,
\label{QED}\eea
where $m_r=Mm_e/(M+m_e)$.  $c_F^{(p)}\equiv Z+\kappa_p=2.792847356$ 
and $c_F^{(e)}=1.00115965$
 (see Eq.~(\ref{cFe})) are the magnetic moments of the proton and electron respectively, which we take exactly, i.e., they include the ${\cal O}(\al)$ corrections. 
 Besides those, there are pure QED recoil corrections of ${\cal O}\left(m_e \al^6 \frac{m_e}{M}\right)$, 
 computed in Ref.~\cite{BY} (we take $Z=1$ in this expression):
 \bea
 \label{recoil}
 \delta E_{\rm hyd,HF}^{\rm QED,recoil}(1S)&=&
 \frac{8m_r^3\al^6}{3m_eM}c_F^{(p)}\frac{m_e}{M}
 \left[
\kappa_p\left(\frac{7}{4} \ln \frac{1}{2 \alpha }-\ln 2+\frac{31}{36}\right)+2 \ln\frac{1}{2 \alpha }-6 \ln 2+\frac{65}{18}\right.\nn\\
&+&\left.\frac{\kappa_p }{\kappa_p+1}\left(-\frac{7}{4} \ln\frac{1}{2 \alpha }+4 \ln 2-\frac{31}{8}\right)
 \right]
 \,.
 \eea 
On top of these, there are recoil corrections of higher orders, as well as corrections due to the hadronic structure of the proton. Similarly to the discussion we had in previous sections for the Lamb shift in regular hydrogen, it would be helpful to organize such computations/results using EFT techniques designed for few-body atomic physics such as NRQED and potential NRQED, as these theories profit from the hierarchy of scales that we have in the problem. Nevertheless, this has not yet been done. What has been done already is to relate this result with the Wilson coefficients that appear in NRQED. This was done in \cite{Peset:2016wjq}. The result can be summarized in the following expression 
\bea
&&
\label{046}
E_{\rm hyd,HF}^{\rm exp}(1S)=\delta E_{\rm hyd,HF}^{\rm QED}(1S)+\delta E_{\rm hyd,HF}^{\rm QED, recoil}(1S)
\\
\nn
&+&\frac{4m_r^3\al^3}{\pi M^2}
\left(c_4^{pe}-\al\left(\frac{M}{m_e}\delta c_4^{pe,(-1)}+\frac{2}{3}\pi\left(2c_F^{(p)}+\frac{7}{4}\kappa_p^2\right)\ln \frac{m_e}{\nu}+\delta K_{\rm hard}\right)\right)+{\cal O}(m\al^8, m\al^7\frac{m_e}{M})
\,,
\eea 
where the second line is the contribution associated with the TPE correction minus its point-like contributions that are already included in the first line of Eq.~(\ref{046}) (see \cite{Peset:2016wjq} for details). 

The comparison of this theoretical result with the experimental number allows to determine (a specific part of) the Wilson coefficient 
of the four-fermion operator of the NRQED Lagrangian (a preliminary number was already obtained in Refs.~\cite{Pineda:2002as,Pineda:2003hb} (see also \cite{Peset:2014jxa})), which we name $\bar c_{4,\rm TPE}^{p e}$:
\bea
\label{c4peHF}
&&
\bar c_{4,\rm TPE}^{p e}\equiv c_4^{pe}-\al^3\left(\frac{M}{m_e}\delta c_4^{pe,(-1)}+\frac{2}{3}\pi\left(2c_F^{(p)}+\frac{7}{4}\kappa_p^2\right)\ln \frac{m_e}{\nu}+\delta K_{\rm hard}\right)
\\
\nn
&&=
c_{4,\rm TPE}^{p e}+\al^3 [K^{pe}-\delta 
K_{\rm hard}]+{\cal O}(\al^4,\al\frac{m_e}{M})
\,.
\eea
It differs from $c_{4,\rm TPE}^{p e}$ by ${\cal O}(\al^3)$ effects (note that $c_{4,\rm TPE}^{p e}$ is of order $\alpha^2$). 

Alternatively, one can determine this Wilson coefficient using dispersion relations. See the discussion in Sec.~\ref{Sec:c4had}.

\subsection{Muonic hydrogen hyperfine splitting}

The EFT for the muonic hydrogen hyperfine splitting is again pNRQED($\mu p$) (see Sec \ref{Sec:pNRQEDmup}). The computation in the EFT was done in \cite{Peset:2016wjq}. Similarly to muonic hydrogen Lamb shift, several of the computations were made before and a summary of these results for the $2S$ hyperfine can be found in \cite{Antognini:2013rsa} (see also~\cite{Indelicato:2012pfa}). Similarly to the previous section, we define\footnote{$K^{p\mu}$ includes all physical effects associated with the chiral and higher scales, except for those we explicitly subtract from point-like computations at the muon mass scale. This is different for $K^{pe}$, which incorporates all the corrections generated by the chiral scale.}
\bea
\bar c_{4,\rm TPE}^{p \mu}&\equiv&
c_{4}^{p \mu}-\al^3\left(
\frac{M}{m_{\mu}}\delta c_4^{p\mu,(-1)}+\frac{2}{3}\pi\left(2c_F^{(p)}+\frac{7}{4}\kappa_p^2\right)\ln \frac{m_{\mu}}{\nu}+\delta K_{\rm hard}
\right)
\\
\nn
&=&
c_{4,\rm TPE}^{p \mu}+\al^3[K^{p\mu}-\delta 
K_{\rm hard}]+{\cal O}(\al^4,\al^3\frac{m_{\mu}}{M})
\,,
\\
\delta \bar E^{\rm TPE}_{p \mu,\rm HF}(nS)&=&\frac{4m_r^3\al^3}{n^3 \pi M^2}\bar c_{4,\rm TPE}^{p \mu}
\,.
\eea
With these definitions, the $1S$  hyperfine splitting energy shift reads
\be
\label{1Smuonichydrogen}
E_{p\mu,\rm HF}^{\rm th}(1S)=183.788(7)\,\mathrm{meV}+(1+0.0040)\delta \bar E^{\rm TPE}_{p \mu,\rm HF}(1S)\,,
\ee
where the error in the first term is the expected size of the ${\cal O}(m_\mu \al^6)$ uncomputed corrections (either 
related to the electron vacuum polarization effects or to higher-order recoil corrections).

For the $2S$ hyperfine splitting energy shift, one has
\bea
\label{2Smuonichydrogen}
&&E_{p\mu,\rm HF}^{\rm th}(2S)=22.9579(8)\,\mathrm{meV}+(1+0.0033)\delta \bar E^{\rm TPE}_{p \mu,\rm HF}(2S)\,,
\eea
where, again, the error in the first term is the expected size of the ${\cal O}(m_\mu \al^6)$ uncomputed corrections (either 
related to the electron vacuum polarization effects or to higher-order recoil corrections).

\section{Observables. Low-\texorpdfstring{$Q^2$}{Q2} limits of lepton-proton scattering}

In this section, we describe low-momentum transfer limits of the elastic lepton-proton scattering ($Q^2 \ll m_{\mu}^2,\; m_{\pi}^2$). We obtain expressions valid in the lepton nonrelativistic limit:
 \be\label{eq:defBeta}
 \beta = \frac{\sqrt{E^2-m^2_{l_i}}}{E}=\frac{|{\bf k}|}{E} \ll 1.
\ee
In this limit, the lepton-proton cross section can be written in terms of the same Wilson coefficients that appear in (muonic) hydrogen spectroscopy. This allows for a smooth and controlled connection with spectroscopy analyses. To cover the kinematic range of modern electron-proton and muon-proton scattering experiments, we relax the condition of nonrelativistic leptons: $\beta \ll 1$. We reproduce known results for point-particle contributions and extend them to a general kinematic setup.

\subsection{Nonrelativistic lepton(muon)-proton scattering}
\label{Sec:NRmupscat}

We consider first the kinematics of the elastic scattering of a lepton on the proton (nucleus) target at rest, which defines the laboratory frame: $p=(M,0)$, $k=(E,{\bf k})$, $p'=(E_p',{\bf k}-{\bf k}')$, $k'=(E',{\bf k}')$, and the lepton scattering angle is $\theta_{\rm lab}$. The lepton-proton scattering cross section at tree level is conveniently expressed as
\be
\label{treelevelPointlike}
\left(\frac{d\sigma_{1\gamma}}{d\Omega}\right)_{\rm point-like}\equiv 
\left(\frac{d\sigma_{0}}{d\Omega}\right)
=\frac{2M\alpha^2}{Q^2}\frac{|{\bf k'}|}{|{\bf k}|}
\frac{\varepsilon}{1-\varepsilon_\mathrm{T}}\frac{1+\frac{\tau_p}{\varepsilon}}{M+E-E'\frac{|{\bf k}|}{|{\bf k}'|}\cos \theta_{\rm lab}},
\ee
with kinematic variables
\be
\varepsilon_\mathrm{T}=\frac{\nu^2-M^4\tau_p(1+\tau_p)(1+2\varepsilon_0)}{\nu^2+M^4\tau_p(1+\tau_p)(1-2\varepsilon_0)}
\,,
\qquad
\quad
\varepsilon=\varepsilon_\mathrm{T}+\varepsilon_0(1-\varepsilon_\mathrm{T})
\,,
\qquad
\quad \varepsilon_0=\frac{2m_\mu^2}{Q^2} \,, 
\qquad
\quad
\tau_p = \frac{Q^2}{4M^2}
\,,
\ee
crossing symmetric variable $\nu=\frac{1}{4}(p+p')\cdot (k+k')=ME-Q^2/4$, and the squared momentum transfer $Q^2 = - (p'-p)^2$. This equation accounts correctly for all relativistic and recoil corrections at tree level. In the static limit of the proton, $M \rightarrow \infty$, we obtain the Mott cross section $\sigma_\text{Mott}$ (see, for instance, \cite{Itzykson:1980rh,Peskin:1995ev})
\begin{align}\label{TLscattering}
\left(\frac{d\sigma_{0}}{d\Omega}\right)_{\rm M \rightarrow \infty}&=\frac{d\sigma_\text{Mott}}{d\Omega}=\frac{\alpha^2}{4{\bf k}^2\beta^2 \sin^4 \frac{\theta_{\rm lab}}{2}}\left(1-\beta^2\sin^2\frac{\theta_{\rm lab}}{2}\right),
\end{align}
with the relativistic lepton velocity $\beta$ in Eq.~\eqref{eq:defBeta}. 
The traditional Mott result corresponds to the scattering of the relativistic lepton on the infinitely heavy source of a Coulomb field. In the 
nonrelativistic limit of the lepton, $\beta \rightarrow 0$, we obtain the traditional Rutherford formula.

The dominant (not suppressed by powers of $\alpha$) structure effects of the spin-1/2 baryon are accounted for by introducing the Sachs electric and magnetic form factors $G_E$ and $G_M$ defined in Sec.~\ref{sec:formfactors} in the differential cross section as \cite{Preedom:1987mx,Tomalak:2018jak}
\be
\label{treelevel}
\left(\frac{d\sigma_{1\gamma}}{d\Omega}\right)_{\rm Sachs}=\frac{2M\alpha^2}{Q^2}\frac{|{\bf k'}|}{|{\bf k}|}
\frac{\varepsilon}{1-\varepsilon_\mathrm{T}}\frac{G_E^2+\frac{\tau_p}{\varepsilon}G_M^2}{M+E-E'\frac{|{\bf k}|}{|{\bf k}'|}\cos \theta_{\rm lab}} = \frac{d\sigma_{0}}{d\Omega} \frac{1}{1+\tau_p} \left( G_E^2+\frac{\tau_p}{\varepsilon}G_M^2 \right).
\ee
To reproduce the point-like particle result of \eq{treelevelPointlike}, one has to make the replacement 
$G_E \rightarrow 1$ and $G_M \rightarrow 1$. 

We now consider the incorporation of ${\cal O}(\alpha)$ corrections. Some of them are already encoded in the Sachs form factors. This makes Eq.~(\ref{treelevel})  to be ill-defined because of the infrared divergences of the proton and/or muon vertices. Such divergences are regulated by the real emission of soft photons from the proton known as soft bremsstrahlung with photons of energy below $\Delta E$. In other words, the observable of $\mu p\to \mu p$ scattering is actually (see, for instance, the discussion in \cite{Peskin:1995ev,Hill:2016gdf}):
\begin{align}\label{sigmameasured}
\left(\frac{d\sigma}{d\Omega}\right)_\text{measured}=\frac{d\sigma}{d\Omega}(p\to p')+\frac{d\sigma}{d\Omega}(p\to p'+\gamma(k_\gamma<\Delta E)).
\end{align}

In the following, we provide the low-momentum transfer expansion, $Q^2 \ll m_{\mu}^2$, of the complete set of ${\cal O}(\al)$ corrections. This means that we will restrict the discussion to the kinematic situation when $\tau_p \ll 1$ and $\tau_{\mu}=\frac{Q^2}{4m_{\mu}^2} \ll 1$. For masses of the muon and proton, $\tau_{\mu} \gg \tau_p$. One can neglect the proton-line contributions with this counting. However, we keep both parameters on the equal footing for generality. To smoothly connect with spectroscopy, we also constrain $\beta \ll 1$ (in other words $|{\bf p}|^2/m_{\mu}^2 \ll 1$). Irrespective of the kinematics, we split the different contributions to the cross section to 
${\cal O}(\alpha)$ in the following way:
\begin{align}
\label{eq:dsigmaTotalmuon}
\left(\frac{d\sigma}{d\Omega}\right)_\text{measured}\hspace{-.5cm}=
Z^2 \left(\frac{d\sigma_{1\gamma}}{d\Omega}\right)_{\rm point-like} 
\hspace*{-.5cm}+
\frac{d\sigma_\text{Mott}}{d\Omega}\left[\delta^{(p)}_{\rm soft}+Z^2 \left( \delta^{(\mu)}_{\rm soft} + \delta_{\rm VP} \right)+Z^3 \left( \delta^{(p\mu)}_{\rm soft} + \delta_{\rm TPE}  \right) +\mathcal{O}(\alpha^2)+\mathcal{O}(\tau^2)\right].
\end{align}
In order to include the electron vacuum polarization correction, we make the replacement~\cite{Mo:1968cg,Bernauer:2013tpr}:
\be
\label{alrep}
\al \rightarrow \frac{\al}{1-\frac{\al}{\pi}\left(\frac{1}{3}\ln\left(\frac{Q^2}{m_e^2}\right)-\frac{5}{9}\right)} 
\simeq \al\left(1+\frac{\al}{\pi}\left(\frac{1}{3}\ln\left(\frac{Q^2}{m_e^2}\right)-\frac{5}{9}\right)\right),
\ee
in the first term of \eq{eq:dsigmaTotalmuon} (this neglects subleading effects in the electron mass 
assuming $Q^2 \gg m^2_e$, which is a good approximation for modern experiments). In the nonrelativistic limit, the different terms in Eq.~\eqref{eq:dsigmaTotalmuon} read
\begin{align}
\delta^{(p)}_{\rm soft} &= \tau_p\left[\beta^2 \left( c_F^{(p)\,2} -Z^2 \right) - Z \left( c_D^{(p)\MS}(\nu) - Z \right)+\frac{4}{3}\frac{Z^4\alpha}{\pi}\left(2\ln \frac{2\Delta E}{\nu} -\frac{5}{3}\right)\right], \label{eq:proton_soft_rad_NR} 
\end{align}
\be
\delta^{(\mu)}_{\rm soft}  \underset{\beta \to 0}{\longrightarrow}
\tau_{\mu}\left[ \beta^2 \left( c_F^{(\mu)\,2} - 1 \right) -  \left( c_D^{(\mu)\MS}(\nu) - 1 \right)+\frac{4}{3}\frac{\alpha}{\pi}\left(2\ln \frac{2\Delta E}{\nu} -\frac{5}{3}\right)\right].
\ee
This computation has been done using NRQED($\mu p$) (see Sec. \ref{NRQED(mu)}). 
Note that, for simplicity, in these two terms we have incorporated the contribution associated with the vertex correction, which in the language of effective theory corresponds to a ``hard'' contribution, and it is incorporated in the Wilson coefficients of the proton and muon. These Wilson coefficients have been computed using dimensional regularization and renormalized in the $\MS$ scheme. This is the same way in which we have done the computation of the soft diagrams drawn in Fig.~\ref{fig:soft}. Some details of such computation are given in Appendix \ref{Sec:DR}.

\begin{figure}[ht]
  \begin{center}
   \includegraphics[width=0.4\textwidth]{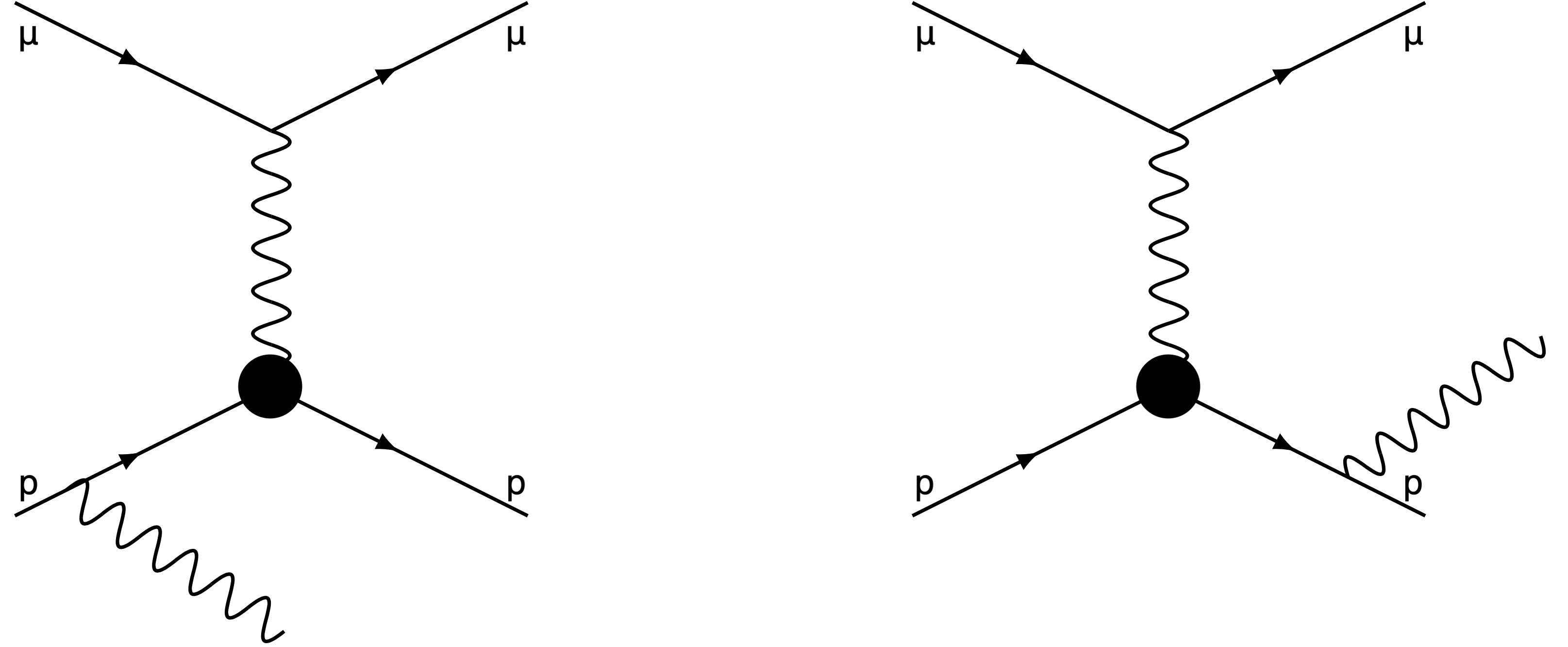}\\
    \includegraphics[width=0.4\textwidth]{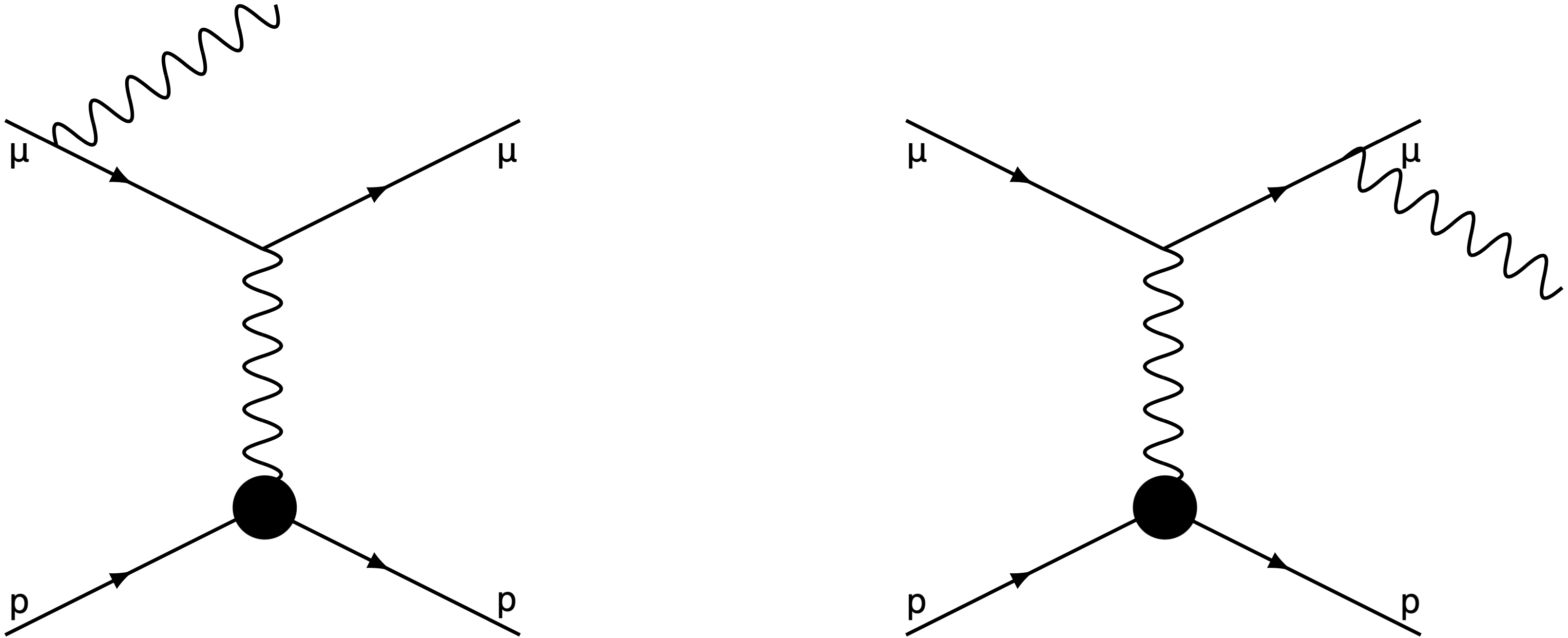}
  \end{center}
  \caption{Leading-order soft diagrams contributing to $\delta^{(p)}_{\rm soft}$ (upper figure) and to $\delta^{(\mu)}_{\rm soft}$ (lower figure).}
  \label{fig:soft}
\end{figure}

\begin{figure}[ht]
  \begin{center}
    \includegraphics[width=0.4\textwidth]{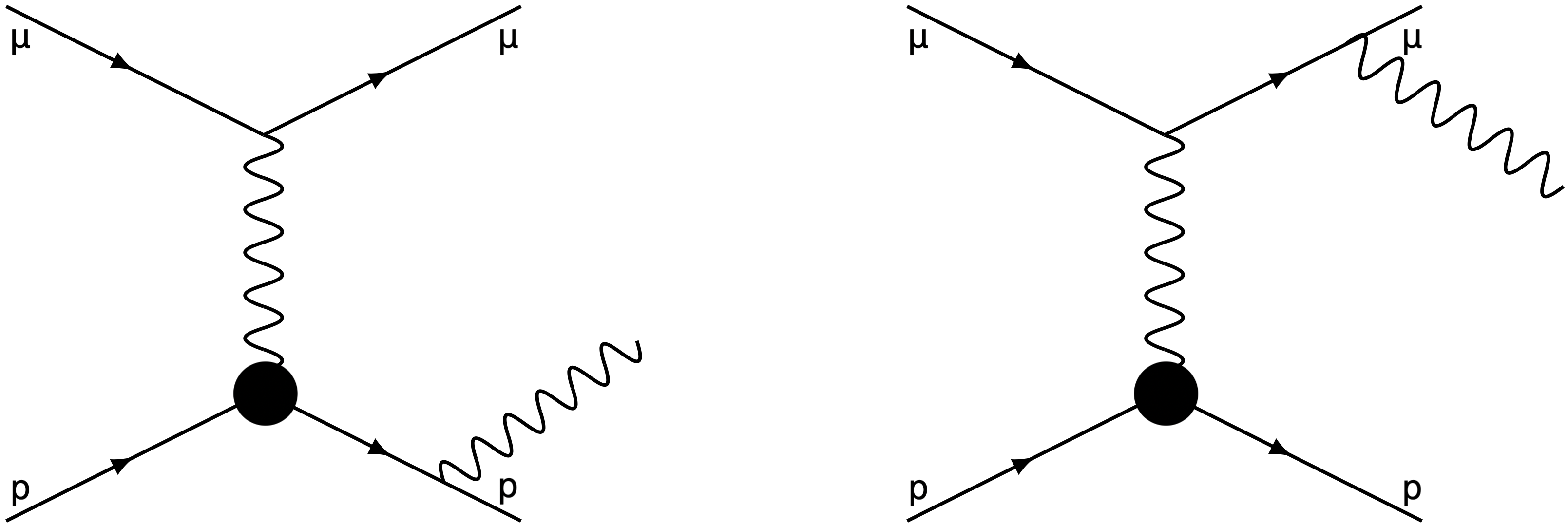}\\
    \includegraphics[width=0.4\textwidth]{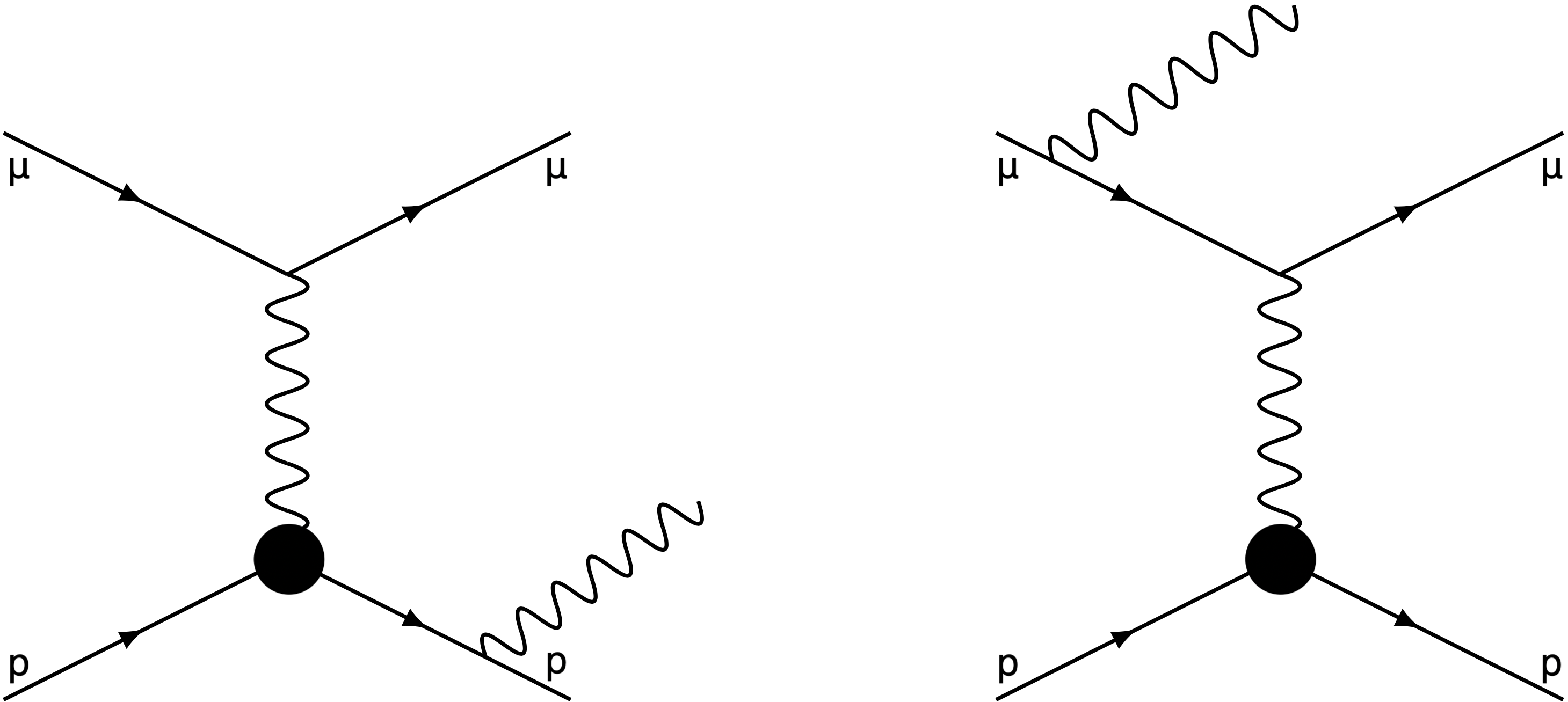}\\
    \includegraphics[width=0.4\textwidth]{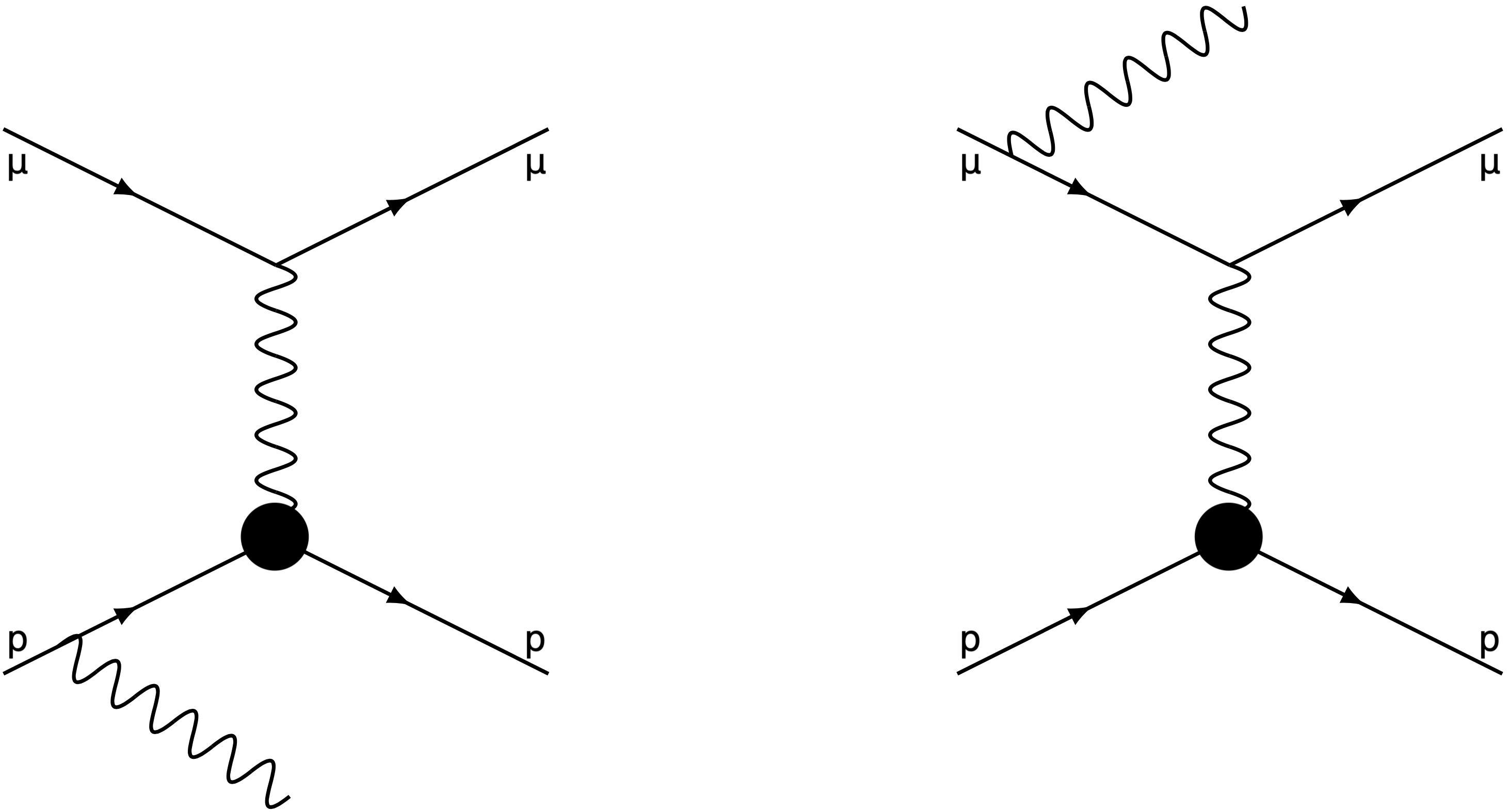}\\
    \includegraphics[width=0.4\textwidth]{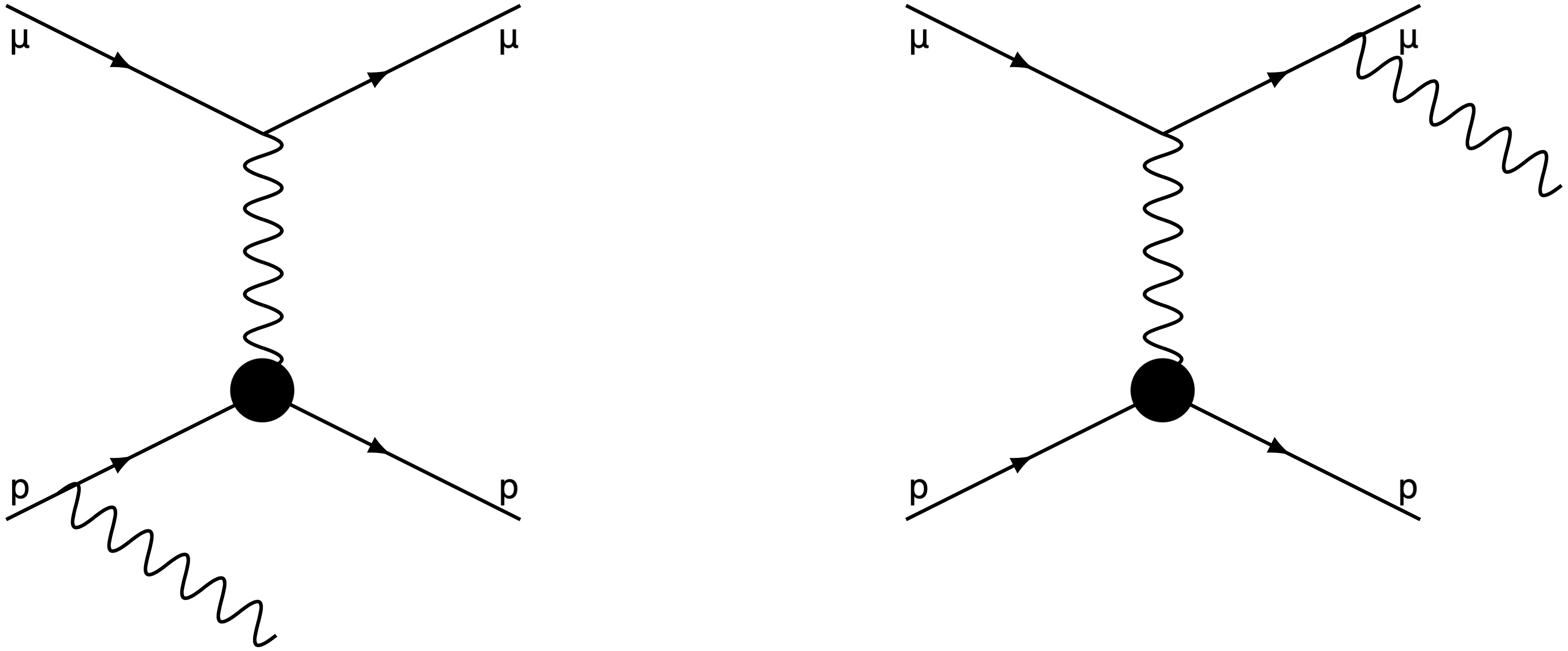}
  \end{center}
  \caption{Leading-order soft diagrams contributing to $\delta^{(p\mu)}_{\rm soft}$.}
  \label{fig:softinterference}
\end{figure}

 The interference term $\delta^{(p\mu)}_{\rm soft}$ in Eq.~\eqref{eq:dsigmaTotalmuon} is obtained from the computation of the diagrams drawn in Fig.~\ref{fig:softinterference}. It is given in the limit $\beta \to 0$ by
\be
\delta^{(p\mu)}_{\rm soft} \underset{\beta \to 0}{\longrightarrow}  \frac{4}{3} \frac{ \alpha}{\pi} \frac{Q^2}{M m_\mu} \left[  \ln \frac{2\Delta E}{\nu} - \frac{5}{6}\right].\label{eq:lepton_proton_soft_rad}
\ee

\begin{figure}[ht]
  \begin{center}
    \includegraphics[width=0.22\textwidth]{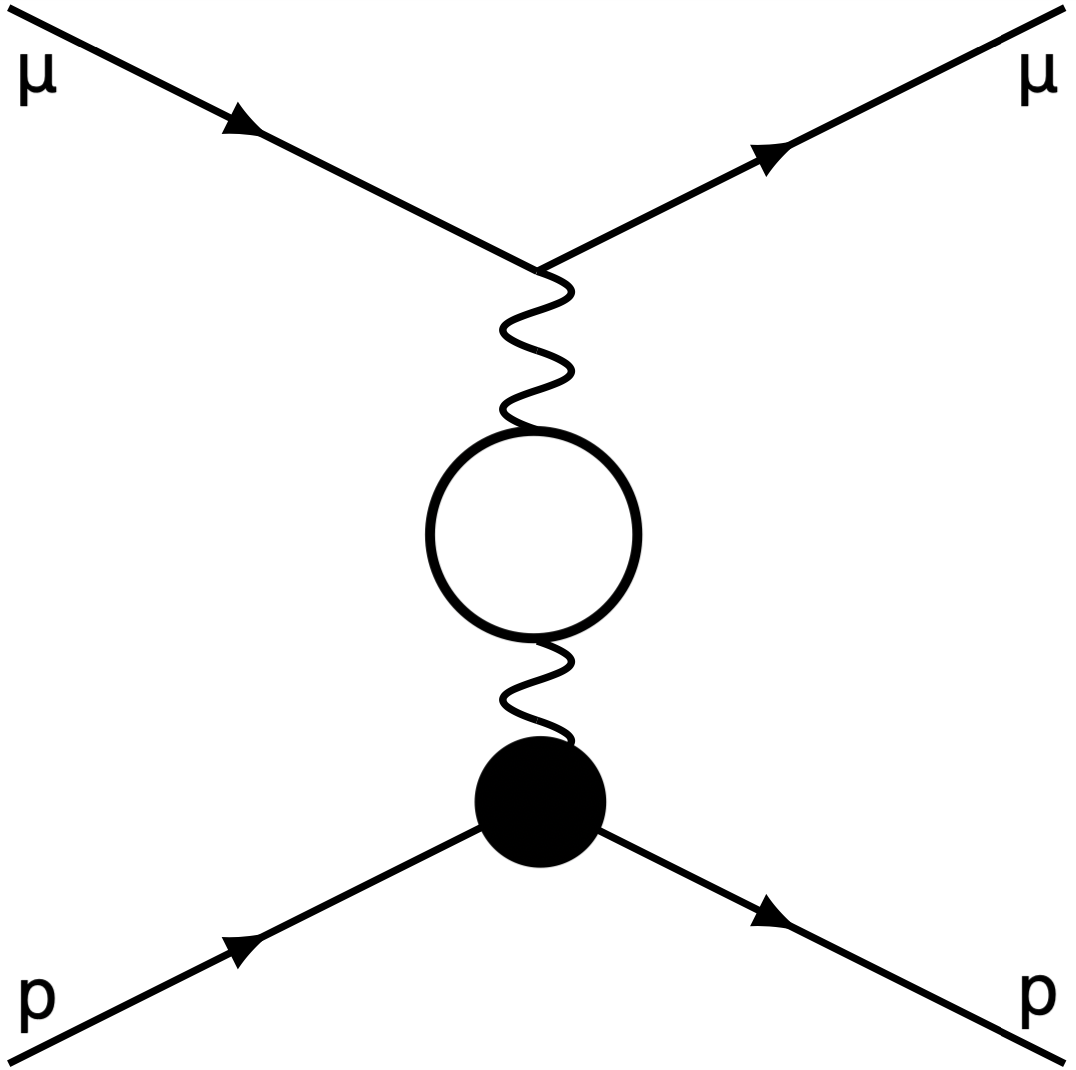}
  \end{center}
  \caption{  Symbolic diagrammatic representation of $\delta_{\rm VP}$.}
  \label{fig:VP}
\end{figure}

The explicit expression for the vacuum polarization correction, beyond the electron vacuum polarization of Eq.~(\ref{alrep}), reads as
\be
\label{deltaVP}
\delta_{\rm VP}= 32 \tau_\mu 
\left[ d_2^{(\mu)} +
\frac{m^2_\mu}{M^2} d_2 + \frac{m^2_\mu}{m_{\tau}^2} d_2^{({\tau})}
\right]
\,.
\ee 
Note that $d_2$ also includes the hadronic vacuum polarization, see \eq{d2p}. We illustrate the diagrammatic representation of such contribution in Fig.~\ref{fig:VP}.

\begin{figure}[ht]
  \begin{center}
    \includegraphics[width=0.65\textwidth]{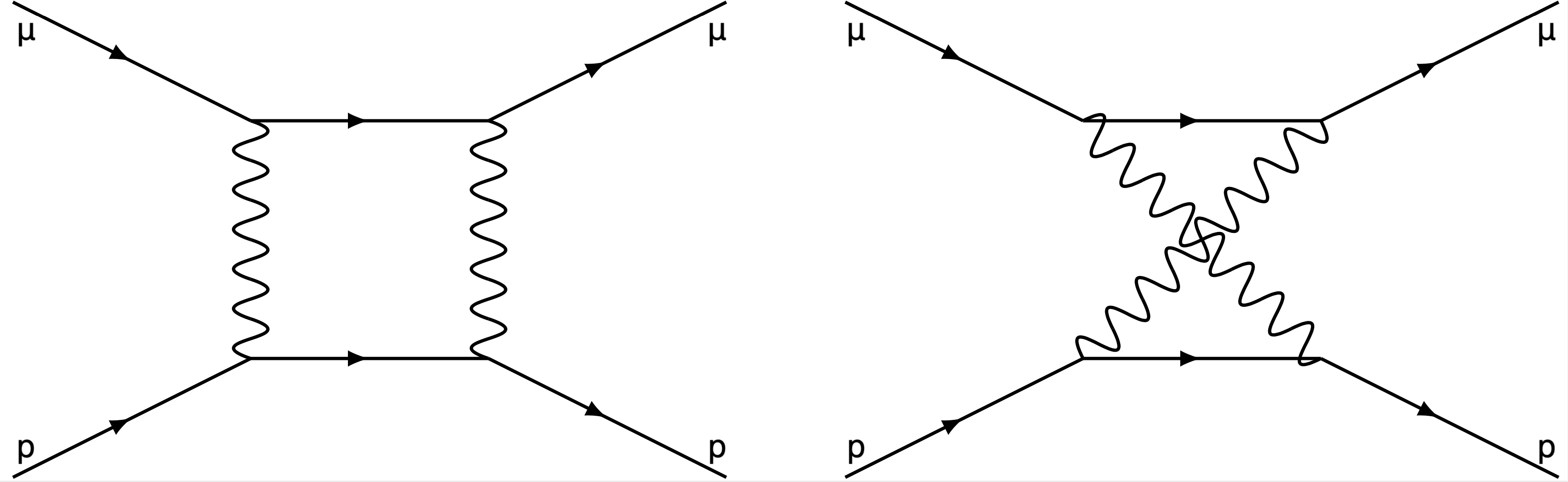}
  \end{center}
  \caption{TPE contributions for point-like particles.}
  \label{fig:TPE}
\end{figure}

For the determination of the TPE contributions, we consider first the case of two point-like spin-1/2 particles,\footnote{For a complete expression of the TPE correction, see the original references~\cite{Eriksson:1961as,Kaiser:2010zz}.} which we draw in Fig.~\ref{fig:TPE}. 
In the $\beta \ll 1$ limit, the leading terms in the $Q^2$-expansion of the TPE correction read 
\be
\delta_{\rm TPE}^{\rm point-like} \underset{\beta \to 0}{\longrightarrow} \frac{\alpha}{\pi} \left(\pi^2  \left( \frac{Q}{2 m_r} - \frac{Q^2}{4  m_r^2 \beta}  \right)+ \frac{7}{3} \frac{Q^2}{M m_\mu} \ln \frac{\nu}{Q}  -\frac{Q^2}{M^2-m^2_\mu} \left( \frac{m_\mu}{M} \ln \frac{M}{\nu} - \frac{M}{m_\mu} \ln \frac{m_\mu}{\nu} \right)\right),
\ee
with the reduced mass $m_r =M m_\mu / \left(M+m_\mu \right)$. The first two terms of this expression correspond to the $\beta \ll 1$ limit of the well-known Feshbach term~\cite{McKinley:1948zz} obtained for the scattering in the static Coulomb field, i.e., when $M \to \infty$:
\be \label{Feshbach1}
\delta_\mathrm{F}= \pi\alpha \beta \frac{\sin \frac{\theta_{\rm lab}}{2}(1-\sin \frac{\theta_{\rm lab}}{2})}{1-\beta^2 \sin^2 \frac{\theta_{\rm lab}}{2}} =  \pi\alpha \frac{Q}{2E} \frac{1 - \frac{Q}{2 |{\bf k}|}}{1-\frac{Q^2}{4E^2}}  \underset{Q^2 \to 0}{\longrightarrow}  \pi \alpha \left(  \frac{Q}{2E} - \frac{Q^2}{4 \beta E^2 }\right) \underset{\beta \to 0}{\longrightarrow}   \pi \alpha \left(\frac{Q}{2m_\mu} - \frac{Q^2}{4m_\mu |{\bf k}|}\right).
\ee
In the language of nonrelativistic EFTs, this term corresponds to a potential loop. Indeed, $\delta_{ \rm TPE}$  could be split into potential, soft and hard contributions (even if it has not been computed in this way):
\be
\delta_{ \rm TPE}^{\rm point-like}=\delta_{\rm pot}+\delta_{\rm soft}+\delta_{\rm hard}^{\rm point-like}.
\ee
The eventual advantage of doing this splitting is that one can deal with one scale at a time. Moreover, the nonperturbative dynamics gets encoded in the hard term. The different terms would then read
\begin{align}
\delta_{\rm pot}&=\delta_\mathrm{F}, \\
\delta_{\rm soft}&= \frac{\alpha}{\pi} \frac{Q^2}{6M m_\mu}\left(14  \ln \frac{\nu}{Q} -1\right), \\
\delta_{\rm hard}^{\rm point-like}&=-\frac{Q^2}{2 M m_\mu} \frac{d_s(\nu)}{\pi \alpha }.
\end{align}
$Q^2$ terms can be absorbed into the Wilson coefficient $d_s(\nu)$, where $d_s(\nu)$ refers to the point-like contribution of Eq.~(\ref{eq:ds}).

Including hadronic TPE effects, only the hard term changes, which now reads
\be
\delta_{\rm hard}^{\rm point-like} \longrightarrow \delta_{\rm hard}=-\frac{Q^2}{2 M m_\mu} 
\left[\frac{d_s(\nu)}{\pi \alpha }-\frac{m_{\mu}}{M}
\frac{c_3^{\rm had}}{\pi \alpha}\right].
\ee 

The expression above has correct relativistic and recoil terms at orders $Q, Q^2 \ln Q^2$ and $Q^2$ for targets with structure. As before, $Q^2$ terms can be absorbed into the Wilson coefficients $d_s(\nu)$ and $c_3^{\rm had}$, where $c_3^{\rm had}$ is dominated by modes with energy of order $m_{\pi}$. The determination of this contribution can be found in \cite{Peset:2014jxa} for muonic hydrogen, and here in Eq.~\eqref{eq:c3TPEep} for regular hydrogen.

These results could be obtained from diagrammatic computations of the EFTs presented in Sec.~\ref{NRQED(mu)} and Sec.~\ref{Sec:pNRQED}. In particular, note that all the hadronic corrections that we have considered appear in the same combination as in $D_d^{\rm had}$, as it should be (see \eq{Dd}). Therefore, the hadronic corrections are the same as in muonic hydrogen spectroscopy. 

Finally, we can perform the same analysis for the scattering of a nonrelativistic proton and a nonrelativistic electron. The computation is identical. We have to eliminate all contributions associated with the electron in the previous computation first, we then replace the muon by the electron, and then reintroduce the vacuum polarization of the muon in Eq.~(\ref{deltaVP}). Again, all hadronic corrections appear in the same combination as in $D_d^{\rm had}$, as it should be (see \eq{Dd}). Therefore, the hadronic corrections are the same as in regular hydrogen spectroscopy. We can also relate these Wilson coefficients with the Wilson coefficients that appear in the muon-proton sector. The Wilson coefficients of ${\cal L}_N$ have the same hadronic content in both regular and muonic hydrogen. Therefore, up to terms dependent on the mass of the lepton (which are suppressed by powers of $\alpha$), they yield the same proton radius. The hadronic vacuum polarization effects are also equal in both theories (with the same caveat as before). Note, on the other hand, that the TPE is different, it depends on the mass of the lepton at the leading non-vanishing order in $\alpha$.

\subsection{Relativistic lepton(muon)-proton scattering}
\label{Sec:relmuon}

The results of the previous section provide a rigorous connection between the non-perturbative effects that appear in spectroscopy and those that appear in lepton-proton scattering. Unfortunately, such kinematic situation is not fulfilled by the  experimental setup of forthcoming muon-proton scattering experiments. For the MUSE experiment~\cite{Gilman:2017hdr},  the three-momentum of the incoming muon is of the order of muon mass, 115-210~MeV, and it will probe the momentum transfer region $Q^2 = 0.0016-0.08~\mathrm{GeV}^2$. Therefore, in this situation, the lepton has to be treated relativistically. In the case of the COMPASS++/AMBER experiment~\cite{Denisov:2018unj}, the lepton is ultrarelativistic ($\beta \simeq 1$), since the energy of the incoming lepton changes in the range 50-200~GeV and the transfer momentum will reach $Q^2 = 0.001-0.02~\mathrm{GeV}^2$. For the general kinematics of the MUSE experiment, the use of HBET applied to the muon-proton sector, as worked out in \cite{Pineda:2002as,Pineda:2004mx}, could be appropriate (see Sec.~\ref{Sec:HBET}). The reason is that the pion mass scale is of the order of the incoming energy of the lepton, and of the order of $Q^2$. Note that this is a problem for the attempts to construct an EFT for the MUSE experiment made in \cite{Dye:2016uep,Dye:2018rgg}, where the pion particle, and other close-by degrees of freedom, are not incorporated in the Lagrangian.

We may consider a specific kinematic setup of the MUSE experiment and assume that, in such corner, $\tau_{\mu} \ll 1$ but still keeping the muon relativistic. This kinematic regime is closer to the one considered in the previous section. In the kinematic regime where $\tau_p$ and $\tau_{\mu}$ are small, we can always keep the linear powers in these parameters and neglect those that are of generic order $\tau^2$. In this particular kinematic setup, the only open channel is the elastic process. This means that the relevant effective theory can be written in terms of the proton and the lepton, even if the lepton needs to be treated relativistically. In principle, one would think that keeping $\tau_{\mu} \ll 1$ would guarantee that all hadronic effects should appear as Wilson coefficients of the appropriate effective theory, as there are no dynamical hadronic degrees of freedom as asymptotic states. Nevertheless, one cannot simply elliminate the pion while keeping the lepton relativistic. Since $E$ is still kept dynamical, there is entanglement with the pion mass. At present, there is no effective theory that profits from the constraint $\tau_{\mu} \ll 1$. We have discussed possible (to-be-constructed) effective theories that could describe such kinematic regime efficiently in Secs. \ref{Sec:Remuon} and \ref{Sec:boosted}. Irrespectively of the construction of the effective theory, it is worth seeing what is known at present. Several terms can be computed with arbitrary $\beta$ without further complications. The corrections corresponding to the real radiation for a general $\beta$ can be found in~\cite{Hill:2016gdf,Yennie:1961ad,tHooft:1978jhc,Maximon:2000hm,Vanderhaeghen:2000ws}. They read 
\begin{align}
\delta^{(p)}_{\rm soft} &= \tau_p\left[\beta^2 \left( c_F^{(p)\,2} -Z^2 \right) - Z \left( c_D^{(p)\MS}(\nu) - Z \right)+\frac{4}{3}\frac{Z^4\alpha}{\pi}\left(2\ln \frac{2\Delta E}{\nu} -\frac{5}{3}\right)\right], \label{eq:proton_soft_rad} \\
\delta^{(\mu)}_{\rm soft} &= \tau_{\mu}\left[\beta^2 \left(  c_F^{(\mu)\,2}  - 1 \right) - \left( c_D^{(\mu)\MS}(\nu) - 1 \right) +\frac{4}{3}\frac{\alpha}{\pi}\left(2\ln \frac{2\Delta E}{\nu} + \frac{1}{2 \beta^2}\left( 1 - \frac{1+3\beta^2}{2\beta} \ln  \frac{1+\beta}{1-\beta}  \right)\right)\right], \label{eq:lepton_soft_rad}\\ 
\delta^{(p\mu)}_{\rm soft}& = \frac{ \alpha}{\pi} \frac{Q^2}{M E \beta^2} \left[  2\left(1- \frac{1-\beta^2}{2\beta} \ln \frac{1+\beta}{1-\beta} \right) \ln \frac{2\Delta E}{\nu} - 2 + \frac{1-\beta^2}{4 \beta} \left( \ln^2 \frac{1+\beta}{1-\beta} + 4 \mathrm{Li}_2 \frac{2\beta}{1+\beta} \right)\right].\label{eq:lepton_proton_soft_rad_2}
\end{align}
These equations have been obtained in the limit $\tau_\mu=Q^2/m_{\mu}^2,~\tau_p=Q^2/M^2\ll 1$ keeping $\beta$ arbitrary otherwise (in particular, $E$ could be much larger than $Q$). In the nonrelativistic limit $\beta \rightarrow 0$, the results of the previous section are recovered. In particular, in the nonrelativistic limit for the muon ($\beta \rightarrow 0$), Eq.~(\ref{eq:lepton_soft_rad}) reduces to Eq.~(\ref{eq:proton_soft_rad_NR}) for $\delta^{(\mu)}_{\rm soft}$. Proton structure effects do not show up at this level of precision in $\delta^{(\mu)}_{\rm soft}$ or in $\delta^{(p\mu)}_{\rm soft}$. Note also that $c_D^{(\mu)\MS}(\nu)$ and $c_F^{(\mu)}$ include ${\cal O}(\alpha)$ radiative corrections. We choose to rewrite them in this way to make the connection with the previous section smooth, but it should be kept in mind that the computation is not performed with an effective theory with nonrelativistic muons and its corresponding Wilson coefficients. 

The TPE correction is more complicated. To simplify the problem, we first consider the point-like contribution. The QED limit of TPE in Eq.~(\ref{eq:dsigmaTotalmuon}) for the scattering of two spin-1/2 particles with masses $M$ and $m_\mu$ is given by\footnote{With an additional condition $\tau_\mu \ll \beta^2 E^2/s$, the QED $Q^2 \to 0$ limit has a much simpler form~\cite{Tomalak:2015hva}
\begin{align}
\delta_{ \rm TPE}^{\rm point-like} &= \frac{ \alpha}{\pi} \left(\pi^2 \frac{M+m_\mu}{2M}\frac{Q}{E}+\frac{2Q^2}{M E \beta^2} \left(1- \frac{1-\beta^2}{2\beta} \ln \frac{1+\beta}{1-\beta} \right) \ln \frac{\nu}{Q}  +\frac{Q^2}{2M E} \left(1- \frac{1}{\beta} \ln \frac{1+\beta}{1-\beta} \right) \ln \frac{Q^2}{M m_\mu}  \right) \nonumber \\
&+ \frac{ \alpha}{\pi} \frac{Q^2}{ME} K , \label{eq:TPE_limit}
\end{align}
with the constant term $K$:
\begin{align}
K &=\frac{M^2 + m_\mu^2}{4M E}\frac{1}{\beta} \left( \mathrm{Li}_2 \left(1 - \frac{m_\mu}{M} \rho\right) + \mathrm{Li}_2 \left( 1 - \frac{M}{m_\mu} \rho\right) - \mathrm{Li}_2 \left( 1 + \frac{m_\mu}{M} \rho\right) - \mathrm{Li}_2 \left( 1 + \frac{M}{m_\mu} \rho\right)- \frac{\pi^2}{2} \right) \nonumber \\
&-\frac{1}{2\beta} \left( \mathrm{Li}_2 \left( 1 - \frac{m_\mu}{M} \rho\right) + \mathrm{Li}_2 \left( 1 - \frac{M}{m_\mu} \rho\right) + \mathrm{Li}_2 \left( 1 + \frac{m_\mu}{M} \rho\right) + \mathrm{Li}_2 \left( 1 + \frac{M}{m_\mu} \rho\right) + \ln^2 \frac{m_\mu}{M} + \frac{\pi^2}{6} \right) \nonumber \\
&-\frac{1}{\beta} \left( \mathrm{Li}_2 \frac{1-\beta}{1+\beta} +\ln\frac{1+\beta}{2\beta} \ln \frac{1+\beta}{1-\beta} -\frac{1}{8} \ln^2 \frac{1+\beta}{1-\beta} \right) +\frac{ \beta M^2 m_\mu^2 \ln \frac{1+\beta}{1-\beta} +\frac{(1-\beta^2)(M^4-m_\mu^4)}{2} \ln \frac{m_\mu}{M}}{(M^2-m_\mu^2)^2-\beta^2(M^2+m_\mu^2)^2}-2, \\
\rho&=\sqrt{ \frac{1-\beta}{1+\beta}}.
\end{align}
In the nonrelativistic limit, the constant $K$ is given by
\begin{align}
K  \underset{\beta \to 0}{\longrightarrow} - \frac{\pi^2 M m_\mu}{4 m_r^2 \beta} + \frac{1}{2} \frac{M^2 + m_\mu^2}{M^2 - m_\mu^2} \ln \frac{m_\mu}{M}.
\end{align}}
\begin{align}
\delta_{ \rm TPE}^{\rm point-like} &= \frac{ \alpha}{\pi} \left(\pi^2 \frac{M+m_\mu}{2M}\frac{Q}{E}+ 2 \left(\frac{\ln \rho}{\beta} - \frac{\ln \rho'}{\beta'} \right) \ln \frac{Q^2}{\nu^2}  +\frac{Q^2}{2M E} \ln \frac{Q^2}{M m_\mu} + \frac{Q^2}{2} \left( \frac{\beta \ln \rho}{s} - \frac{\beta' \ln \rho'}{u} \right) \right) \nonumber \\
&+\frac{ \alpha}{\pi} \left( 1 + \frac{m_\mu}{M} + \frac{M}{m_\mu} \right) \frac{Q^2}{4 |{\bf{k}}|^2 - s \frac{Q^2}{M^2}} \left( \pi^2 \frac{M+2m_\mu}{2M}\frac{Q}{M}+ \frac{Q^2}{M^2} \left(  \frac{3}{2} \ln \frac{Q^2}{M m_\mu} - \frac{1}{4} \ln \frac{m^2_\mu}{M^2} - 3 \right)\right)\nonumber \\
&+\frac{ \alpha}{\pi} \frac{Q^2}{M^2} \left( 1 + \frac{m_\mu}{M} + \frac{M}{m_\mu} \right) \frac{Q^2}{4 |{\bf{k}}|^2 - s \frac{Q^2}{M^2}}  \frac{\left( 3 s - M^2 + m_\mu^2 \right) f \left( Q^2 \right) - \left( 3 u - M^2 + m_\mu^2 \right) f' \left( Q^2 \right)}{8 M m_\mu} \nonumber \\
&+\frac{ \alpha}{\pi} \frac{Q^2}{M E} \left( \frac{s f \left( Q^2 \right) - u f'\left( Q^2 \right)}{4 M E} + \frac{M^4-m_\mu^2}{4 s u} \ln \frac{m_\mu^2}{M^2} - 2 \right), \label{eq:TPE_limit0}
\end{align}
with the following definitions:
\begin{align}
f \left( Q^2 \right) &= \frac{1}{\beta}\left( \ln \rho \ln \frac{Q^2}{M m_\mu} + \frac{1}{2} \ln  \frac{\rho m_\mu }{M}  \ln  \frac{\rho M}{m_\mu}  + 2 \ln \frac{1+\beta}{2 \beta} \ln \rho - \mathrm{Li}_2(\rho^2) - \frac{\pi^2}{3} \right) \nonumber \\
&- \frac{1}{\beta }\mathrm{Re} \left( \mathrm{Li}_2 \left( 1 +  \frac{\rho m_\mu}{M}\right) + \mathrm{Li}_2 \left( 1 +  \frac{\rho M}{m_\mu}\right) \right), \\
f^\prime \left( Q^2 \right) &= \frac{1}{\beta^\prime}\left( \ln \rho^\prime \ln \frac{Q^2}{M m_\mu} + \frac{1}{2} \ln  \frac{\rho^\prime m_\mu}{M} \ln  \frac{\rho^\prime M}{m_\mu}  + 2 \ln \frac{1+\beta^\prime}{2 \beta^\prime} \ln \rho^\prime - \mathrm{Li}_2 \left( \rho'^2 \right) + \frac{\pi^2}{6} \right) \nonumber \\
&- \frac{1}{\beta^\prime} \left( \mathrm{Li}_2 \left( 1 - \frac{ \rho^\prime m_\mu}{M}\right) + \mathrm{Li}_2 \left( 1 - \frac{\rho^\prime M}{m_\mu}\right) \right), \\
\rho&=\sqrt{ \frac{1-\beta}{1+\beta}}, \qquad \rho' =\sqrt{ \frac{1-\beta'}{1+\beta'}},
\end{align}
Mandelstam invariants $s = M^2 + m_\mu^2 + 2 M E,~u = 2M^2 + 2m_\mu^2 -s + Q^2$, and the final lepton velocity $\beta'$. In the heavy-proton limit, the corresponding expression is given by 
\begin{align}
\delta_{ \rm TPE}^{\rm point-like} &\underset{M \to \infty}{\longrightarrow} \frac{ \alpha}{\pi} \left(\pi^2 \frac{M+m_\mu}{2M}\frac{Q}{E} - \pi^2 \frac{Q^2}{4 \beta E^2 } +\frac{2Q^2}{M E \beta^2} \left(1- \frac{1-\beta^2}{2\beta} \ln \frac{1+\beta}{1-\beta} \right) \ln \frac{\nu}{Q}  
\nn
\right.
\\ 
&
\qquad\qquad
\left.
+\frac{Q^2}{2M E} \left(1- \frac{1}{\beta} \ln \frac{1+\beta}{1-\beta} \right) \ln \frac{Q^2}{M m_\mu}  \right) +\frac{ \alpha}{\pi} \frac{Q^2}{2 ME} \left( \ln \frac{m_\mu}{M} -4 \right)\nonumber \\
&+\frac{ \alpha}{\pi}  \frac{Q^2}{4 |{\bf{k}}|^2 - Q^2} \left( \pi^2 \frac{M+3m_\mu}{M}\frac{Q}{2m_\mu} - \pi^2 \frac{M+m_\mu}{M}\frac{Q^2}{4 \beta m^2_\mu}+ \frac{Q^2}{M m_\mu} \left(  \frac{3}{2} \ln \frac{Q^2}{M m_\mu} - \frac{1}{4} \ln \frac{m^2_\mu}{M^2} - 3 \right)\right) \nonumber \\
&-\frac{ \alpha}{\pi}  \left(\frac{m_\mu}{M} \frac{Q^2}{E^2}+ \frac{Q^2}{4 |{\bf{k}}|^2 - Q^2} \frac{Q^2}{ M m_\mu} \right) \frac{1}{2\beta} \left(\rho \left( 1 - \ln \frac{ \rho m_\mu}{M} \right) - \frac{1}{\rho} \left( 1+\ln \frac{\rho M}{m_\mu}\right) \right)  \nonumber \\
&+ \frac{ \alpha}{\pi} \frac{Q^2}{ME} \left( \frac{1}{4} \ln \frac{m_\mu^2}{M^2} - 2 - \frac{1}{\beta} \left( -\ln^2 \rho +2 \ln \rho \ln \frac{2\beta}{1+\beta} + \ln \rho \ln \frac{m_\mu}{M} + \mathrm{Li}_2 \left( \rho^2 \right) +\frac{\pi^2}{3}\right) \right). \label{eq:TPE_nonrel_limit}
\end{align}
The leading term represents the Feshbach correction of Eq.~(\ref{Feshbach1}) and can be expressed as
\begin{align}
\delta_{ \rm TPE}^{\rm point-like} &\underset{M \to \infty}{\longrightarrow} \alpha \pi \frac{ \frac{Q}{2E}}{ 1+\frac{Q}{2|{\bf{k}}|}} . \label{eq:TPE_nonrel_limit2}
\end{align}

The expressions above have correct relativistic and recoil terms at orders $Q$ and $Q^2 \ln Q^2$ for targets with structure. The $Q^2$ term contains structure effects and contributions from magnetization of the proton, as well as from all inelastic intermediate states. It can be written as~\cite{Tomalak:2017owk,Tomalak:2018jak,Tomalak:2015hva,Tomalak:2016kyd}
\begin{equation}
\delta_{ \rm TPE}^{\rm str} =  \frac{f^{2\gamma}_+ \left( E \right)}{e^2} \frac{Q^2}{2 M E},
\end{equation}
where the forward unpolarized amplitude $f^{2\gamma}_+ \left( E \right)$ is evaluated over the kinematic coverage of the proton Dirac and Pauli form factors and the unpolarized proton structure functions defined in Sec.~\ref{Sec:Structurefunctions} and can be expressed as~\cite{Tomalak:2017owk,Tomalak:2016kyd}
\begin{equation}
f^{2\gamma}_+ \left(E \right)  =  - \frac{4 \alpha^2}{M^2} \int \frac{i  \mathrm{d}^4 q}{\pi^2}  \frac{  M^2 \left( k \cdot q \right)^2 \left( 2 \mathrm{S}_1 - \mathrm{S}_2 \right) + q^2 \left( M^2 m^2 \mathrm{S}_1 -  \left( k \cdot p \right)^2  \mathrm{S}_2 \right) + 2  \left( k \cdot p \right) \left( k \cdot q \right) \left( p \cdot q \right) \mathrm{S}_2}{ \left( q^4 - 4 \left(k \cdot q \right)^2 \right) (q^2)^2 }.  \label{fplus_amplitude}
\end{equation}
Note that in the structure functions $S_1$ and $S_2$, one subtracts the point-like contribution. 

In the lepton nonrelativistic limit, $f_+$ reproduces the Wilson coefficient $c^{\mathrm{had}}_3$ described above:
\begin{equation}
f^{2\gamma}_+ \left( E = m_\mu \right) =  4 \frac{m_\mu}{M} c^{\mathrm{had}}_3.
\end{equation}
It would be interesting to derive $\delta_{ \rm TPE}^{\rm str}$ from the suitable effective theory. We emphasize that it is in this, four-fermion part, where there are differences with respect to the spectroscopy computation, whereas the bilinear part (where the proton radius is) remains the same. 

\subsection{Ultrarelativistic lepton(electron)-proton scattering}
\label{Sec:ultra}
We now repeat the discussion of the previous section replacing the scattered lepton from muon to electron. In this case, we have an opposite relation between the scale of momentum transfer and the lepton mass. We kept $Q^2 \ll m^2_\mu$ in the previous section, whereas now we are in the opposite situation: $Q^2 \gg m_e^2$ (but still keeping $Q^2 \ll m_{\mu}^2,\; m_{\pi}^2$). Electrons are typically ultrarelativistic in electron-proton scattering experiments. For instance, high-statistics Mainz data~\cite{Bernauer:2010wm,Bernauer:2013tpr} was taken with beam energies between 180 and 855~MeV covering the momentum transfer range $Q^2=0.004-1~\mathrm{GeV}^2$, while the recent PRad experiment at JLab~\cite{Xiong:2019umf} was performed at beam energies 1.1 and 2.2 GeV covering $Q^2=0.00021-0.06~\mathrm{GeV}^2$. Two other experiments are going to perform measurements at lower beam energies: $30-70~\mathrm{MeV}$ with $Q^2=10^{-5}-3\times10^{-4}~\mathrm{GeV}^2$ by ProRad experiment~\cite{Hoballah:2018szw,Faus-Golfe:2017kcu}, and $20-60~\mathrm{MeV}$ with $Q^2=0.0003-0.008~\mathrm{GeV}^2$ by ULQ2 experiment at Tohoku University~\cite{Suda:2018xxx}. Nevertheless, we still can profit from several results obtained in the previous sections. The point-like result of \eq{treelevelPointlike} is the same and can be simplified noting that $\varepsilon_0 \to 0,~\varepsilon \to \varepsilon_\mathrm{T}$ as
\begin{align}
\varepsilon \to \varepsilon_\mathrm{T} \to  \frac{\nu^2-M^4\tau_p(1+\tau_p)}{\nu^2+M^4\tau_p(1+\tau_p)} = \frac{1}{1+2(1+\tau_p) \tan^2 \frac{\theta_{\rm lab}}{2}},
\end{align}
where the same substitution applies also after the inclusion of the proton form factors as in \eq{treelevel}.

Similarly to Sec.~\ref{Sec:NRmupscat}, we organize the cross section and all ${\cal O}(\alpha)$ corrections in the following way:
\begin{align}
\label{dsigmaTotalelectron}
\left(\frac{d\sigma}{d\Omega}\right)_\text{measured}\hspace{-.5cm}=
 Z^2 \left(\frac{d\sigma_{1\gamma}}{d\Omega}\right)_{\rm point-like}
\hspace{-.5cm}+
\frac{d\sigma_\text{Mott}}{d\Omega}\left[\delta^{(p)}_{\rm soft}+Z^2 \left( \delta^{(e)}_{\rm soft} + \delta_{\rm VP} + \delta^{(e)}_{\rm vert} \right)+Z^3 \left( \delta^{(p e)}_{\rm soft}+  \delta_{\rm TPE} \right)+\mathcal{O}(\alpha^2,\tau_p^2)\right].
\end{align}
The structure of the ${\cal O}(\alpha)$ corrections is similar to the one before, though in some cases the explicit expressions are quite different since now we take the lepton to be ultrarelativistic. 

To include the electron vacuum polarization correction, we also make the replacement of Eq.~(\ref{alrep}) in the first term of \eq{dsigmaTotalelectron}. The vacuum polarization correction from heavier particles, $\delta_{\rm VP}$, is the same as in the muon-proton scattering, i.e., it is given by \eq{deltaVP}. 

The vertex correction on the proton side $\delta^{(p)}_\mathrm{soft}$ can be obtained from Eq.~(\ref{eq:lepton_soft_rad}), replacing the muon by the electron and taking the  ultrarelativistic limit $\beta \to 1$ for the electron:
\be
\delta^{(p)}_{\rm soft}=\tau_p\left[\left( c_F^{(p)\,2} -Z^2 \right) - Z \left( c_D^{(p)\MS}(\nu) - Z \right)+\frac{4}{3}\frac{Z^4\alpha}{\pi}\left(2\ln \frac{2\Delta E}{\nu} -\frac{5}{3}\right)\right]
\,.
 \label{eq:proton_soft_rad_electron}
\ee 

The QED vertex correction on the electron side in the limit $Q^2 \gg m_e^2$ cannot be described by the Wilson coefficients of the nonrelativistic theory as in Eq.~(\ref{eq:lepton_soft_rad}). It is now given by the following expression~\cite{Hill:2016gdf,Bernauer:2013tpr,Vanderhaeghen:2000ws}:
\be
 \delta^{(e)}_{\rm vert} = \frac{\alpha}{\pi} \left[ 2 \left(  \ln \frac{Q^2}{m_e^2} - 1\right) \ln \frac{\nu}{m_e}   - \frac{1}{2} \ln^2 \frac{Q^2}{m_e^2} + \frac{3}{2} \ln \frac{Q^2}{m_e^2} - 2 + \frac{\pi^2}{6} \right],
\ee
while the radiation of soft photons from the electron line $\delta^{(e)}_\mathrm{soft}$ is described by
\be
\delta^{(e)}_\mathrm{soft}=\frac{\alpha}{\pi} \left[ 2 \left(  \ln \frac{Q^2}{m_e^2} - 1\right) \ln \left( \frac{2\Delta E}{\nu}  \right) + f \left( E, Q^2\right)\right],
\ee
where
\be
f \left( E, Q^2\right) =  \left( 1-\ln \frac{2E}{m_e} \right)\ln \frac{2E}{m_e} +  \left( 1-\ln \frac{2E'}{m_e} \right)\ln \frac{2E'}{m_e} - \mathrm{Li}_2 \left( 1 - \frac{4 E E'}{Q^2} \right)- \frac{\pi^2}{3}, \qquad E' = E - \frac{Q^2}{2M},
\ee
up to terms suppressed by powers of the mass of the electron. If we further take the limits $Q^2/M^2\ll 1$ and 
$Q^2/E^2\ll 1$, we obtain the following leading expressions:
\be
\delta^{(e)}_{\rm soft} = \frac{\alpha}{\pi} \left[ 2 \left(  \ln \frac{Q^2}{m_e^2} - 1\right) \ln \left( \frac{\Delta E}{E} \frac{m_e}{\nu} \right) +  \frac{1}{2} \ln^2 \frac{Q^2}{m^2_e} -\frac{\pi^2}{6} + \frac{Q^2}{2M E} \left( \ln \frac{Q^2}{m^2_e} + \frac{M}{E}\ln \frac{Q}{2E} +\frac{M}{2E}-1 \right) \right] , \label{eq:electron_soft_rad}
\ee
\be
\delta^{(pe)}_{\rm soft}= 2 \frac{ \alpha}{\pi} \left[ \frac{Q^2}{M E} \ln \frac{2\Delta E}{\nu} - \frac{Q^2}{M E}\right].\label{eq:electron_proton_soft_rad}
\ee
Note that all these results for soft-photon emission are independent of the proton structure. 

For the determination of the TPE contribution, it is convenient to consider first the case of two point-like spin-1/2 particles. 
The low momentum transfer limit of the TPE correction in the scattering of two point-like spin-1/2 particles with mass $M$ and vanishing mass $m_e$, i.e., $Q^2 \gg m^2_e$, is given by~\cite{Brown:1970te,Tomalak:2014sva,Tomalak:2015aoa}
\begin{align}
\delta_{ \rm TPE}^{\rm point-like} &= \frac{ \alpha}{\pi} \left(\pi^2 \frac{Q}{2E}+2\frac{Q^2}{M E } \ln \frac{\nu}{Q}  +\frac{Q^2}{M E} \left(1+  \ln \frac{Q}{2E} \right) \ln \frac{Q}{2E}  +\frac{Q^2}{ME} K_e  \right), \label{eq:TPE_electron}
\end{align}
with the constant term $K_e$:
\begin{align}
K_e &= -\frac{1}{2} \left( \ln^2 \frac{2E}{M} + \left( 1-\frac{M}{2E}\right) \mathrm{Li}_2 \left( 1 - \frac{M}{2E} \right)+ \left( 1+\frac{M}{2E}\right) \mathrm{Li}_2 \left( 1 + \frac{M}{2E} \right) \right) \nonumber \\
&+ \frac{2E^2 +M^2}{4E^2-M^2} \ln \frac{2E}{M} + \frac{\pi^2}{12} \left(1-\frac{3M}{2E} \right)- 1.
\end{align}
When the heavy proton limit is considered, $K_e$ describes the exchange of Coulomb photons only
\begin{align}
K_e & \underset{M \to \infty}{\longrightarrow} -\frac{\pi^2}{4} \frac{M}{E}.
\end{align}
The expressions above have correct coefficients at orders $Q, Q^2 \ln^2 Q^2$ for targets with structure. Note that we could also reproduce the vertex corrections, radiation of soft photons and infrared pieces of the TPE term by taking the low-$Q^2$ limit of expressions from \cite{Hill:2016gdf}, where the author formulates the computation of the radiative corrections in a SCET approach.

Inelastic intermediate states contribute to $Q^2 \ln Q$ term as well~\cite{Brown:1970te,Tomalak:2015aoa,Gorchtein:2014hla}. This correction, $\delta_{ \rm TPE}^{\rm inel}$, is expressed as the photon energy $E_\gamma$ integral over the total photoabsorption cross section on the proton $\sigma_\mathrm{T} \left( E_\gamma \right)$:
\begin{align}
\delta_{ \rm TPE}^{\rm inel} &= C \left( E \right) \frac{Q^2}{M E} \ln \frac{Q}{2E}, \nonumber \\
C \left( E \right) & = \frac{M E}{2\pi^3} \int \limits_{m_\pi + \frac{m_\pi^2}{2M}}^\infty \frac{\mathrm{d} E_\gamma}{E_\gamma} \left( \frac{E_\gamma}{E} - \frac{E_\gamma}{E} \ln \Bigg | 1 - \frac{E^2}{E^2_\gamma} \Bigg | + \left( 1 + \frac{E_\gamma^2}{2E^2}\right)\ln \Bigg |\frac{E-E_\gamma}{E+E_\gamma}\Bigg | \right) \sigma_\mathrm{T} \left( E_\gamma \right), \label{eq:TPE_electron_inelastic}
\end{align}
starting from the pion production threshold. To obtain the resulting correction up to $Q^2 \ln Q$ terms, $\delta_{ \rm TPE}^{\rm inel}$ should be added to $\delta_{ \rm TPE}^{\rm point-like}$ in Eq.~(\ref{dsigmaTotalelectron}), i.e., $\delta_{ \rm TPE} = \delta_{ \rm TPE}^{\rm point-like} + \delta_{ \rm TPE}^{\rm inel} + \mathcal{O}(\tau_p)$. 

\section{Determination of the low-energy constants}

In the previous sections, we have given theoretical expressions for the Lamb shift in regular and muonic hydrogen. We have also given theoretical expressions for the electron-proton and muon-proton elastic scattering in some specific kinematic situations. In these theoretical expressions, the nonperturbative physics associated with hadronic scales is encoded in coefficients that can be related to Wilson coefficients of the appropriate EFT. Later, expressions valid in kinematic configurations realizable in modern or near-future elastic lepton-proton scattering experiments (not directly obtained from EFTs) were derived. We now review some determinations of the proton radius (but we also briefly discuss other Wilson coefficients). The relative precision of these determinations of the proton radius is of order $\al$.

\begin{figure}[ht]
  \begin{center}
\hspace*{-4cm}   \includegraphics[width=0.75\textwidth]{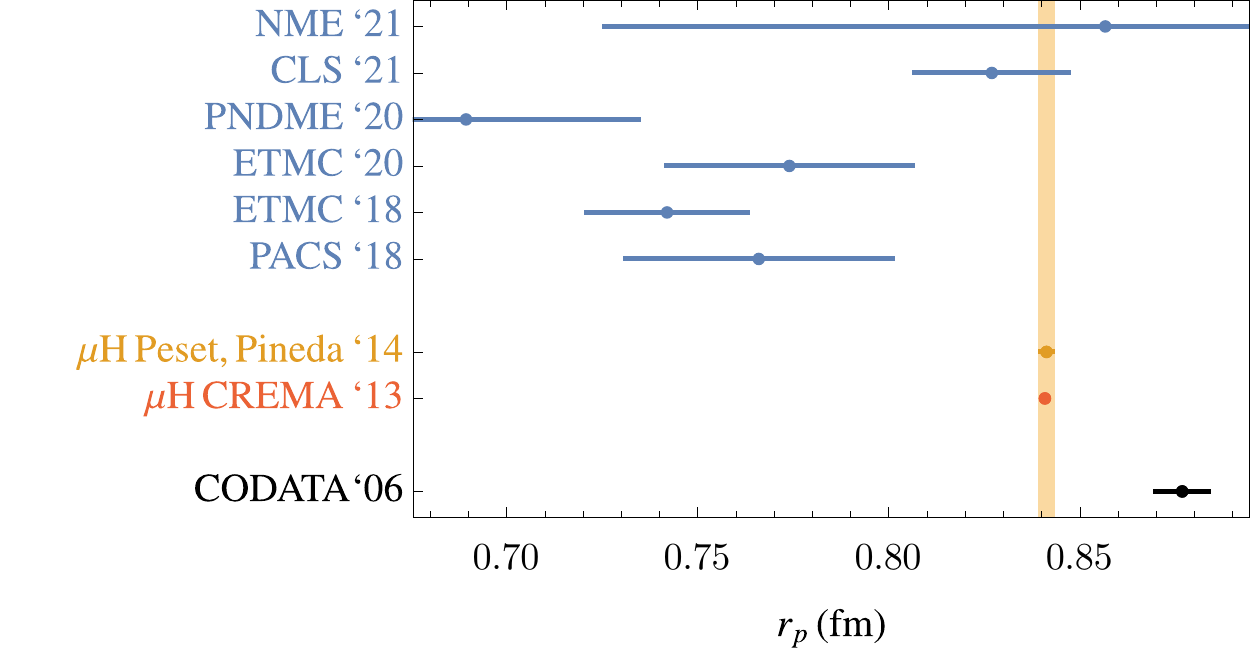}
  \end{center}
  \caption{ 
  Values of $r_p$ in units of fm obtained from recent lattice simulations in comparison with the determinations from muonic hydrogen (in blue). Following historical order, we have (errors are combined errors in quadrature when necessary): PACS'18 $r_p=0.766(35)$ fm \cite{Shintani:2018ozy} (Erratum); ETMC'18 $r_p=0.742(21)$ fm, and ETMC'20 $r_p=0.774(32)$ fm \cite{Alexandrou:2018sjm,Alexandrou:2020aja}; PNDME'20 $r_p=0.69(5)$ fm \cite{Jang:2019jkn}; CLS'21 $r_p=0.827(20)$ fm \cite{Djukanovic:2021cgp}; and NME'21 $r_p= 0.86(13)$ fm \cite{Park:2021ypf}. In the case of \cite{Jang:2019jkn,Djukanovic:2021cgp,Park:2021ypf}, we obtain the proton radius by combination of the isovector result, $\langle r^2\rangle_E^{(v)}$, with the experimental measurement for the neutron radius $\langle r^2\rangle_E^{(n)}=-0.1161(22)$ fm \cite{Zyla20}  so that $\langle r^2\rangle_E=\langle r^2\rangle_E^{(v)}+\langle r^2\rangle_E^{(n)}$. We compare to the values obtained in muonic hydrogen from an EFT analysis $r_p=0.8413 (15)$ fm \cite{ Peset:2014yha} (orange point) and the CREMA determination $r_p=0.84087 (39)$ fm \cite{Antognini:1900ns} (red point), and to the averaged value reported by CODATA06 $r_p=0.8768 (69)$ fm \cite{Mohr:2008fa} (black point). 
  }
  \label{fig:rplattice}
\end{figure}

As a matter of principle, it should be possible to derive these Wilson coefficients directly from QCD. Thus, they should be pure functions of the light quark masses and $\Lambda_{\rm QCD}$. Nevertheless, at present, it is not possible to obtain accurate numbers for them from first-principle computations. Numerical lattice simulations of QCD, though rapidly evolving, are still
in their early stages, and produce numbers with large errors for these Wilson coefficients. We show a set of recent determinations of the electromagnetic proton radius in Fig.~\ref{fig:rplattice}, and in Fig.~\ref{fig:rM2} for the magnetic proton radius squared, $\langle r_M^2 \rangle$. Note that some of these determinations are not genuine determinations of the electromagnetic proton radius, since they only compute the isovector part, and take the isoscalar part from the experiment to produce predictions for the proton radii. They also suffer (with the exception of \cite{Alexandrou:2020aja}) from the same problems that direct fits to elastic electron-proton data: one has to perform extrapolations to $Q^2 \rightarrow 0$, which is one of the main sources of (uncontrolled) uncertainty in these analyses (see the discussion in forthcoming sections). Another issue that limits the accuracy of present lattice determinations, in comparison with the determinations below, is that these lattice determinations of the proton radius do not incorporate ${\cal O}(\alpha)$ corrections. Therefore, they cannot compete, at present, with the precision of theoretical expressions one has for spectroscopy and lepton-proton scattering.

\subsection{Chiral perturbation theory}
\label{Sec:chiralPT}
Once one is giving up direct QCD computations, the next most rigorous approach is to compute the Wilson coefficients using chiral perturbation theory. Nevertheless, the accuracy one can reach within this framework is limited. It can only predict the nonanalytic dependence in the light-quark masses (and the nonanalytic dependence in $N_c$, the number of colors, if working in the large $N_c$ limit). For the case of the electric proton charge radius, the leading contribution is the chiral logarithm, but the constant term cannot be fixed by the theory. Therefore, chiral perturbation theory alone can only give the order of magnitude of the proton radius. For other radii, the situation is potentially better. If we are below the two-pion production threshold in the $t$ channel, the form factors are analytic functions in $Q^2$. Therefore, the series is convergent and truncating the series by a polynomial is not a problem if the error of the unknown coefficients can be estimated. The high-order coefficients of the Taylor expansion are determined (with relative) higher and higher accuracy by the first singularity closest to the origin (this is due to a theorem in complex variable by Darboux (1878)). The leading scaling in the pion mass of these radii is the following (we remind that $\Delta=M_{\Delta}-M$):
\be
\langle r^{2k} \rangle_E \sim \frac{1}{F_{\pi}^2} \frac{1}{m_{\pi}^{2k-2}}\left[1+{\cal O}\left(\frac{m_{\pi}}{\Delta}\right)\right]
\,,\ee
\be
\langle r^{2k} \rangle_M \sim \frac{M}{F_{\pi}^2} \frac{1}{m_{\pi}^{2k-1}}\left[1+{\cal O}\left(\frac{m_{\pi}}{\Delta}\right)\right]
\,,
\ee
up to single logarithms. 

\begin{figure}[ht]
  \begin{center}
   \hspace*{-3.5cm}\includegraphics[width=0.9\textwidth]{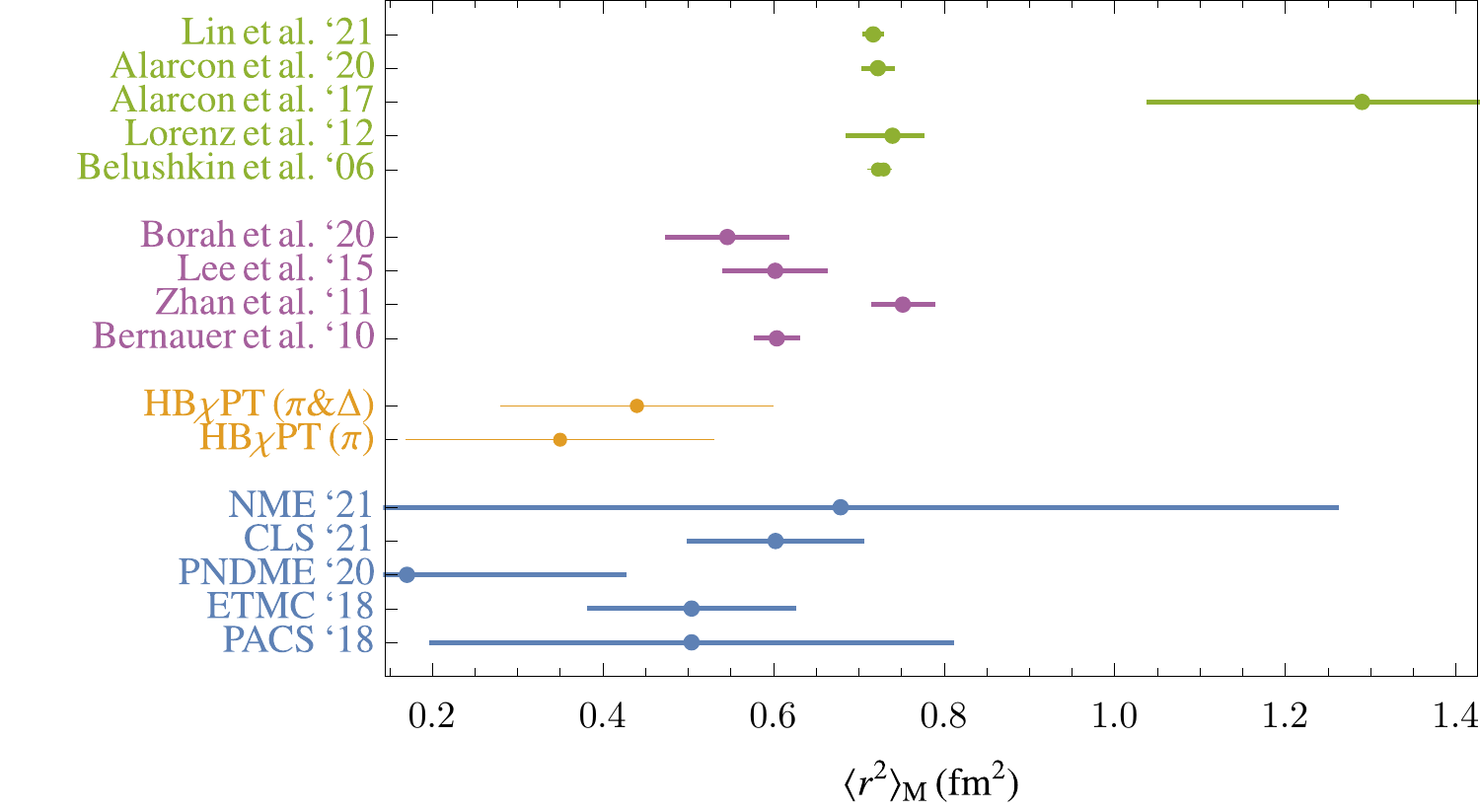}
  \end{center}
 \caption{ Values of 
$\langle r^2 \rangle_{\rm M}$ in units of fm$^2$ 
from different methods. From bottom to top: The first numbers (in blue) are recent predictions by lattice simulations: PACS'18 $0.50(30)$ fm$^2$ \cite{Shintani:2018ozy} (Erratum); ETMC'18 $0.50(12)$ fm$^2$ \cite{Alexandrou:2018sjm}; PNDME'20 $0.17(25)$ fm$^2$ \cite{Jang:2019jkn}; CLS'21 $0.60(10)$ fm$^2$ \cite{Djukanovic:2021cgp}; NME'21 $0.7(6)$ fm$^2$ \cite{Park:2021ypf}. In the case of \cite{Jang:2019jkn,Djukanovic:2021cgp,Park:2021ypf}, we obtain the proton radius by combination of the isovector result with the experimental measurement for the neutron radius $\langle r^2\rangle_M^{(n)}=0.864^{+0.009}_{-0.008}$ fm and neutron magnetic moment $
\mu^{(n)}=-1.9130427(5)$ \cite{Zyla20} so that $\langle r^2\rangle_M=(\mu^{(v)}\langle r^2\rangle_M^{(v)}+\mu^{(n)}\langle r^2\rangle_M^{(n)})/(\mu^{(v)}+\mu^{(n)})$.
The 2nd set of numbers (in orange) is first the pure chiral prediction 
(with only pions): $0.35(18)$ fm$^2$,
and then  the result after the inclusion 
of the effects associated with the Delta particle: $0.44(16)$ fm$^2$. The next points (in lilac) are Bernauer et al.'10 $0.604(24)$ fm$^2$ \cite{Bernauer:2010wm}, Zhan et al.'11 $0.752(35)$ fm$^2$ \cite{Zhan:2011ji}, Lee et al.'15 $0.602(59)$ fm$^2$ \cite{Lee:2015jqa}, and Borah et al.'20 $0.546(70)$ fm$^2$ \cite{Borah:2020gte}. These points come from fits using general functions over a large range of energies. Finally (in green), we include Belushkin et al.'06 0.729(9) and 0.723$_{-0.012}^{+0.003}$ fm$^2$ \cite{Belushkin:2006qa}; Lorenz et al.'12 $0.740(34)$ fm$^2$\cite{Lorenz:2012tm}; Alarcon et al.'17 $1.29(25)$ fm$^2$ \cite{Alarcon:2017lhg}, Alarcon et al.'20 $0.723(17)$ fm$^2$ \cite{Alarcon:2020kcz}; and Lin et al.'21 0.717(10) fm$^2$ \cite{Lin:2021umk}, which incorporate dispersion relation constraints.}
  \label{fig:rM2}
\end{figure}

In Fig.~\ref{fig:rM2}, and in Table~\ref{tab:rn}, we give the values of these radii as predicted by pure chiral perturbation theory, and after including the contributions associated with the Delta particle (which can be motivated by a large $N_c$ analysis). In order to estimate the error, the following scaling was taken \cite{Peset:2014jxa}: $m_{\pi} \sim \sqrt{\lQ m_q}$ and $\Delta \sim \frac{\lQ}{N_c}$. One then has the double expansion 
$\frac{m_{\pi}}{\lQ} \sim \sqrt{\frac{m_q}{\lQ}}$ and $\frac{\Delta}{\lQ} \sim \frac{1}{N_c}$. To determine the 
relative size between $m_{\pi}$ and $\Delta$, one observes that  $\frac{m_{\pi}}{\Delta} \sim N_c \sqrt{\frac{m_q}{\lQ}} \sim 1/2$.
Therefore, a 50\% uncertainty is associated with the pure chiral computation. Leaving aside the Delta, the splitting with the next resonances suggests a mass gap of order $\lQ \sim$ 500-770 
MeV depending on whether one considers the Roper resonance or the $\rho$. Similarly, it is also taken $m_K \sim \sqrt{\lQ m_s} \sim 500$ MeV of order $\lQ$.  
Therefore, when including the effect of the $\Delta$, one assigns $\frac{m_{\pi}}{\lQ} \sim 1/3$ and $\frac{\Delta}{\lQ} \sim 1/2$, as the uncertainties of the pure chiral 
and the Delta-related contribution respectively, and adds these errors linearly for the final error. This gives the expected size of the uncomputed corrections, which is taken as the error of the computation. A similar analysis yields the error quoted in Eq.~(\ref{eq:c3had}) below. 

For $\langle r^{2k} \rangle _{M/E}$ with large $k$, it is likely that the error is smaller than the one quoted in Table~\ref{tab:rn}.  This error is estimated using the size of the ratios of the pion and/or Delta-Nucleon mass difference with respect to higher resonances~\cite{Peset:2014jxa}. Nevertheless, it is expected that there should be some kind of suppression in $k$, the power of the radii. This is worth being investigated further in the future. Indeed, the importance of the Delta resonance becomes smaller and smaller as $k$ increases.
General analytic expressions for these coefficients to arbitrary order can be found in \cite{Peset:2014jxa}. These coefficients are derived from the Sachs form factors computed in \cite{Bernard:1992qa,Fearing:1997dp,Bernard:1998gv}. The numerical values at low orders can be found in \cite{Horbatsch:2016ilr}. Some of the numerical values of these coefficients at higher order were communicated by one of us to one of the authors of \cite{Alarcon:2017lhg} before its publication. 

Note that these predictions for the radii do not include electromagnetic corrections. Therefore, they cannot reach ${\cal O}(\alpha)$ precision (in practice, the hadronic error is larger and masks possible ${\cal O}(\alpha)$ effects). 

\begin{table}[h]
\addtolength{\arraycolsep}{0.2cm}
$$
\begin{array}{|l||c|c|c|c|c|}
\hline
  & \pi & \pi\&\Delta  & \cite{Alarcon:2017lhg} & \cite{Borah:2020gte}& \cite{Distler:2010zq}
\\ \hline\hline
  \langle r^4 \rangle_{\rm E} &  0.71(36)  & 0.60(29) &   (1.16,1.70)&1.08 (29) & 2.59(19)(04)
\\ \hline\hline
\langle r^6 \rangle_{\rm E} & 5.4(2.7) & 5.0(2.0) & (7.59,9.00)&&29.8(7.6)(12.6)
\\ \hline\hline
\langle r^8 \rangle_{\rm E} & 104(52) & 99(37) & (121,129)&&
\\ \hline\hline
\langle r^{10} \rangle_{\rm E} \times 10^{-3}& 3.7(1.8) & 3.5(1.3) & (3.86,3.94)&&
\\ \hline\hline
\langle r^{12} \rangle_{\rm E} \times 10^{-3}& 201(100) & 196 (69) & (202,204)&&
\\ \hline\hline
\langle r^{14} \rangle_{\rm E} \times 10^{-5}& 159(79) & 155(54) & (155,156)&&
\\ \hline\hline
\langle r^{16} \rangle_{\rm E} \times 10^{-7}& 169(85) & 166(58) & (161,162)&&
\\ \hline\hline
\langle r^{18} \rangle_{\rm E} \times 10^{-9}& 235(112) & 232(80) & (220,221)&&
\\ \hline\hline
\langle r^{20} \rangle_{\rm E} \times 10^{-12}& 41(21) & 41(14) & (37.9,38.0)&&
\\ \hline\hline
\langle r^4 \rangle_{\rm M} & 0.71(35) & 0.79(28) & (2.14,2.81)&-2.0 (1.9)&
\\ \hline\hline
\langle r^6 \rangle_{\rm M} & 6.3(3.2) & 6.9(2.4) & (12.9,15.2)&&
\\ \hline\hline
\langle r^8 \rangle_{\rm M} & 127(63) & 137(47) & (194,213)&&
\\ \hline\hline
\langle r^{10} \rangle_{\rm M} \times 10^{-2}& 44(22) & 47(16) &&&
\\ \hline\hline
\langle r^{12} \rangle_{\rm M} \times 10^{-3}& 239(120) & 254(87) &&&
\\ \hline\hline
\langle r^{14} \rangle_{\rm M} \times 10^{-5}& 184(92) & 194(66) &&&
\\ \hline\hline
\langle r^{16} \rangle_{\rm M} \times 10^{-7}& 191(96) & 201(69) &&&
\\ \hline\hline
\langle r^{18} \rangle_{\rm M} \times 10^{-9}& 258(129)  & 271(92) &&&
\\ \hline\hline
\langle r^{20} \rangle_{\rm M} \times 10^{-12}& 44(22)  & 46(16) &&&
\\ \hline
\end{array}
$$
\caption{{Values of 
$\langle r^n \rangle_{\rm E}$ and 
$\langle r^n \rangle_{\rm M}$ in units of {\rm fm}$^n$. The first two columns follow from chiral perturbation theory. 
The first column is the pure chiral prediction 
(with only pions),
and the second column is the result after the inclusion 
of the effects associated with the Delta particle.
Uncertainties in the last two digits are shown 
in parentheses. The third column are the intervals predicted by chiral motivated dispersion relation determination \cite{Alarcon:2017lhg}. The 4th column \cite{Borah:2020gte} are fits to data using the muonic hydrogen proton radius value. The 5th column is taken from the analysis of scattering data in \cite{Distler:2010zq}.}}
\label{tab:rn}
\end{table}

The odd powers of the moments of the charge distribution of the proton are obtained (defined) through the relation:
\be
\label{rMoments}
\langle r^{2k+1}\rangle
\equiv
\frac{\pi^{3/2}\Gamma(2+k)}{\Gamma(-1/2-k)}2^{4+2k}
\int \frac{d^3q}{(2\pi)^3}\frac{1}{{\bf q}^{2(2+k)}}
\left[
G_E({\bf q}^2)-\sum_{n=0}^k\frac{{\bf q}^{2n}}{n!} \left(\frac{d}{d{\bf q}^2}\right)^nG_E({\bf q}^2)\Big|_{{\bf q}^2=0}
\right]
\,,
\ee
and for $G^2_E$ the analogous expression reads
\be
\langle r^{2k+1}\rangle_{(2)}
=
\frac{\pi^{3/2}\Gamma[2+k]}{\Gamma[-1/2-k]}2^{4+2k}
\int \frac{d^3q}{(2\pi)^3}\frac{1}{{\bf q}^{2(2+k)}}
\left[
G^2_E(-{\bf q}^2)-\sum_{n=0}^k\frac{{\bf q}^{2n}}{n!} \left(\frac{d}{d{\bf q}^2}\right)^nG^2_E(-{\bf q}^2)\Big|_{{\bf q}^2=0}
\right]
\,.
\ee
By using dimensional regularization, one can eliminate all the terms proportional to integer even powers of ${\bf q}^2$.  For $k>1$, $
\langle r^{2k+1}\rangle$ is dominated by the chiral result and can be approximated by
\be
\langle r^{2k+1}\rangle
\simeq
\frac{\pi^{3/2}\Gamma(2+k)}{\Gamma(-1/2-k)}2^{4+2k}
\int \frac{d^{D-1}q}{(2\pi)^{D-1}}\frac{1}{{\bf q}^{2(2+k)}}
G^{(2)}_E({\bf q}^2)
\,,
\ee
and 
$\langle r^{2k+1}\rangle_{(2)}
\simeq
2\langle r^{2k+1}\rangle$ at leading order in chiral perturbation theory. 
Analytic expressions of these quantities obtained from chiral perturbation theory can be found in Eq.~(4.12) of \cite{Peset:2014jxa}.  

We show some numbers for these moments in Table~\ref{tab:rnZemach} both in the effective theory with 
only pions and in the effective theory with pions and Deltas. We also compare with results obtained from different parameterizations of the experimental data for the form factors. 
In Table~\ref{tab:rnZemach}, we compare odd moments with the standard dipole ansatz  \cite{Janssens:1965kd}, 
and with different determinations using experimental data of the electric Sachs form factor fitted to more sophisticated functions 
\cite{Kelly:2004hm,Distler:2010zq}.\footnote{The agreement with \cite{Kelly:2004hm} for $n=7$ is accidental. We have checked that the
growth with $n$ is different with respect to the chiral prediction.} The latter fit claims to be the most accurate. Nevertheless, we observe large differences, bigger than the errors. This is especially worrisome for large $n$, since the chiral prediction is expected to 
give the dominant contribution of $\langle r^n \rangle$ for $n \geq 3$. In this respect, the chiral result may help to shape the appropriate fit function and, thus, to discriminate between different options, as well as to assess uncertainties. The impact of 
choosing different fit functions can be fully appreciated, for instance, in the different values of the electric proton charge radius 
obtained in Ref. \cite{Bernauer:2010wm} versus Ref. \cite{Lorenz:2012tm} from direct fits to the $ep$ scattering data. Such values differ by around 3 standard deviations. Even more worrisome is the fact that the larger the $n$, the more sensitive is the determination of $\langle r^{2n+1}\rangle$ to the subtraction terms included to render these objects to be finite for odd powers of  $2n+1$. This should be contrasted with the fact that, on general grounds, one expects 
the charge/Zemach moments will be more and more sensitive to the chiral region for $n \rightarrow \infty$. We stop the discussion here, but the reason for such large discrepancies should be further investigated. 

\begin{table}[ht]
\addtolength{\arraycolsep}{0.2cm}
$$
\begin{array}{|l||c|c|c|c|}
\hline
  & \langle r^3 \rangle   & \langle r^5 \rangle  & \langle r^7 \rangle &  \langle r^3 \rangle_{(2)}
\\ \hline\hline
\pi & 
0.50(25)
& 
1.62(81)
& 
20.1(10.5)
& 
1.00(50)
\\
\pi\&\Delta & 
0.41(21)
& 
1.52(59)
& 
20.2(7.3)
& 
0.81(42)
\\ \hline
\cite{Janssens:1965kd}& 0.7706  &  1.775  &7.006  & 2.023 \\
\cite{Kelly:2004hm}  & 
0.9838
& 
3.209
 & 
19.69
& 
2.526
\\
\cite{Distler:2010zq}  & 1.16(4)  & 8.0(1.2)(1.0) & ---& 2.85(8) \\
     \hline
\end{array}
$$
\caption{{ Values of $\langle r^{2n+1} \rangle$ in fm units. The first two rows give the prediction from chiral perturbation theory: the first 
row for the effective theory with only pions and the second for the theory with pions and Deltas. The third row 
corresponds to the standard dipole fit of Ref.~\cite{Janssens:1965kd} with $\langle r^2 \rangle=0.6581$ fm$^3$. 
The fourth and fifth rows correspond to different parameterizations of experimental data \cite{Kelly:2004hm,Distler:2010zq}, 
with the latest fit being based on Mainz data. For completeness, 
we also quote $\langle r^3 \rangle_{(2)}=2.71$ fm$^3$ from Ref.~\cite{Friar:2005jz}.}}
\label{tab:rnZemach}
\end{table}

The other Wilson coefficients that appear in the observables we are considering in this review are $c_3$ and $c_4$ (see Eqs.~(\ref{c3}) and (\ref{c4pe})). $c^{\rm had}_3$ encodes all the hadronic effects to the spin-independent four-fermion Wilson coefficient. 
At ${\cal O}(\al^2)$, it is generated by the TPE contribution. Since $c_3^{\rm had}$ depends linearly on the lepton mass, it is dominated by the infrared dynamics and diverges linearly in the chiral limit. This produces an extra
$m_{l_i}/m_{\pi}$ suppression with respect to its natural size, and allows us to compute the leading 
pure-chiral and Delta-related effects in a model-independent way. The complete matching computation 
between HBET and NRQED was made in Ref.~\cite{Peset:2014jxa} to which we refer for details (partial results can be found in \cite{Pineda:2004mx,Nevado:2007dd}, and in Ref.~\cite{Alarcon:2013cba} in the context of relativistic baryon effective theory). Overall, in Ref.~\cite{Peset:2014jxa} the following result was obtained for muonic hydrogen
\bea
\nn
c^{\rm had}_{3} &\sim& \al^2 \frac{m_{\mu}}{m_{\pi}}\left[1+\#\frac{m_{\pi}}{\Delta}+\cdots\right]
+{\cal O}\left(\al^2 \frac{m_{\mu}}{\lQ}\right)
\\
&=&\al^2 \frac{m_{\mu}}{m_{\pi}} \begin{cases}\displaystyle 
47.2(23.6)&(\pi),\\
56.7(20.6)&(\pi+\Delta), \end{cases}
\label{eq:c3had}
\eea
where the upper and lower numbers refer to the matching computation with only pions, or with pions and the Delta particle, respectively. The error was obtained following the same procedure as  for the numbers given in Fig.~\ref{fig:rM2}, and in Table~\ref{tab:rn}. 

Eq.~(\ref{eq:c3had}) is a pure prediction of the effective theory. It is the most precise expression that can be obtained in a model-independent way, since ${\cal O}(\alpha^2\frac{m_{\mu}}{\lQ})$ effects are not controlled by the chiral theory and would require new counterterms.
This number is only marginally bigger than $c^{\rm had}_{3}=\al^2 \frac{m_{\mu}}{m_{\pi}}54.4(3.3)$, the number used in Ref.~\cite{Antognini:1900ns}, which follows from the analysis in Ref.~\cite{Birse:2012eb}. 
This number is obtained as the sum of 
the elastic and inelastic terms from Ref.~\cite{Carlson:2011zd} and the subtraction term from 
Ref.~\cite{Birse:2012eb}. Note that this evaluation is model dependent due to unavoidable modelling of the subtraction function. It needs to split $c_3^{\rm had}$ into three contributions:\footnote{There is some degree of arbitrariness in the definition of the different terms so that some contributions can move from one term to the other, in particular between the Born and polarizability term. This has to be taken into account when comparing with the literature.}
\be \label{eq:c3split}
c_3^{\rm had}=c_3^{\rm Born}+c_3^{\rm pol}=c_3^{\rm Born}+c_3^{\rm sub}+c_3^{\rm  inel}
\,.
\ee
The inelastic contribution reads
\bea
c_3^{\mathrm{inel}} &=& \frac{\alpha^2}{\pi} \frac{M}{m_\mu} \int_0^\infty \frac{dQ^2}{Q^2} \int_{\nu^{\mathrm{inel}}_{\mathrm{thr}}}^\infty d\rho \left(	\frac{ \widetilde\gamma_1(\tilde{\tau},\tau_\mu) \mathrm{Im} S_1(\rho,Q^2)}{\rho} +	\frac{ \rho \widetilde\gamma_2(\tilde{\tau},\tau_\mu) \mathrm{Im} S_2(\rho,Q^2)}{Q^2} \right), 
\eea
with kinematic notation:
\be
  \tilde{\tau} = \frac{\rho^2}{Q^2},  \label{taus}
\ee
and the weighting functions $\tilde{\gamma}_1$ and $\tilde\gamma_2$ are given by \cite{Carlson:2011zd}
\bea
\tilde{\gamma}_1(\tau_1,\tau_2) &=& \frac{\sqrt{\tau_2} \gamma_1(\tau_2) - \sqrt{\tau_1} \gamma_1(\tau_1)}{\tau_2 - \tau_1},\nonumber\\
\tilde\gamma_2(\tau_1,\tau_2) &=& \frac{1}{\tau_2 - \tau_1}\left(\frac{ \gamma_2(\tau_1)}	{\sqrt{\tau_1}} - \frac{\gamma_2(\tau_2)}{\sqrt{\tau_2} } \right).
\eea
The Born contribution at the leading order can be expressed as
\be
c_{3,\rm Born}^{pl_i}=\frac{\pi}{3}\al^2M^2m_{l_i}\langle r^3 \rangle_{(2)}
\,,
\qquad
\langle r^3 \rangle_{(2)}= \frac{48}{\pi}
\int_0^{\infty} {d Q \over Q^4}
\left(
G_E^2(Q^2)-1+\frac{Q^2}{3} r_p^2
\right)\,.
\ee
Note that the terms proportional to "1" and $r_p^2$ vanish in dimensional regularization. Finally \cite{Pachucki:1999zza}, 
\bea
c_{3,\rm sub}^{pl_i}
&=&
- e^4 M m_{l_i}\int {d^4k_E \over (2\pi)^4}{1 \over k_E^4}{1 \over
k_E^4+4m_{l_i}^2k_{0,E}^2 }
\nn
(3k_{0,E}^2+{\bf k}^2)S_1(0,k_E^2)
\\
&=&
-\frac{\al^2 M}{2m_{l_i}}
\int_0^{\infty}\frac{d Q^2}{Q^2}
\left\{
1+\left(1-\frac{Q^2}{2 m_{l_i}^2}\right)\left(\sqrt{\frac{4m^2_{l_i}}{Q^2}+1}-1\right)
\right\}
S_1(0,Q^2)
\,.
\eea

These terms were computed using different techniques. $c_3^{\rm Born}$ is proportional to $\langle r^3 \rangle_{(2)}$ at leading order. To compute this object accurately requires a very precise knowledge of the elastic form factors, as a very precise cancellation of the point-like contributions should occur (for some determinations of this quantity see \cite{Pachucki:1999zza,Carlson:2011zd}). The attempts to compute $c_3^{\rm pol}$ using dispersion relations require the introduction of a momentum-dependent subtraction function to make the dispersion relation integrals convergent (assuming Regge behavior as discussed in Section~\ref{Sec:Structurefunctions}). Such function cannot be deduced from experiment, nor theory, without extra assumptions~\cite{Gasser:1974wd,Gorchtein:2013yga,Tomalak:2015hva,Caprini:2016wvy}, and, therefore, introduce some model dependence, which is difficult to quantify, as emphasized
in Ref. \cite{Hill:2010yb}. The determination in Ref.~\cite{Birse:2012eb}, used in \cite{Antognini:1900ns}, suffers from this problem. Even though the 
low-energy behavior of the forward virtual Compton tensor was computed to ${\cal O}(p^4)$, 
this does not reflect in an improved determination of the polarizability correction, since 
an effective dipole form factor is used, not only at the $\rho$ mass scale, but also at the chiral 
scale. This problem also introduces a model dependence in the error estimate of the subtraction term. For comparison, we show different values for the subtraction term obtained in the literature in Table~\ref{Table:Sub}.\footnote{Note that this object is divergent in HBET once the Delta particle is introduced \cite{Peset:2014jxa}, and very large in relativistic B$\chi$PT \cite{Alarcon:2013cba}.} 
\begin{table}[htb]
\centering
$$
\begin{array}{|c|cccccc|c|}                          
\hline
(\mu {\rm eV})  &   \cite{Pachucki:1999zza}   &  \cite{Martynenko:2005rc}  & \cite{Carlson:2011zd}   & \cite{Birse:2012eb}  &  \cite{Gorchtein:2013yga}    &  \cite{Tomalak:2015hva,Tomalak:2018uhr}           &  \cite{Alarcon:2013cba}  \\
\hline
\Delta E^{(\mathrm{sub})} &  -1.8  &  -2.3  &  -5.3(1.9)  & -4.2(1.0) & 2.3 (4.6)^{(1)} & 2.3(1.3)                               
& $3.0$  \\
\hline
\end{array}
$$
\caption{ Contributions to the Lamb shift generated by the subtraction term that can be found in the literature. $^{(1)}$This number is the adjusted value of Ref.~\cite{Gorchtein:2013yga}, given in \cite{Alarcon:2013cba}. 
\label{Table:Sub}}
\end{table}
 
The dispersion relation motivated determinations of $c_{3}^{\rm pol}$ use the computations for the subtraction term discussed above. The analysis of Ref.~\cite{Alarcon:2013cba} has a different status. 
In this reference, the polarizability correction was computed using $ {\rm B}\chi{\rm PT}$ with only pions.  Such computation treats the baryon relativistically. The result incorporates some subleading effects, which are sometimes used to give an estimate of higher-order effects in $ {\rm HB}\chi{\rm PT}$. Nevertheless, the computation also assumes that a theory with only baryons and pions is appropriate at the proton mass scale. This should be taken with due caution. Still, it would be desirable to have a deeper theoretical understanding of the difference to the HBET (with only pions) result, which may signal that relativistic corrections are important for the polarizability contribution. In any case, the  $ {\rm B}\chi{\rm PT}$ computation differs from the pure chiral HBET result by around 50\% (this means around 1.5 times the error used for the chiral contribution in HBET, once effects associated with the Delta particle are incorporated in the calculation), which is reasonable. We show these results in Table~\ref{Table:Epol} (for convenience we directly write the correction to the Lamb shift). 

\begin{table}[ht]
\centering
$$
\begin{array}{|c|ccccc|c|cc|}                          
\hline
(\mu {\rm eV})  & {\rm DR+Model}& \cite{Pachucki:1999zza}   &  \cite{Martynenko:2005rc}  & \cite{Carlson:2011zd}  &  \cite{Gorchtein:2013yga}   &  {\rm B}\chi{\rm PT} \cite{Alarcon:2013cba}  (\pi) & {\rm HBET} \cite{Nevado:2007dd} (\pi) & \cite{Peset:2014yha}(\pi\&\Delta)\\
\hline
\Delta E_{\mathrm{pol}} &  & 12(2)       &    11.5     &   7.4  (2.4)   &   15.3(5.6)    & 8.2 (^{+1.2}_{-2.5}) & 18.5(9.3) & 26.2(10.0)
\\
\hline
\end{array}
$$
\caption{Predictions for the polarizability contribution to the $n=2$ Lamb shift. The first four entries use dispersion relations for the inelastic term and different modeling
functions for the subtraction term. The number of the fourth entry has been taken from \cite{Alarcon:2013cba}. 
The 5th entry is the prediction obtained using B$\chi$PT. The last two entries are the predictions of HBET: 
The 6th entry is the prediction at leading order (only pions) and the last entry is the 
prediction at leading and next-to-leading orders (pions and Deltas). 
\label{Table:Epol}}
\end{table}

For the electron-proton sector, the suppression factor $\frac{m_{e}}{m_{\pi}}$ (though logarithmic enhanced) makes the TPE contribution to be barely negligible for determinations of the proton radius from regular hydrogen\footnote{Strictly speaking, this is only true if the Rydberg constant has already been determined. In the determination of the Rydberg constant from the very precise 1S-2S energy difference measurement \cite{PhysRevLett.107.203001}, such contribution has to be taken into account.}, and eventual electron-proton scattering at very low momentum transfer. Nevertheless, we profit from the theoretical setup developed in \cite{Peset:2014jxa} for muonic hydrogen, to obtain the equivalent prediction from chiral perturbation theory of Eq.~\eqref{eq:c3had} for regular hydrogen
\bea
c^{\rm had}_{3} 
&=&\al^2 \frac{m_{e}}{m_{\pi}} \begin{cases}\displaystyle 
115.5(57.8)&(\pi),\\
140.4(47.1)&(\pi+\Delta). \end{cases}
\label{eq:c3TPEep}
\eea 
We can split the contribution to the fine-splitting energy into polarizability and Born contributions, as in Eq.~\eqref{eq:c3split}, to find separate predictions for each term.\footnote{The Born contribution was actually also computed in \cite{Pineda:2004mx}. The result is correct for the pure chiral case but incorrect for the correction including the Delta particle. The correct expression can be found in \cite{Peset:2014jxa}.}
The result for the polarizability contribution approaches very reasonably the logarithmic result obtained in \cite{Pineda:2004mx} where it is found (using the same normalization as in Table~\ref{tab:ETPEep}) $\Delta E_\mathrm{TPE}=-66.5$ Hz for only pions, and $\Delta E_\mathrm{TPE}=-77.6$ including the Delta (to compare with the two first entries in the second column of Table~\ref{tab:ETPEep}). In Table~\ref{tab:ETPEep}, we quote the result for the polarizability from \cite{Khriplovich:1997fi} and for the Born contribution from~\cite{Friar:2005jz} which are collected together in Ref.~\cite{Eides:2007exa}. The reference~\cite{Friar:2005jz} is quoted for the polarizability in \cite{Mohr:2012tt}  where, however, no estimate is given for the Born correction. Other estimates for just the polarizability term can be found in \cite{Faustov:1999ga,Rosenfelder:1999px,Martynenko:2005rc} obtaining similar numbers.\footnote{Note that the polarizability term suffers from the same drawbacks as in the case of muonic hydrogen, in particular the uncontrolled dependence on the subtraction function.} A more recent determination of the TPE in hydrogen using a dispersion relation analysis of the Mainz data without modeling of the $Q^2$ dependence of the unpolarized proton structure functions and using the proton charge radius from muonic hydrogen can be found in \cite{Tomalak:2018uhr}. 
In Table~\ref{tab:ETPEep}, we display these results. We observe that, just as in muonic hydrogen, while each contribution is separately quite different from the determination in the chiral theory, the total TPE are perfectly compatible among the different determinations.
\begin{table}[ht]
\addtolength{\arraycolsep}{0.2cm}
$$
\begin{array}{|l||c|c|c|}
\hline
  & \Delta E_\mathrm{Born}\,[\mathrm{Hz} ]  & \Delta E_\mathrm{pol}\,[\mathrm{Hz}] &  \Delta E_\mathrm{TPE}\,[\mathrm{Hz}]
\\ \hline\hline
\pi & -14.7(7.4) & -87.2(43.6)  & -101.9(51.0) \\
\pi\&\Delta & -12.0(5.7)& -111.9(38.5)& -123.9(41.6)\\
\hline
\cite{Faustov:1999ga}& --& -99(10) & --\\
\cite{Rosenfelder:1999px}& --& -95(7) & --\\
\cite{Eides:2007exa}& -40(19)  &  -70(13)  &-110(23)  \\
\cite{Tomalak:2018uhr} &-39.9(6.8)   &-65.1(7.2) & -105.0(9.9)\\
     \hline
\end{array}
$$
\caption{Predictions for the TPE energy in Hz and its splitting into polarizability and Born contributions (normalized by a factor $\tfrac{\delta_{\ell 0}}{n^3}$) for regular hydrogen. The first two lines correspond to the results from the chiral theory, for only pions, and after also including the Delta, presented in this review. The third and fourth line correspond to predictions of the polarizability term from Refs. \cite{Faustov:1999ga} and \cite{Rosenfelder:1999px} respectively. The fifth line corresponds to the result presented in \cite{Eides:2007exa}, which adds the result for the polarizability from \cite{Khriplovich:1997fi} and the Born contribution from~\cite{Friar:2005jz}, where the Born contribution is approximated by the leading term in the nonrelativistic expansion, i.e., the Zemach radius. The last line is the most recent determination using dispersion relations and a subtraction function in \cite{Tomalak:2018uhr}.}
\label{tab:ETPEep}
\end{table}

\subsection{\texorpdfstring{$r_p$}{rp}. Dispersion relation motivated fits of form factors}

As the precision one may reach with pure chiral perturbation theory computations is limited (if one tries to increase the precision, new counterterms appear that cannot be determined by theory), one may think of alternative approaches. All of them need extra experimental input. One of them is the use of dispersion relations to describe the form factors.  In other words, to use parameterizations of the form factors that satisfy Eqs.~(\ref{eq:F1F2DR}). This ensures that such parameterizations fulfill unitarity, analyticity and crossing symmetry by construction. Note that chiral perturbation theory computations also enjoy these properties, within the accuracy of the computation, for small momentum transfers. If the parameterizations used for the dispersion relation implemented chiral symmetry, they should coincide  (up to the accuracy chiral symmetry has been implemented in those dispersion-relation motivated parameterizations) with chiral perturbation theory computations at low momentum transfer. Even if chiral symmetry is implemented, the freedom one has in parameterizing the form factors is enormous. In practice, physically-motivated (but not model independent) parameterizations of the imaginary part of the form factors are used. In order for such analyses to be useful, they need experimental input to determine the parameters that are introduced in such parameterizations. One such analysis can be found in \cite{Belushkin:2006qa}, where fits are performed simultaneously to space-like and time-like experimental data. The experimental data in the space-like region corresponds to the real part of the form factors. It is obtained from elastic electron-proton scattering, whereas the imaginary parts are constrained by the proton-proton annihilation. The result obtained for the electric proton charge radius in Ref. \cite{Belushkin:2006qa} was $r_p=0.844_{-0.004}^{+0.008}$ or $r_p=0.830^{+0.005}_{-0.008}$ depending on the parameterization of the
 high-energy part. Indeed, the difference between both numbers gives a lower bound of the error associated with the form chosen for the parameterization of the data at high energies. 

Since then, several analyses have been performed along similar lines but trying to include more chiral structure in physically-motivated parameterizations of the imaginary part and/or fitting to more recent data of the form factors. In these more recent analyses, the experimental data in the time-like region is not directly used in the fits but only the one in the space-like region, which is more precise. Updates by the Bonn group can be found in \cite{Lorenz:2012tm}, where the value $r_p=0.84_{-0.1}^{+0.1}$ fm was obtained after fitting the dispersion-relation motivated parameterizations to the Mainz data. A more recent update can be found in \cite{Lin:2021umk}, where the PRad data was added to the previous fit and the following values were reported: $r_p =0.838^{+0.005}_{-0.004}{}^{+0.004}_{-0.003}$ fm and $\sqrt{\langle r^2 \rangle_M} =0.847 \pm 0.004 \pm 0.004$ fm. 

Other groups have also applied combinations of chiral symmetry and dispersion relations to fit Mainz data, like the one in \cite{Alarcon:2017lhg}. In this reference, they did not dare to try to determine the lowest radii, i.e., $r_p$, but only higher radii. In this reference, it was said that the higher-order derivatives depend on the $\rho$ and $\omega$ mass and the two-pion threshold. We argue that these higher radii should mainly depend on the singularity closest to the origin, the two-pion threshold in this case. This seems to be confirmed by Table~\ref{tab:rn}. In more recent analysis \cite{Alarcon:2018zbz}  by the same group, the authors gave indeed a value for the proton radius: $r_p=0.844(7)$ fm
and in \cite{Alarcon:2020kcz} for its magnetic counterpart: $\sqrt{\langle r^2 \rangle_\mathrm{M}} = 0.850 \pm 0.001 \pm 0.010$ fm.

\begin{figure}[ht]
  \begin{center}
   \hspace*{-2.7cm}\includegraphics[width=0.95\textwidth]{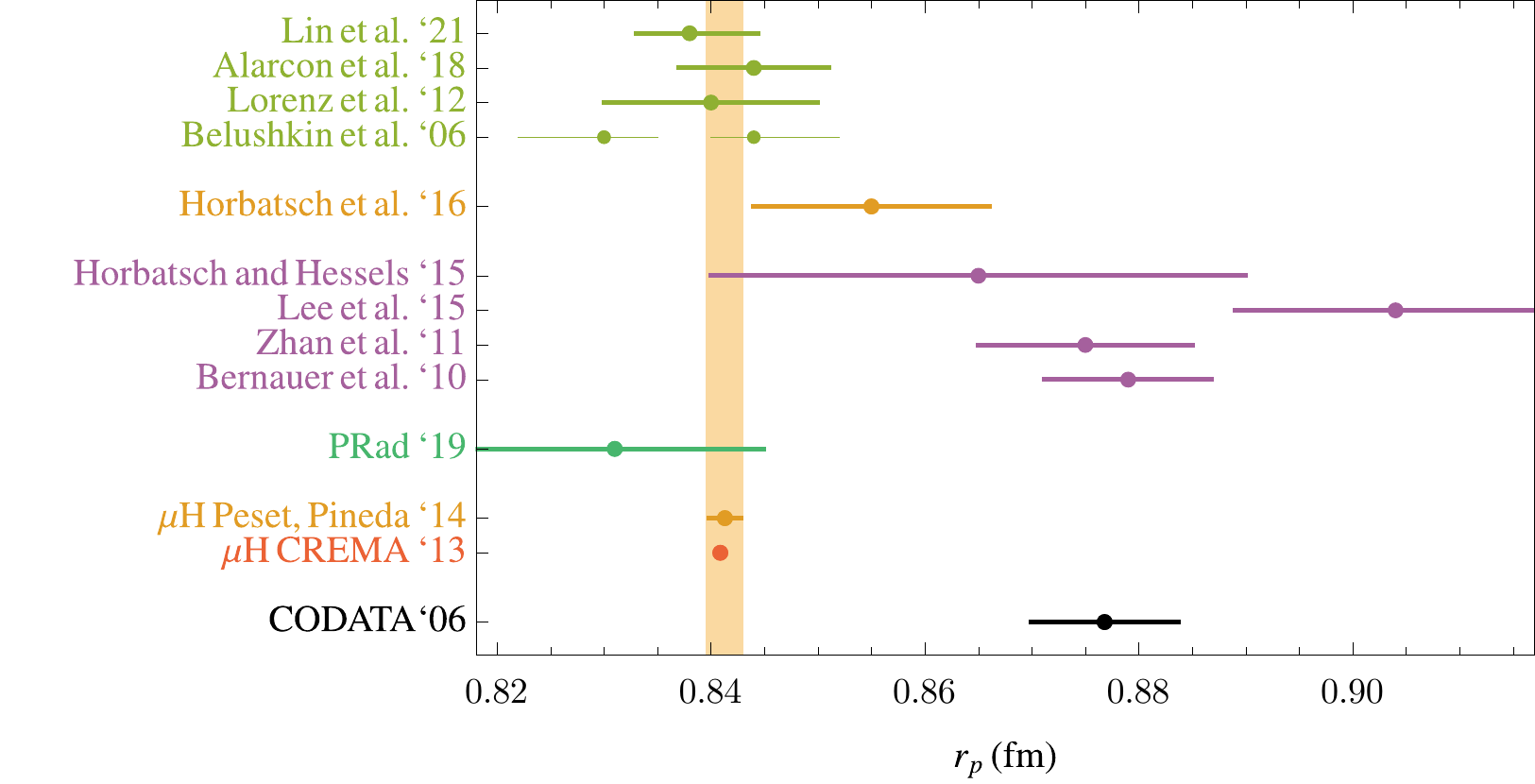}
  \end{center}
    \caption{ Values of $r_p$ in units of fm obtained from fits to electron-proton data in comparison with determinations from muonic hydrogen. From top to bottom (when necessary we have combined errors in quadrature).  
    Using dispersion relations (in green): Belushkin et al. `06
   $r_p=0.844_{-0.004}^{+0.008}$ fm or $r_p=0.830^{+0.005}_{-0.008}$ fm \cite{Belushkin:2006qa}; Lorenz et al. `12 $r_p=0.84_{-0.1}^{+0.1}$ fm \cite{Lorenz:2012tm}; Alarcon et al. `18 $r_p=0.844(7)$ fm \cite{Alarcon:2018zbz}; and Lin et al `21 $r_p=0.838_{-0.05}^{+0.06}$ fm \cite{Lin:2021umk}. 
   Using chiral perturbation theory: Horbatsch et al. `16 $r_p=0.855(11)$ fm \cite{Horbatsch:2016ilr} (in orange).
    Using fits to functions (in purple): Bernauer et al. `10 $r_p=0.879(8)$ fm \cite{Bernauer:2010wm}; Zhan et al. `11 $r_p=0.875(10)$ fm \cite{Zhan:2011ji}; Lee et al. `15 
   $r_p=0.904(15)$ fm \cite{Lee:2015jqa}; Horbatsch and Hessels `15 $r_p=0.865(25)$ fm \cite{Horbatsch:2015qda}. 
    The determination from the PRad experiment \cite{Xiong:2019umf} (in turquoise).
     We compare to the values obtained in muonic hydrogen from an EFT analysis $r_p=0.8413 (15)$ fm \cite{ Peset:2014yha} (in orange) and the CREMA determination $r_p=0.84087 (39)$ fm \cite{Antognini:1900ns} (in red), and to the averaged value reported by CODATA06 $r_p=0.8768 (69)$ fm \cite{Mohr:2008fa} (in black).
   }
  \label{fig:rpfits}
\end{figure}

We show these numbers for the proton radius in Fig.~\ref{fig:rpfits} in comparison with the muonic hydrogen determinations. The values for $\langle r^2_M \rangle$ reviewed in this section can be found in Fig.~\ref{fig:rM2}.

\subsection{\texorpdfstring{$r_p$}{rp}. Pure fits of electron-proton scattering data without dispersion-relation motivated parameterization}

The main problems in determinations of the proton radius from fits to electron-proton scattering data are choosing the functional form, and the range of energies, to be used in the fit. At present, experiments have not reached $Q^2$ low enough to directly fit the proton radius, and one has to extrapolate to reach the $Q^2 \rightarrow 0$ limit. In the previous section, the way out from this problem has been the use of dispersion relations and the use of physically motivated parameterizations for the imaginary part of the form factors. These parameterizations, however, unavoidably introduce some degree of model dependence. This has the problem that it is not easy, or possible, to assess the error to model-dependent analyses. If one restricts to direct fits to electron-proton data, and without extra constraints, the parameterization of the form factors is arbitrary, and so is the range of $Q^2$ used for the fits. Nevertheless, one can find strong arguments for why one should focus on the low-$Q^2$ part of the data in order to extract the proton charge radius~\cite{Griffioen:2015hta,Cui:2021vgm}.
The charged-pion production threshold at 
$Q^2=- 4 m_\pi^2 \approx -0.078 \ \rm GeV^2$ 
results in a branch cut in the analytically continued 
form factor. Thus, 
one can seriously doubt attempts \cite{Bernauer:2010wm,Zhan:2011ji,Lee:2015jqa} at fitting 
(by polynomials or other functions, 
such as splines) 
data beyond the value of $Q^2=0.078 \ \rm GeV^2$, 
and having confidence in an accurate determination
of the slope of $G_{\rm E}(Q^2)$ at $Q^2=0$. Note that this problem also affects most lattice determinations discussed above. Another point of concern is that at higher $Q^2$ the model dependence of the TPE contribution is expected to affect more the fits. 
Fits that include higher-$Q^2$ MAMI data 
\cite{Bernauer:2010wm}
also require floating 31 normalization constants, 
and the floating of these constants leads to considerable
flexibility in the fits, 
which also makes determination of the higher-order moments
more difficult. However, 
concentrating only on low-$Q^2$ data
($Q^2<0.078  \ \rm GeV^2$)
has not allowed for an accurate determination of $r_{\rm p}$, 
since this data cannot determine the necessary higher moments
(in particular, $\langle r^2 \rangle_{\rm M}$
and
$\langle r^4 \rangle_{\rm E}$)
to sufficient accuracy
to allow for a precise extrapolation to $Q^2$=0 
of the required first derivative of $G_{\rm E}(Q^2)$.
Thus, 
out of necessity, 
many attempts have been made to 
fit scattering data up to higher $Q^2$ 
to determine $r_p$ 
while simultaneously determining the higher-order
moments. 

In Ref.~\cite{Horbatsch:2015qda}, 
it was shown that values of $r_{\rm p}$
ranging from 0.84 to 0.89 fm are possible from 
acceptable fits of the MAMI data, 
with the value of $r_{\rm p}$ 
depending on the functional forms of 
$G_{\rm E}$ and $G_{\rm M}$ that are  
used for the extrapolation to $Q^2$=0. 
In particular, 
the higher moments assumed by the 
different functional forms lead to a
wide range of $r_{\rm p}$ values.
The implication of that work~\cite{Horbatsch:2015qda}
is that a precise value of $r_{\rm p}$ cannot be obtained from
electron-proton elastic scattering, 
unless precise lower-$Q^2$ data becomes available,
or unless there are external constraints on the 
functional form of $G_{\rm E}(Q^2)$ and 
$G_{\rm M}(Q^2)$. 
The latter 
(external constraints on the functional forms) can be obtained from chiral perturbation theory. This was the approach followed in \cite{Horbatsch:2016ilr}.  If we are below the two-pion production threshold in the $t$ channel, the function is analytic. Therefore, the series is convergent and truncating the series by a polynomial is not a problem if the error of the unknown coefficients can be estimated, which is the case using the prediction from chiral perturbation theory. The first
derivative of $G_{\rm M}(Q^2)$ and 
the second derivative of $G_{\rm E}(Q^2)$ at $Q^2$=0 
(which are proportional to the magnetic and electric 
moments 
$\langle r^2 \rangle_{\rm M}$  
 and 
$\langle r^4 \rangle_{\rm E}$,
respectively)
are of particular importance 
in this extrapolation to $Q^2$=0.
The result of the fit in \cite{Horbatsch:2016ilr} can be found in Fig.~\ref{fig:rpfits}. 

It would be interesting to see how the situation changes with recent measurements. On the one hand, we have a recent measurement at MAMI using the radiative return method \cite{Mihovilovic:2019jiz}. Nevertheless, at present, the errors are pretty large. The situation is quite different with the recent measurement by the PRad collaboration \cite{Xiong:2019umf}. 
This is the first electron scattering experiment to use a magnetic-spectrometer-free method along
with a windowless hydrogen gas target, which overcomes several limitations of previous electron-proton scattering experiments and has
reached unprecedentedly small scattering angles. The value for the proton radius obtained in \cite{Xiong:2019umf} was $r_p=0.831(7)_{\rm stat}(12)_{\rm syst}$. This number was obtained using a simple Pade[1,1] rational function for $G_E^p$:
\be
G_E^p(Q^2)=\frac{1+p_1Q^2}{1+p_2Q^2}
\;,
\ee
with fit parameters $p_1$ and $p_2$.  
This determination has been criticized by Paz in \cite{Paz:2020prs} and by Horbatsch in \cite{Horbatsch:2019wdn}. In the former the statistical error is claimed to be larger. The latter mentions that the PRad data would be more appropriated to fit higher moments. Also, in  \cite{Borah:2020gte}, it is mentioned that the experiment does not provide complete uncertainty correlations and uses too small number of coefficients in z-expansion fits. It is also worth mentioning this last reference for an alternative fit of higher moments of the Sachs form factors. In this reference, the authors take the proton radius obtained from the muonic hydrogen Lamb shift as the accepted value, and determine the higher moments using this constraint in fits to elastic electron-proton scattering data. The incorporation of the value of the proton radius from muonic hydrogen in the fits seems to bring higher moments $\langle r^4 \rangle_{\rm E}$ and $\langle r^2 \rangle_{\rm M}$ (for $\langle r^4 \rangle_{\rm M}$ the errors are too large) closer to chiral perturbation theory predictions if applying the fit methods and parameterizations used in ~\cite{Borah:2020gte} to MAMI data. They agree within one sigma. See Table~\ref{tab:rn} and Fig.~\ref{fig:rM2}.

Finally, we would like to mention a series of proposed experiments that could measure the proton radius or find violations of lepton universality. One example of the latter is the measurement of lepton-pair photoproduction on the hydrogen target~\cite{Pauk:2015oaa,Heller:2018ypa,Heller:2019dyv}. Another possibility, to which not much attention has been paid, is the use of inverse kinematics experiments (see for instance \cite{Gakh:2016xby}) that would allow to reach $Q^2$ values four orders of magnitude smaller than nowadays experiments. Finally, two other experiments are going to perform measurements at lower beam energies: $30-70~\mathrm{MeV}$ with $Q^2=10^{-5}-3\times10^{-4}~\mathrm{GeV}^2$ by ProRad experiment~\cite{Hoballah:2018szw,Faus-Golfe:2017kcu} and $20-60~\mathrm{MeV}$ with $Q^2=0.0003-0.008~\mathrm{GeV}^2$ by ULQ2 experiment at Tohoku University~\cite{Suda:2018xxx}. For these experiments, it is worth emphasizing that a very high precision should be achieved as we approach $Q^2=0$, otherwise it will not be possible to measure the slope but only the absolute normalization of the proton charge. 
 
\subsection{\texorpdfstring{$r_p$}{rp}. Fits to regular hydrogen}

\begin{figure}[ht]
  \begin{center}
 \hspace*{-1cm} \includegraphics[width=1.05\textwidth]{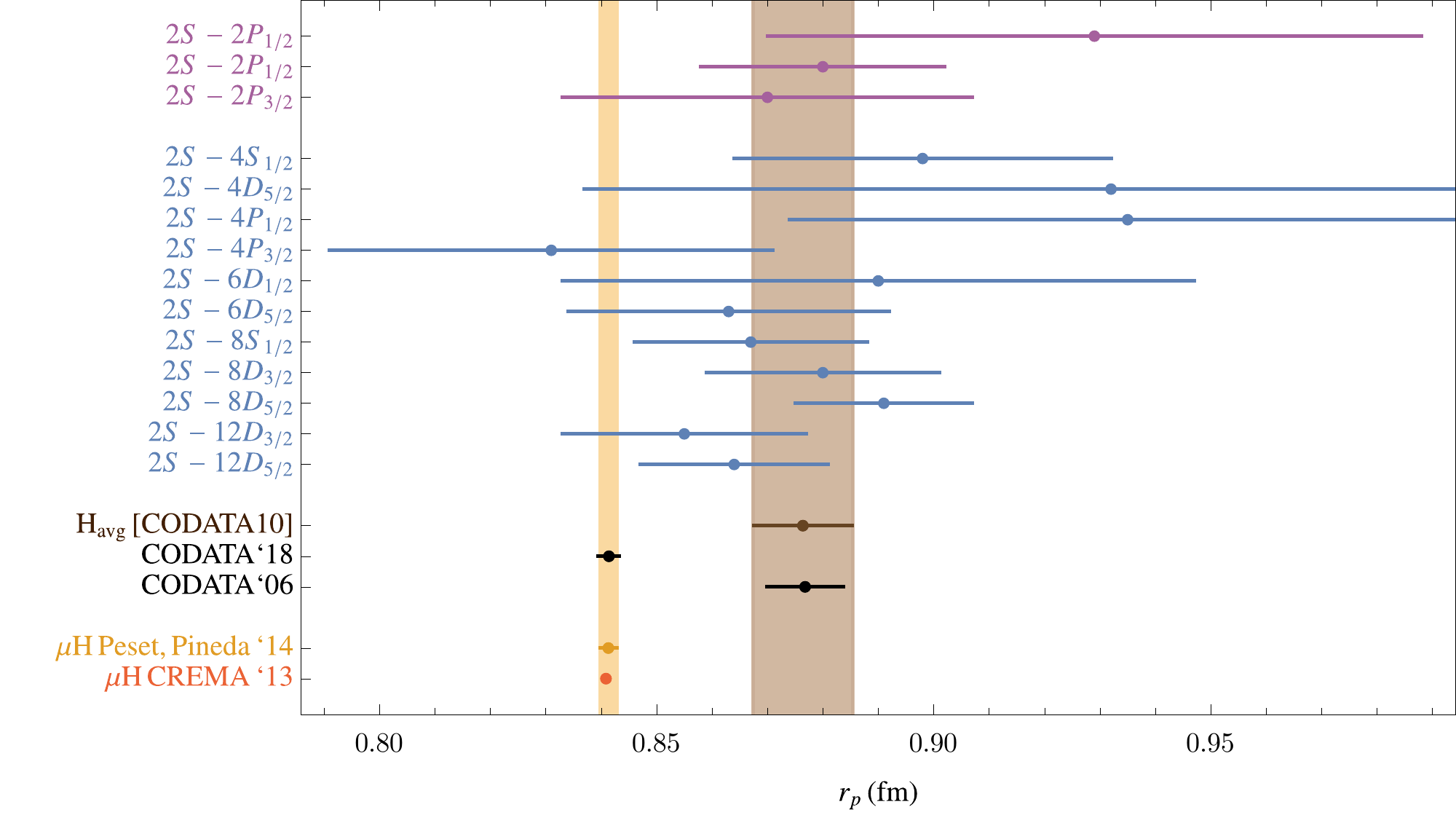}
  \end{center}
    \caption{Pre-2012 determinations of the proton radius from hydrogen and its 2012 average for the CODATA \cite{Mohr:2012tt}. From top to bottom, the first three determinations from Lamb-shift like splittings come from \cite{Newton:1979,Lundeen:1981zz,Hagley:1994zz} (in purple). The next eleven values (in blue) are obtained from combining the measurement of the 1S-2S interval \cite{Fischer:2004} with \cite{Weitz:1995zz} for the first two, \cite{Berkeland:1995} for the next two, and \cite{Bourzeix:1996}, \cite{Beauvoir:1997}, \cite{Schwob:1999} for the $n=6,8,12$ level measurements respectively. In brown the average from hydrogen spectroscopy $r_p=0.8764 (89)$ fm in \cite{Mohr:2012tt}. 
     In black the values given by the CODATA06 $r_p=0.8768 (69)$ fm \cite{Mohr:2008fa}, and CODATA18 $r_p=0.8768 (69)$ fm. We also compare to the values obtained in muonic hydrogen from an EFT analysis $r_p=0.8413 (15)$ fm \cite{ Peset:2014yha} (in orange) and the CREMA determination $r_p=0.84087 (39)$ fm \cite{Antognini:1900ns} (in red).}
  \label{fig:rphissummaryold}
\end{figure}

\begin{figure}[ht]
  \begin{center}
  \hspace*{-3cm}  \includegraphics[width=.9\textwidth]{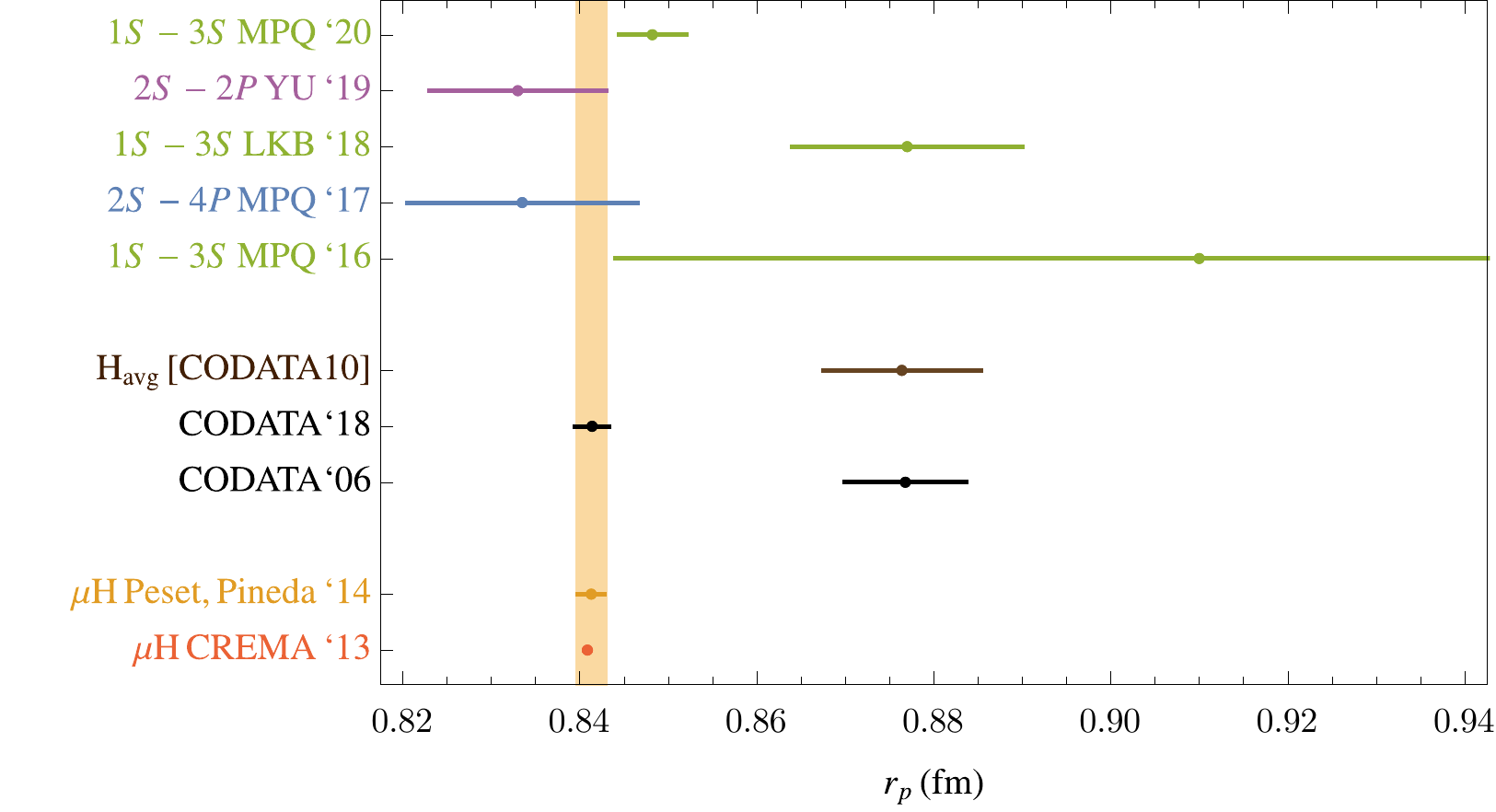}
  \end{center}
  \caption{ Summary of determinations of the proton radius from post-2013 measurements of hydrogen energy shifts. In green, measurements of the 1S-3S energy difference, in purple 2S-2P, and in blue 2S-4P. From top to bottom: 1S-3S MPQ `20 $r_p=0.8482 (38)$ fm \cite{Grinin1061}; YU `19 $r_p=0.833 (10)$ fm \cite{Bezginov:2019mdi}; LKB `18
   $r_p=0.877(13)$ fm \cite{Fleurbaey:2018fih}; MPQ `17
   $r_p=0.8335 (130)$ fm \cite{Beyer:2017gug}; MPQ `16 $r_p=0.910 (66)$ fm \cite{Yost:2016}. In brown, the average from hydrogen spectroscopy $r_p=0.8764 (89)$ fm in \cite{Mohr:2012tt}. 
     In black, the values given by the CODATA06 $r_p=0.8768 (69)$ fm \cite{Mohr:2008fa}, and CODATA18 $r_p=0.8768 (69)$ fm. We also compare to the values obtained in muonic hydrogen from an EFT analysis $r_p=0.8413 (15)$ fm \cite{ Peset:2014yha} (in orange) and the CREMA determination $r_p=0.84087 (39)$ fm \cite{Antognini:1900ns} (in red).}
  \label{fig:rpsummaryhyd}
\end{figure}

An alternative to direct determinations of the form factors from lepton-proton scattering is the use of spectroscopy, in particular of energy differences that can be measured with enough precision to be sensitive to the proton radius. The use of hydrogen spectroscopy has the advantage that $Q^2$ is very small. This line of research has been quite successful, and the accurate measurements obtained in a series of experiments have produced competitive determinations of the proton radius (see references in Figs.~\ref{fig:rphissummaryold} and \ref{fig:rpsummaryhyd}). The existing situation before the appearance of the proton radius puzzle is summarized in Fig.~\ref{fig:rphissummaryold}. Such measurements had large individual errors but after the averaging of \cite{Mohr:2012tt} the error shrank considerably. The resulting average favoured a larger value for the proton radius than the result one obtains from muonic hydrogen Lamb shift.  The situation has changed since. New measurements have been performed. We summarize the new measurements, and their comparison with the 2012 average, in Fig.~\ref{fig:rpsummaryhyd}. There are two measurements of the 1S-3S shift  by the MPQ group \cite{Grinin1061}. The second measurement overshadows the first, being considerably more precise. This measurement is particularly interesting as there is also a recent measurement of the same energy shift by the Paris group \cite{Fleurbaey:2018fih}.
These two measurements disagree with each other at the two sigma level. The measurement by the MPQ group \cite{Grinin1061} would confirm the result from muonic hydrogen (though slightly bigger by a little bit more than one standard deviation), whereas the Paris group experiment \cite{Fleurbaey:2018fih} would confirm the value obtained by the 2012 hydrogen average.\footnote{A funny comment often mentioned in conferences is that several members of the MPQ group participated in the muonic hydrogen measurement, whereas the Paris group is the most important group in the set of experiments that contributed to the 2010 hydrogen average.} This may point to a systematic problem in the set of measurements made by the Paris group. This view is strongly reinforced by the two recent and independent experiments of the 2S-4P \cite{Beyer:2017gug}, 
and 2S-2P \cite{Bezginov:2019mdi} 
energy shifts. These two measurements are in perfect accordance with 
the muonic hydrogen result and disagree with the CODATA06 value.


As we have mentioned in Sec.~\ref{Sec:pNRQED}, the EFT analysis makes explicit that the combination of hadronic effects that are measured in the Lamb shift of hydrogen or muonic hydrogen is Eq.~(\ref{Dd}). Therefore, we have a combination of the proton radius, the TPE correction, and the hadronic vacuum polarization effects. Nevertheless, the TPE correction and the hadronic vacuum polarization effects are ${\cal O}(\alpha)$ suppressed. On top of that, the hadronic vacuum polarization can be obtained from other sources with enough precision (dispersion relations). For the case of the TPE correction, the fact that it is ${\cal O}(m_{l_i})$ suppressed makes such contribution negligible for the case of regular hydrogen. This is the reason why Lamb shift measurements of regular hydrogen can then be thought of as genuine measurements of the proton radius with ${\cal O}(\alpha)$ precision (provided the Rydberg constant is known). Indeed, the influence of the value taken for the Rydberg constant in the determination of the proton radius from regular hydrogen spectroscopy should not be underestimated.  For instance, if one takes the proton radius value from the muonic hydrogen determination for granted, one can use this value to determine the Rydberg constant from the regular hydrogen 1S-2S energy difference with high precision (as it is made in \cite{PhysRevA.93.022513}). In turn, this Rydberg constant can be used in the recently measured 1S-3S energy shifts \cite{Fleurbaey:2018fih,Grinin1061}. Out of this exercise, one gets smaller values of the proton radius, closer to the muonic hydrogen result. The same outcome is obtained if one averages over all available determinations of the proton radius (as the muonic hydrogen Lamb shift result weights the most).

\subsection{\texorpdfstring{$r_p$}{rp}. Fits to muonic hydrogen Lamb shift}

At present, there have been two measurements of the muonic hydrogen Lamb shift \cite{Pohl:2010zza,Antognini:1900ns}. Both measurements have been performed by the same group and with the same experimental setup. Similarly to what happens with regular hydrogen, the EFT analysis makes explicit that the combination of hadronic effects that are measured in the Lamb shift of muonic hydrogen is Eq.~(\ref{Dd}). Therefore, we have a combination of the proton radius, the TPE correction, and the hadronic vacuum polarization effects. Again, the TPE correction and the hadronic vacuum polarization effects are ${\cal O}(\alpha)$ suppressed, and the hadronic vacuum polarization can be obtained from other sources with enough precision (dispersion relations). What is different now is that the TPE correction is much larger than in regular hydrogen. The fact that the TPE correction is of ${\cal O}(m_{\mu})$ makes this contribution not suppressed compared with hadronic contributions of the order of (or close to) the pion mass. Therefore, to use the muonic hydrogen Lamb shift for determinations of the proton radius, it is compulsory to take the TPE correction from other sources. The value $c^{\rm had}_{3}=\al^2 \frac{m_{\mu}}{m_{\pi}}54.4(3.3)$, which follows from the analysis in Ref.~\cite{Birse:2012eb}, was used in Ref.~\cite{Antognini:1900ns}. An alternative determination only using chiral perturbation theory can be found in Ref.~\cite{Peset:2014jxa}, and yields $c^{\rm had}_{3} = \al^2 \frac{m_{\mu}}{m_{\pi}}
56.7(20.6)$. An extensive discussion about the derivation of these numbers can be found in Sec.~\ref{Sec:chiralPT}. The corresponding predictions for the proton radius can be found in the Figures.

Finally, we would like to remark that if we could get the proton radius from another place, the muonic hydrogen would be an ideal place to determine $c_3^{\rm had}$, the TPE hadronic contribution. 

\subsection{\texorpdfstring{$c_4^{\rm had}$}{c4} from hydrogen and muonic hydrogen}
\label{Sec:c4had}

For completeness, we also consider determinations of $c_4^{pe}$ and $c_4^{p \mu}$. $c^{\rm had}_4$ encodes all the hadronic effects to the spin-dependent four-fermion Wilson coefficients. At present, there are no lattice predictions for these quantities. With respect to chiral perturbation theory, similarly to $c^{\rm had}_3$, this coefficient diverges in the chiral limit. Nevertheless, it only does so logarithmically (unlike in $c_3^{\rm had}$, where the divergence was linear). Such computation can be found in Ref.~\cite{Pineda:2002as}, and gives the right order of magnitude of the Wilson coefficient, around 2/3 of the experimental number if setting the renormalization scale $\nu=m_{\rho}$.\footnote{Strictly speaking one should distinguish between the electron-proton and the muon-proton case, but the contribution proportional to the logarithm of the mass of the lepton is small, even in the electron case, and does not alter significantly the estimate.} 
One cannot improve over this estimate with chiral perturbation theory alone, up to the price of introducing extra counterterms. An exception to this is the determination of the relation between the $c_4^{pe}$ and $c_4^{p \mu}$. The difference between them can indeed be determined with high accuracy using chiral perturbation theory:
\bea
\label{evsmuon}
\bar c_{4,\rm TPE}^{p\mu}
&=&\bar c_{4,\rm TPE}^{pe}+(\bar c_{4,\rm TPE}^{p\mu}-\bar c_{4,\rm TPE}^{pe})
\\
\nn
&=&\bar c_{4,\rm TPE}^{pe}+\left([c_{4,\rm TPE}^{p{\mu}}-c_{4,\rm TPE}^{p{e}}]+\al [K^{p\mu} -K^{pe}]\right)
+{\cal O}\left(\al\frac{m_{\mu}}{M}\right)
\,.
\eea
The first term can be determined from the hyperfine energy shift measurement with high precision. The constants $K^{p\mu}$ and $K^{pe}$ that appear in the above equation are unknown at present.\footnote{See \cite{Peset:2016wjq} for details.} They introduce an error of around 1\%, which is added to the error budget. One then approximates
Eq.~(\ref{evsmuon}) as
\be
c_{4,\rm TPE}^{p \mu}=c_{4,\rm TPE}^{p e}
+[c_{4,\rm TPE}^{p{\mu}}-c_{4,\rm TPE}^{p{e}}]
+{\cal O}(\al^3)
\,.
\ee
The key point is that the second term within parenthesis can be determined using chiral perturbation theory accurately.  This was done in Ref. \cite{Peset:2016wjq}, and the following result was obtained:
\be
c_{4,\rm TPE}^{p{\mu}}-c_{4,\rm TPE}^{p{e}}=\alpha^2 \, 3.68(72)
\;.
\ee

\begin{figure}[ht]
\begin{center}
\hspace*{-4.cm}\includegraphics[width=0.8\textwidth]{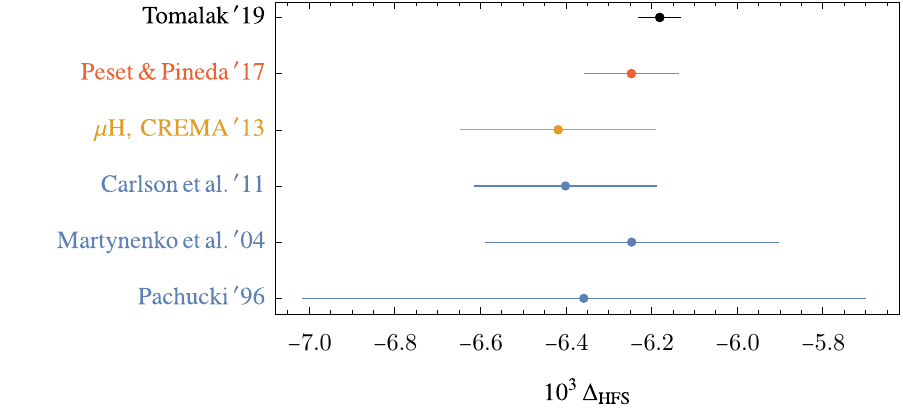}
\caption {Different predictions for the relative TPE correction to the hyperfine splitting of the 2S energy level in muonic hydrogen $\Delta_\mathrm{HFS}$. We follow historical order from bottom to top. The first three results are determinations using dispersion relations: \cite{Pachucki1,Martynenko:2004bt,Carlson:2011af}. The fourth one is the direct determination from the measurement \cite{Antognini:1900ns} of the hyperfine splitting of the muonic hydrogen as computed in Ref. \cite{Peset:2016wjq}. The fifth one is the determination from the combination of the measurement of hyperfine splitting of hydrogen and chiral perturbation theory \cite{Peset:2016wjq}. The last one is the determination from the combination of the measurement of hyperfine splitting of hydrogen and dispersion relations \cite{Tomalak:2017lxo,Tomalak:2018uhr}.}
\label{fig:hyperfine_TPE_muonic_hydrogen}
\end{center}
\end{figure}

For experiment-based determinations of these Wilson coefficients, the situation is the following. At present, lepton-proton scattering data is not precise enough to determine these Wilson coefficients, which have to be determined from spectroscopy, or using dispersion relations. For hydrogen, the direct determination from spectroscopy yields
\be
\label{c4peHFnum}
\bar c_{4,\rm TPE}^{p e}=-\alpha^2 48.69(3)
\,.
\ee
The analogous number for muonic hydrogen can be found (properly rescaled) in Fig.~\ref{fig:hyperfine_TPE_muonic_hydrogen}. This number is taken from Ref. \cite{Peset:2016wjq}.

Alternative determinations were obtained from dispersion relations in
\cite{Pachucki1,Martynenko:2004bt,Carlson:2011af} (see also \cite{Eides:2000xc,Faustov:2001pn,Carlson:2011zd,Tomalak:2017npu}). In Refs. \cite{Tomalak:2017lxo,Tomalak:2018uhr}, the author considers the difference between the hydrogen and muonic hydrogen TPE corrections to reduce uncertainties propagated from experimental measurements of the proton spin structure. Note that, the relativistic expressions were used in this case and the high-energy behavior is suppressed, which should improve the convergence, as it also does in the chiral computation. This explains the most accurate number quoted. We collect these determinations in Fig.~\ref{fig:hyperfine_TPE_muonic_hydrogen}. The dispersion relation formulas used in these references were given by~\cite{Drell:1966kk,Carlson:2008ke,Tomalak:2017owk}
 \begin{equation}
\Delta_{\mathrm{HFS}} = \frac{\alpha}{\pi^2 c_F^{(p)}} \int \limits^{\infty}_{0} \frac{\mathrm{d} Q^2}{Q^2} \int \limits^{\infty}_{\nu^{\mathrm{inel}}_{\mathrm{thr}}} \frac{\mathrm{d}\rho}{M}  \left( \frac{\left( 2 + \sigma\left(\tau_\mathrm{l}\right) \sigma\left(\tilde{\tau}\right) \right) \mathrm{Im} A_1 \left(\rho, Q^2 \right)}{  \sqrt{\tilde{\tau}} \sqrt{1+\tau_\mathrm{l}} + \sqrt{\tau_\mathrm{l}} \sqrt{1+\tilde{\tau}}   } - 3 \frac{Q^2}{M \rho}  \frac{ \sigma\left(\tau_\mathrm{l}\right) \sigma\left(\tilde{\tau} \right) \mathrm{Im} A_2 \left(\rho, Q^2 \right) }{  \sqrt{\tilde{\tau}} \sqrt{1+\tau_\mathrm{l}} + \sqrt{\tau_\mathrm{l}} \sqrt{1+\tilde{\tau}}   } \right) ,
\end{equation}
in terms of the relative correction to the Fermi hyperfine splitting,
where $\sigma \left( \tau \right) = \tau - \sqrt{\tau \left( 1 + \tau \right)}$ and Coulomb contributions from wave functions are subtracted. It can be expressed in terms of Wilson coefficients as 
\begin{equation}
\Delta_{\mathrm{HFS}} = \frac{3}{2 \pi \alpha} \frac{m_{l_i}}{M} \frac{c_4^{pl_i}}{c_F^{(p)}}.
\end{equation}

\section{Conclusions}

We have reviewed the theoretical expressions used in different experiments for the determination of the proton charge radius. The observables we have considered are the elastic electron-proton scattering, the elastic muon-proton scattering, and the Lamb shift in regular and muonic hydrogen. With the help of EFTs, we have presented a unified vision of these observables. This guarantees that the same Wilson coefficients are used for those observables. For spectroscopy, this goal is completely achieved, as we have EFTs at our disposal, see Secs. \ref{NRQED(mu)}, \ref{NRQED(e)}, and \ref{Sec:pNRQED}. These EFTs can also describe the scattering of a nonrelativistic lepton with a nonrelativistic proton. Therefore, for this specific set of observables, we can guarantee that the very same proton radius is measured in all these experiments (at least with relative ${\cal O}(\alpha)$ precision). Actually, the very same discussion applies to other (quasi-)low-energy constants that appear in these, or analogous, low-energy experiments: one can relate them with Wilson coefficients of the EFT in such a way that it can be guaranteed that we are talking about the same quantity.

The theoretical expressions for the elastic muon-proton scattering are provided in \eq{eq:dsigmaTotalmuon}. At the end of that section, the set of changes that have to be made to obtain the theoretical expression for the case of the elastic electron-proton scattering is also explained. In \eq{El3}, we present the theoretical expression for the muonic hydrogen Lamb shift $E(2P_{1/2})-E(2S_{1/2})$. For the case of regular hydrogen, there are different theoretical expressions depending on the energy difference one considers. The theoretical expressions in this case can be found in \Sec{Sec:regularhyd}. 
The hadronic quantities that appear in these expressions can be related to the Wilson coefficients of the effective theory in such a way that it is guaranteed that we are talking about the same quantity. These hadronic contributions can also be related to matrix elements of QCD (expectation values of QCD operators), which therefore, could eventually be determined from lattice simulations. In all cases, 
the relative precision (with respect to the proton radius) is of ${\cal O}(\alpha)$, and the relative theoretical error is of ${\cal O}(\alpha^2)$. This precision requires the knowledge of the hadronic TPE contribution with ${\cal O}(\alpha^2)$ precision (the leading non-vanishing contribution), at least for the case of the muon-proton sector.

The above analysis clarifies the connection between (quasi-)low-energy constants that appear in physical observables and the Wilson coefficients that appear in the EFT. By posing the problem in terms of Wilson coefficients, it is easy to realize that definitions of the proton radius, and some other quasi-low-energy constants like $\langle r^2 \rangle_M$ and $\langle r^4 \rangle_E$ (for the explicit discussion see Sec. \ref{sec:formfactors}) are ambiguous once electromagnetic corrections are switched on. To elliminate such ambiguity, it is necessary to specify the renormalization scheme and the cutoff of the effective theory. With nowadays precision, this is only relevant for the proton radius. But in the future, it could be an issue for other low-energy constants as well. 
For the particular case at hand, EFTs evidence the relevance of working in a minimal basis of operators. For the above set of experiments, it is impossible to determine, from experiment, the proton radius isolated from the TPE correction and the hadronic vacuum polarization. This is better seen in pNRQED, where these three terms combine in the Wilson coefficient of the delta potential, and therefore, they always appear in the same way in observables. 

The kinematic regime of the lepton-proton sector where we have EFTs at our disposal is limited, and cannot be applied in most elastic lepton-proton scattering experiments available at present, or in the near future. In Secs. \ref{Sec:relmuon} and \ref{Sec:ultra}, we have presented theoretical expressions for kinematic regions that are of relevance in those experiments. Our main constraint is that $Q^2 \ll m_{\pi}^2$. These expressions cannot be directly derived from EFTs, but we have tried to connect to some extent with those EFTs. In particular, we have seen that the proton radius is the same but the TPE correction is different. We have given explicit expressions for the elastic muon-proton scattering in a kinematic situation that can be of relevance for the forthcoming experiments MUSE \cite{Gilman:2017hdr} and COMPASS \cite{Denisov:2018unj} (see \eq{eq:dsigmaTotalmuon}). As far as we know,  the different pieces of this expression have never been assembled together in a single paper. By using such theoretical expressions in these experiments, it would be guaranteed that the same definition of the proton radius is used (and measured) as in the other experiments that are used for the determination of the proton radius. For the case of electron-proton scattering with a relativistic electron, the resulting expression can be found in \eq{dsigmaTotalelectron}. We have also discussed possible ways to construct EFTs that would efficiently deal with such kinematic situations in Secs. \ref{Sec:HBET}, \ref{Sec:Remuon}, \ref{Sec:boosted}, and \ref{Sec:rele}. 

After these theoretical discussions, we have reviewed determinations of the proton radius and other low-energy constants. Lattice determinations are maturing fast, but they are not yet precise enough. The precision of chiral computations is also limited because of the appearance of counterterms as one tries to increase the accuracy of such computations. This is particularly so for the proton radius where chiral perturbation theory can only predict the logarithmic dependence in the pion mass (on the other hand, for other radii chiral perturbation theory works better). These two approaches guarantee model independence. Alternatively, one can use experimental data to determine the proton radius. Several different experiments are available at present. By far, the most precise determination comes from the muonic hydrogen Lamb shift. The limit in the precision of this determination is given by the error of the spin-independent TPE correction to the Lamb shift, which we have discussed in the review. The initial tension between the determination of the proton radius from the muonic hydrogen Lamb shift with determinations from electron-proton scattering and hydrogen spectroscopy has faded over time. It soon became clear that previous determinations of the proton radius from elastic electron-proton scattering could not predict the proton radius with the claimed accuracy. It is now believed that the pre-muonic-hydrogen proton-radius determinations from electron-proton scattering did not take into account the flexibility in the possible functional form of the form factors (at least in the error analysis). 
The use of fits up to high energies could distort the extrapolation to $Q^2 \rightarrow 0$. Therefore, it is more sensible to restrict fits to the small $Q^2$ region, but then one needs to fix the higher terms in the $Q^2$ expansion by other means, like chiral perturbation theory, which, within one standard deviation, agree with the muonic hydrogen number. Alternative determinations fitting scattering data using dispersion relations also agree with the muonic hydrogen number. In this respect, in retrospect, one lesson learned from the muonic hydrogen Lamb shift result was that the errors associated with form factor parameterizations used for the CODATA08 average were underestimated.

The experimental situation of lepton-proton scattering has improved by the recent result in \cite{Xiong:2019umf} (though somewhat disputed by some authors), and more experiments are expected to come both in the electron-proton and in the muon-proton sector. These last ones would be the first muon-proton scattering experiments with percent level of accuracy, and will help check universality of the lepton interactions. 

With respect to the determinations from hydrogen spectroscopy, the accumulated new experimental data strongly favours the proton radius value obtained from muonic hydrogen Lamb shift. The only remaining problem is some hydrogen measurements by the Paris group, which are at the 3-sigma level of tension at most. In this respect, the 1S-3S energy shift is particularly compelling. We now have two measurements of this quantity by two different experimental collaborations, one by the Paris group \cite{Fleurbaey:2018fih} and the other by the MPQ collaboration \cite{Grinin1061}. This provides a venue that may help disentangling this remaining issue. Note also that the proton radius and the Rydberg constant are strongly correlated. If one takes the proton radius value from the muonic hydrogen determination for granted, one can use this value to determine the Rydberg constant from the 1S-2S energy difference with high precision. In turn, this Rydberg constant can be used in the 1S-3S energy shifts recently measured. Out of this exercise, one gets smaller values of the proton radius, closer to the muonic hydrogen result. The same outcome is obtained if one averages over all available determinations of the proton radius (as the muonic hydrogen Lamb shift result weights the most). Therefore, taking all accumulated evidence together, we cannot talk of a proton radius puzzle anymore but, at most, of a remaining tension between some experiments, which should, nevertheless, be clarified.  

Finally, it is worth emphasizing that the research activity associated with the proton radius puzzle has highlighted  the existence of a large variety of experiments that measure with high precision the different low-energy constants associated with the electromagnetic interaction of the proton. This asks for a unified theoretical setup to handle them. In this respect, we have also discussed in this review the determination of the spin-dependent TPE effects, which can be directly determined from spectroscopy. Other radii have also been briefly discussed.

\medskip
\noindent
{\bf Acknowledgments}\\
\noindent
This work was supported in part by the Spanish grants FPA2017-86989-P and SEV-2016-0588 from the ministerio de Ciencia, Innovaci\'on y Universidades, and the grant 2017SGR1069 from the Generalitat de Catalunya. This project has received funding from the European Union's Horizon 2020 research and innovation programme under grant agreement No 824093. IFAE is partially funded by the CERCA program of the Generalitat de Catalunya. The work of OT was supported by the U.S. Department of Energy, Office of Science, Office of High Energy Physics, under Awards DE-SC0019095 and DE-SC0008475. Fermilab is operated by Fermi Research Alliance, LLC under Contract No. DE-AC02-07CH11359 with the United States Department of Energy. OT acknowledges support by the Visiting Scholars Award Program of the Universities Research Association. The work of CP is supported by the DFG within the Emmy
Noether Programme under grant DY130/1-1 and by the NSFC and the DFG
through the funds provided to the
Sino-German Collaborative Research Center TRR110 ``Symmetries and the
Emergence of Structure in QCD'' (NSFC
Grant No. 12070131001, DFG Project-ID 196253076 - TRR 110).

\appendix

\section{Soft-photon emission in dimensional regularization}
\label{Sec:DR}
We describe here how to compute the soft emission contribution to the cross section using dimensional regularization and $\MS$ renormalization scheme. 
We expand \eq{treelevel} at low $Q^2$ to find ($D=4+2\epsilon$)
\begin{align}\label{TLscatteringlowQ2}
\frac{d\sigma}{d\Omega}(p\to p')&=
&=\frac{d\sigma_\text{Mott}}{d\Omega}\left[Z^2+\tau\left((c_F^{(p)})^2-Z c_{D,\MS}^{(p)}(\nu)\right)-Z^4_p\frac{4}{3}\frac{\alpha}{\pi}\frac{1}{\hat\varepsilon}+\mathcal{O}(\alpha,\tau^2)\right]
\,,
\end{align}
where $1/\hat \epsilon=1/\epsilon+\gamma_E-\ln(4\pi)$ and $ c_D^{(p)\MS}$ is the Wilson coefficient renormalized in the $\MS$ scheme, which can be found in \cite{Manohar:1997qy,Peset:2018ria}.

The second term in the RHS of \eq{sigmameasured} can be extracted in dimensional regularization and in the nonrelativistic limit from the result obtained in \cite{Gastmans:1973uv}. We find
\begin{align}\label{softemissionscattering}
\frac{d\sigma}{d\Omega}(p\to p'+\gamma(k<\Delta E))&=\frac{d\sigma_\text{Mott}}{d\Omega}Z^4\left[\frac{4}{3}\frac{\alpha}{\pi}\tau\left(\frac{1}{\hat\varepsilon}+2\ln\left(\frac{2\Delta E}{\nu}\right)-\frac{5}{3}\right)\right]
\,.
\end{align}
Putting together \eq{TLscatteringlowQ2} and \eq{softemissionscattering}, we find for the soft part of \eq{sigmameasured}
\begin{align}
\left(\frac{d\sigma}{d\Omega}\right)_\text{measured}\hspace{-.4cm}=\frac{d\sigma_\text{Mott}}{d\Omega}\left[1+\tau\left[(c_F^{(p)})^2-c_D^{(p)\MS}(\mu)+\frac{4}{3}\frac{\alpha}{\pi}\left(2\ln\left(\frac{2\Delta E}{\nu}\right)-\frac{5}{3}\right)\right]+\mathcal{O}(\alpha)+\mathcal{O}(\tau^2)\right]
\,,
\end{align}
for $Z=1$.

\bibliographystyle{JHEP}
\bibliography{bib-protonradius}

\end{document}